\documentclass[acmsmall,screen]{acmart}
\usepackage{lineno}
\usepackage[normalem]{ulem}
\usepackage{hyperref}
\usepackage{multirow}
\useunder{\uline}{\ul}{}
\usepackage{float}
\usepackage{ragged2e}
\usepackage{booktabs}
\usepackage{enumitem}
\usepackage[justification=centering]{caption}
\usepackage[utf8]{inputenc}
\usepackage{enumitem}
\usepackage{caption} 
\usepackage[most]{tcolorbox}
\usepackage{longtable}
\usepackage{tablefootnote}
\usepackage{placeins}

\AtBeginDocument{%
  \providecommand\BibTeX{{%
    Bib\TeX}}}

\setcopyright{rightsretained}
\copyrightyear{2024}
\acmYear{2024}    
\acmDOI{10.1145/3637366}

\acmJournal{PACMHCI}
\acmVolume{8}
\acmNumber{CSCW1}
\acmArticle{89}
\acmMonth{4}

\makeatletter
\gdef\@copyrightpermission{
  \begin{minipage}{0.2\columnwidth}
   \href{https://creativecommons.org/licenses/by/4.0/}{\includegraphics[width=0.90\textwidth]{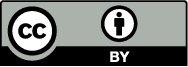}}
  \end{minipage}\hfill
  \begin{minipage}{0.8\columnwidth}
   \href{https://creativecommons.org/licenses/by/4.0/}{This work is licensed under a Creative Commons Attribution International 4.0 License.}
  \end{minipage}
  \vspace{5pt}
}
\makeatother
\begin{document}

\title{Linguistically Differentiating Acts and Recalls of Racial Microaggressions on Social Media}

\author{Uma Sushmitha Gunturi}
\orcid{0000-0002-2045-0792}
\affiliation{%
  \institution{Department of Computer Science, Virginia Tech}
  \country{USA}}
\email{umasushmitha@vt.edu}

\author{Anisha Kumar}
\orcid{0000-0002-1159-2553}
\affiliation{%
  \institution{Department of Computer Science, Virginia Tech}
  \country{USA}}
\email{anishak@vt.edu}

\author{Xiaohan Ding}
\orcid{0009-0003-2679-3344}
\affiliation{%
  \institution{Department of Computer Science, Virginia Tech}
  \country{USA}}
\email{xiaohan@vt.edu}

\author{Eugenia H. Rho}
\orcid{0000-0002-0961-4397}
\affiliation{%
  \institution{Department of Computer Science, Virginia Tech}
  \country{USA}}
\email{eugenia@vt.edu}
\renewcommand{\shortauthors}{Uma Sushmitha Gunturi, Anisha Kumar, Xiaohan Ding, \& Eugenia H. Rho}
\begin{abstract}
In this work, we examine the linguistic signature of online racial microaggressions (acts) and how it differs from that of personal narratives recalling experiences of such aggressions (recalls) by Black social media users. We manually curate and annotate a corpus of acts and recalls from \textit{in-the-wild} social media discussions, and verify labels with Black workshop participants. We leverage Natural Language Processing (NLP) and qualitative analysis on this data to classify (RQ1), interpret (RQ2), and characterize (RQ3) the language underlying acts and recalls of racial microaggressions in the context of racism in the U.S. Our findings show that neural language models (LMs) can classify acts and recalls with high accuracy (RQ1) with contextual words revealing themes that associate Blacks with objects that reify negative stereotypes (RQ2). Furthermore, overlapping linguistic signatures between acts and recalls serve functionally different purposes (RQ3), providing broader implications to the current challenges in content moderation systems on social media.
 \end{abstract}




\maketitle
\textbf{Offensive Content Warning:} This paper contains offensive language and content that may cause distress for readers.
\section{Introduction}

Experiences of interpersonal racism - whether implicit or explicit - are still a regular part of life for most Black individuals living in the United States.  While overt forms of racism against Blacks may have subsided compared to the decades past \cite{rho_class_2017}, race scholars argue that racism in modern society has not gone away; rather, it has morphed into more implicit and covert forms of expressions and subconscious acts that manifest in everyday life \cite{sue_racial_2007}. This \textit{modern racism}, which is also referred to as symbolic racism or racial microaggressions, is often “highly disguised, invisible, and takes on subtle forms that lie outside the level of conscious awareness” \cite{sue_racial_2007, williams_after_2021-1}.  In this work, we examine the linguistic signature of online racial microaggressions and how it differs from that of personal narratives recalling experiences of such aggressions shared by Black social media users. 

Racial microaggressions are subtle acts of racism that often leave the victims questioning the intent of the aggressor, as the line of offense is often blurred, not immediately recognizable, and masked through humor or seemingly harmless intentions \cite{williams_after_2021-1}. While microaggressions can occur in both on- and offline settings, our work focuses on racial microaggressions in online discussion communities. In recent years, it has been reported that an increasing number of Black individuals feel neither safe nor comfortable discussing race-related issues or experiences of racism and other potentially sensitive topics on social media \cite{guynn_2020, duggan_2020}. Despite the vocal prominence of public figures and influencers who are increasingly becoming more visible at the forefront of race-related conversations on the internet \cite{noor_2020}, about 43\% of regular Black users report feeling anxious when it comes to discussing race-related matters publicly online \cite{duggan_2020}. Fearing risks of harassment, hate speech, and invalidation of lived experiences \cite{10.1145/3479610}, some even choose to self-censor their views on racism or keep personal experiences of racism strictly private, especially on social media \cite{bar-tal_self-censorship_2017-1, bar-tal_self-censorship_2017}.  Further, it is not just backlash from others, but also content moderation and hate speech policies by social media companies that seem to stifle conversations about race and racism for Black users \cite{10.1145/3479610,kew_2018, guynn_2020}. For instance, Facebook and Nextdoor have been reported to remove posts shared by Black users regarding their personal experiences with racism either through automated filtering or through moderation \cite{vincent_2020, dwoskin_tiku_timberg_2021, allyn_2020}. Some Black users have reported being banned from the platform altogether \cite{gebel_2020} or locked out of their accounts for several hours or even days, a punishment known as being sent to ``Facebook Jail" for sharing their views or experiences on race-related topics \cite{guynn_2020}. 

Human-Computer Interaction (HCI) scholarship in content moderation has shown that both human moderators and moderation systems disproportionately and often erroneously chastise users who often post about issues related to their marginalized identities, leading to false positives in content moderation decisions \cite{10.1145/3479610, 10.1145/3290605.3300372}. At the same time, moderation systems often fail to detect race- or gender-based microaggressions that target marginalized users \cite{doi:10.1177/2053951719897945}, allowing harmful content to remain online as false negatives \cite{10.1145/2858036.2858248, Park2022MeasuringTP}. This is understandable given the significant topical and linguistic overlap between false positives and negatives, which can make it challenging for content moderation systems and human moderators to distinguish between the two. Such challenge underlies the premise of our work. In this work, we motivate the need to examine both false positives and false negatives in tandem, specifically in the context of acts and recalls of racial microaggressions. Through this work, we examine the linguistic signature of online racial microaggressions (acts) and how it differs from that of personal narratives recalling experiences of racial aggressions (recalls) shared by Black social media users, by asking the following questions:  

\begin{itemize}[noitemsep]
    \item \textbf{RQ1:} \textit{How can we leverage state-of-the-art language models to differentiate acts and recalls?}  
    \item \textbf{RQ2:}\textit{What are the similarities and differences between acts and recalls in terms of:} \\
        \textbf{a) Themes:} \textit{what themes best characterize acts vs. recalls?}\\
        \textbf{b) Contexts:} \textit{what contextual words are most predictive of acts vs. recalls?}
    \item \textbf{RQ3:} \textit{What are the similarities and differences in the linguistic signature of acts and recalls?}
\end{itemize}

We answer these questions in the context of racism in the United States from the perspective of Black individuals. Through this work, we present a manually curated corpus of acts (2000 posts) and recalls (1264 posts), which were hand-annotated and iteratively verified through a workshop with Black participants. We use techniques in natural language processing (NLP) along with in-depth qualitative analyses to examine the language underlying acts and recalls of racial microaggressions with an aim to comparatively understand the lexical patterns that differentiate the two.  Our findings show that state-of-the-art neural classifiers are able to distinguish acts and recalls with relatively high accuracy (95.4\%) while more traditional language models based on n-grams features do so less efficiently (RQ1). While acts and recalls are thematically, contextually (RQ2), and linguistically (RQ3) distinct, they also share certain themes (appearance, criminality, ability, personality, and sexual exoticism) and key linguistic signatures (use of first person pronouns and out-group language) that can make it challenging for platforms and human moderators to distinguish between acts and recalls.

Our findings represent an initial step towards better understanding semantic differences between acts and recalls of racial microaggressions on social media platforms and a re-evaluation of whether and how current socio-technical systems are able to differentiate between false positives and false negatives. Together, we aim to understand how Computer-Supported Cooperative Work (CSCW) research can best support members of marginalized groups. We argue that it is crucial to differentiate between acts (false negatives) and recalls (false positives) and utilize this knowledge to build and enhance online systems. By doing so, we can create constructive and safe online spaces where users can discuss, share, and learn from dialogues on race and racism with others. This understanding is vital in enacting inclusive and supportive online environments. Our contributions are as follows:  

\begin{itemize}[topsep=5pt]
    \item We provide an in-depth characterization of the themes and linguistic signatures that underlie acts and recalls of racial microaggressions on social media communities, which has not been empirically examined at-scale by prior research to the best of our knowledge. Our insights show both distinct and shared themes and language patterns across acts and recalls, highlighting  key challenges faced by content moderation systems and human moderators in their effort to distinguish between false positives and false negatives.
    \item Unlike the publicly available, generic off-the-shelf toxicity detectors that merely provide a score without an underlying explanation, in our comparative analysis of acts and recalls, we go beyond just classifying between the two, by leveraging interpretation techniques in deep learning (DL) to explain our classification results. Specifically, we address the lack of interpretive insight, typically associated with large pretrained language models by using Integrated Gradients \cite{Sundararajan2017AxiomaticAF} to identify contextual words that are most predictive of acts and recalls of racial microaggressions associated with Black users. By doing so, we overcome the black-box nature of DL language models by making our classification results explainable – a practice we believe is contextually crucial when working with textual corpora such as ours that contain semantically nuanced and subjective content.
    \item We complement recent efforts in the NLP community to capture and surface implicit hate speech and microaggressions online \cite{breitfeller_finding_2019, elsherief_latent_2021-1}, by providing a dataset that is more specific to the context of race and racism in the U.S. Our data is hand-annotated and validated by Black participants whose labels we used as gold-standard truths to resolve any discrepancies between non-Black annotators. We make our dataset publicly available for the wider research community in hopes that it would serve as a benchmark in examining language associated with racial biases and microaggressions.
    \item We provide insights from our workshop discussions with Black participants to further inform and validate our findings. Given their racial identity as Black individuals and the contextual familiarity with the content of our data, we corroborate our findings with the rich insights from our participants. By doing so, we aim to strive towards the goal of directly incorporating the experiences and views of the marginalized in CSCW and HCI research.
\end{itemize} 
\section{Related Work}
\subsection{Racial Microaggressions and Implicit Biases}
Microaggressions are often subconscious \cite{essed_understanding_1991, sue_racial_2007} or even unintentional \cite{nadal_impact_2014, sue_microaggressions_2020, williams_after_2021-1}, meaning that they are driven by an individual’s implicit biases toward people who are not members of one’s own in-group \cite{article-66, lepri-paper}. In the context of racial microaggressions, social psychologists argue that most people generally do not deliberately exhibit or act on racial biases all the time \cite{phd_biased_2020, 10.1145/3274342, 10.1145/3313831.3376874}. Instead, most racial biases today often take implicit forms and manifest through social conditions to which people are exposed to, or through which they interact with others \cite{phd_biased_2020}. Nonetheless, implicit racial biases can have consequential damages. For example, researchers found that teachers were more likely to discipline Black students on their first rather than second offense, implying that instructors were quicker to see so-called “patterns” of bad behavior in Black children compared to those of other races \cite{okonofua_vicious_2016, riddle_racial_2019}. In another study, people who saw images of Black families tended to associate them with poorer and less safe neighborhoods, despite how middle-class those families appeared in the pictures \cite{bonam_polluting_2016}. Even when people generally do not consciously harbor racist views, research has shown that they tend to subconsciously link criminality and primitiveness with Blackness \cite{bruner_perceptual_1957, goff_not_2008, rattan_role_2010, simons_gorillas_1999}. In essence, people’s cognition can subconsciously interact with the conditions they are exposed to when determining responses to other people, especially in the context of race \cite{riddle_racial_2019}.  Such a premise suggests that social conditions shape the nature of interactions, and this is not exclusive to offline realms.

\subsection{Racial Microaggressions in Offline vs. Online Settings}
While racial microaggressions that occur face-to-face may be difficult to respond to, people can still immediately call out the transgression as it presently occurs (or has just happened), especially when the offender is visibly identified and present. In fact, there are a countless number of guidelines providing recommendations on how to respond to racial microaggressions in various contexts \cite{doi:10.1177/2332649220933307, Marshall2021RespondingAN, to_they_2020}. However, such guidelines are specifically tailored to in-person, offline interactions, which occur in settings much different from those of online environments. According to \cite{sue_racial_2007}, an example of an offline racial microaggression would be a White person checking their wallet or clutching their bag as a Black man approaches or passes them, insinuating a sense of fear that Black people are most likely to be criminals. Online, racial microaggressions often surface on social media as posts and comments reacting to content posted by Black users (e.g., ``You're too pretty for a Black girl").

The different affordances of on- vs. offline settings in which racial microaggression occur can potentially shape how victims experience or react to such transgressions. For example, online, offenders can hide behind the anonymity afforded by throwaway accounts, which can sometimes disinhibit bad behavior \cite{andalibi_understanding_2016}. While de-identified settings are necessary and crucial in circumstances where users disclose sensitive personal experiences \cite{andalibi_understanding_2016} or exchange support in stigmatized contexts \cite{rho_class_2017}, the affordance of online anonymity can make hate speech and harassment effortless \cite{banks2010regulating, 10.1145/3287098.3287107}. Further, throughout networked publics on social media, single users can easily connect to a congregation of thousands of others, meaning the scale of interaction and exposure can be one-to-many \cite{papacharissi_social_2010, boyd2010social}. While there are advantages to such scalability \cite{dwoskin_tiku_timberg_2021, khan_social_2014, utz_informational_2016}, this also means that users sharing or recalling personal experiences of racism online, can potentially face an army of aggressors who can instantaneously pile on their post by flagging or downvoting content or trolling in the comment threads. This can lead to an uptick in engagement metrics that may trigger content moderation systems to automatically flag the post for further review or removal for potentially violating platform policies \cite{10.1145/3290605.3300372}. Hence, scaled interactions in online settings can make it difficult for users to share recalls of racial microaggressions, or even personally respond to acts without the burden of risking oneself against mob harassment \cite{10.1145/3274424, nova2021facebook}. Furthermore, unlike racial microaggressions that occur in person, online acts of racial microaggressions are rarely directed at individuals. Instead, the stereotyping language of acts often targets racial groups as a whole in generalized expressions \cite{10.1145/3148330.3152697, 10.1145/3512901, 10.1145/3290605.3300372}, making it harder for users to directly call out the aggressor’s offense based on personal grounds beyond the context of one’s race.

\subsection{Language as a Condition of Online Discourse}
Scholars in CSCW and HCI studying online discourse have shown early on that the conditions through which social interactions take place inevitably shape the nature of such interactions  \cite{rho_fostering_2018, rho_hashtag_2019, rho_political_2020}. It is therefore not just the design of platform features or algorithmic content-ranking, but also the language users regularly encounter through others that characterize the conditions of how people come across and talk about certain topics \cite{rho_political_2020}. For example, when people process contentious issues online, negatively charged affective words induce more negative conclusions in ensuing discussions \cite{rho_fostering_2018}.  As such, linguistic patterns can effectuate negativity biases toward the subject of discourse among users \cite{baumeister_bad_2001, cavazza_swearing_2014, moore-berg_empathy_2022, stieglitz_emotions_2013, utych_negative_2018}; language that characterizes online discussions on race or racism is not an exception \cite{rho_hashtag_2019, rho_political_2020}. Further, while it is important to recognize that racism expressed through language can and does take extreme and overt forms, our present study focuses primarily on the more subtle and implicit expressions of racial biases in the context of online racial microaggressions targeting Black users. We do so for several reasons. First, research has shown that racial microaggressions can profoundly impact people’s physical and mental health, self-esteem, and academic performance \cite{aw_relationship_2012, hall_its_2015, keels_psychological_2017, nadal_impact_2014}. However, the implicit nature of racial microaggressions can be camouflaged across everyday life \cite{sue_microaggressions_2020, williams_after_2021-1}, such that the impact of offense and harm is often unrecognizable or perceived as insignificant \cite{10.1145/3411764.3445590}, especially by the offenders (and bystanders), while receivers are relegated to self-doubt and distress \cite{sue2019disarming, AckermanBarger2020}. Second, while automated detection of explicit racial profanity (e.g., via customized lexicons or regular expressions) is currently possible and widely implemented across online platforms, detecting the more nuanced and inconspicuous language around racial microaggressions masked through everyday language is not yet systematically feasible at-scale \cite{stieglitz_emotions_2013,lepri-paper}. Finally, human content-moderation too can be predisposed to the moderator’s own unconscious biases and subjective understandings of racism and race-related matters, which makes drawing the line between acts and recalls of racial microaggressions difficult. As a result, racially marginalized users repeatedly face the burden of navigating and resolving situations where they are censored, locked-out, or banned from their accounts for sharing personal views or experiences of racism or race-related issues \cite{guynn_2020}. Such experiences aggregated over time can invalidate or elicit self-doubt towards the user’s lived experience as a racial minority \cite{assari_depressive_2018, haynes_three_2016}. Such issues are precisely the challenges we bring attention to and aim to address through this work.

\subsection{Understanding False Positives and False Negatives in Tandem in Content Moderation Decision-Making}
Content moderation research in CSCW and HCI has shown that  people from marginalized communities are disproportionately affected compared to other users \cite{10.1145/3479610, 10.1145/3290605.3300372, 10.1145/3411764.3445279, feuston2020conformity}. For example, trans and Black users are more likely to become victims of content moderation false positives, meaning that their comments are mis-classified and censored as harmful even when they do not violate platform policies \cite{10.1145/3479610}. Likewise, false negatives (toxic content that violate platform policies, but are left undetected) and their impact on users has been well documented in prior HCI literature \cite{10.1145/2858036.2858248, 10.1145/3526113.3545616, 10.1145/3549498}.
For example, \cite{Park2022MeasuringTP} shows that content moderators removed only one in 20 comments violating macro Reddit moderation norms in 2016, and one in 10 violating comments in 2020, highlighting that some categories of violation were more likely to slip through the cracks, leaving most anti-social behaviors unmoderated. This is because false negative content, which often includes microaggressions or implicit hate speech, are rarely explicit \cite{10.1145/3517428.3544801, doi:10.1177/2056305120975716, doi:10.1177/2332649220933307} and tend to be subtly disguised through humor \cite{Espaillat2019AnES}, insider expressions, and neologisms \cite{cscw-humour}. As a result, this makes it harder for language models, let alone even human moderators, to identify false negatives \cite{caselli-etal-2020-feel}.


Although content moderation researchers in CSCW and HCI have examined both false positives \cite{10.1145/3290605.3300372, 10.1145/3359294, 10.1145/3479610} and negatives \cite{10.1145/3549498, Park2022MeasuringTP}, as well as the impact they have on users \cite{gillespie2018custodians, feuston2020conformity}, such studies have generally examined the two discretely rather than in tandem. Our research motivates the need to examine false positives and negatives in conjunction. The significant topical and linguistic overlap between recalls and acts of racial microaggressions \cite{10.1145/3479610, 10.1145/3411764.3445279} makes it difficult for content moderation systems and human moderators to detect the two apart \cite{10.1145/3403676.3403691, 10.1145/3290605.3300372}, often leading the former to be censored as a false positive and the latter to remain on the platform as a false negative \cite{10.1145/3479610, 10.1145/3491102.3501999}. Incorrect content flagging \cite{10.1145/3313831.3376178, chandrasekharan2018internet}, or the inability of moderation systems and human moderators to differentiate when someone is critiquing racism (false positive) versus being racist (false negative) \cite{chandrasekharan2019crossmod, jhaver2019does}, can be a gate-keeping practice that can evolve into a form of digital gentrification \cite{feuston2020conformity, lingel_2019}, further exacerbating disparities between platform members and content moderators \cite{99-sarah}. Our work aims to address such challenges highlighted by prior CSCW work in content moderation by examining the intertwined relationship between false positives and false negatives, specifically in the context of race-related social media posts. In so doing, this work takes a step towards building content moderation practices that aims to distinguish between false positives (recalls of racial microaggressions) and false negatives (acts of racial microaggressions) on social media. 

\par Furthermore, most commercial toxicity models (e.g., Perspective API\footnote{https://perspectiveapi.com/}) used in social media content moderation systems, are not designed to distinguish false positives and false negatives in the first place, but to merely assign toxicity scores. As a result, content moderation algorithms that rely on these toxicity scores have difficulty differentiating online acts of racial microaggressions (false negatives) and discussions recalling experiences of racial microaggressions (false positives), often penalizing users they are supposed to protect \cite{fitzsimmons_2021, 10.1145/3185593}. We explicitly address this gap by training a language model to comparatively learn the linguistic and topical nuances that are intertwined between false positives and the false negatives.

\subsection{Lack of Explanations in Content Removal Decisions on Social Media}
One major concern highlighted by prior research examining marginalized user's experiences with online content moderation is that most  moderation systems fail to explain content removal decisions \cite{jhaver2019does, 10.1145/3403676.3403691}. Users who frequently experience content removals are often given vague (e.g., violated terms of service) or no explanations at all \cite{10.1145/3479610, lepri-paper}. As a result, they are left with very little knowledge as to what part of their language may have caused their post to be removed in the first place \cite{recast-2, doi:10.1177/1461444818773059}. Such lack of transparency and context around content removals \cite{looking_glass_tanu, 10.1145/3173574.3173677} make users feel frustrated and helpless as they are left unsure of how to subsequently engage on the platform \cite{doi:10.1177/1461444818773059}. Prior research has argued that enhancing the explainability of moderation decisions can not only empower users \cite{sundar2015toward, 10.1145/3411764.3445188}, but also moderators as well \cite{quteprints126386, 10.1145/3411764.3445092}. Moderators who rely on semi-automated crowdsourcing \cite{10.1145/3369457.3369491, Link2016AHA, 10.1145/1993574.1993599} or AI-led moderation \cite{10.1145/3491102.3501999, 10.1145/3415173} experience less cognitive burden, stress, and reduced symptoms of Post Traumatic Stress Disorder (PTSD) \cite{10.1145/3479561, Das2020FastAA,Cook2022AweVA} compared to those who do not \cite{10.1145/3411764.3445092}. However, the semantic and topical overlap between recalls and acts of racial microaggressions can make it difficult for content moderation systems and human moderators to detect the two apart, leading to false positives and negatives. Decision-making around grey content areas \cite{10.1145/3479610, 10.1145/3173574.3174014}, such as microaggressions, are also heavily subject to the moderator's personal and often limited understanding of what is and is not microaggressions \cite{10.1145/3479610}, which can often lead to arbitrary and inconsistent removal justifications \cite{ doi:10.1177/14614448221109804}. 

We address this challenge by providing a computational approach that makes classification decisions around false positives (recalls) and false negatives (acts) interpretable. In our work, we not only build a model that can distinguish between false negatives (acts) and false positives (recalls), but also provide a computational approach that helps contextualize as to what contributed most to the model's classifications decisions. Such insight can not only help moderators learn and understand the nuanced differences between acts and recalls, but also inform their decision-making as well. In our work, we use a deep learning interpretation technique to demonstrate which input feature (word) of a given post contributes most to the output decision as to why that post is most likely to be a false positive or a false negative. In so doing, we provide a scaled approach in help contextualizing and explaining model decisions in classifying between acts and recalls.

\subsection{Challenges in Detecting Toxicity Through Language Models}

CSCW and HCI scholars have demonstrated early on, both the effectiveness and limitations of prior approaches in identifying and curbing online toxicity, such as crowdsourcing \cite{10.1145/3411764.3445279, lees_capturing_2021, 10.1145/3479525}, nudging user behavior (e.g turning off comments) \cite{schneider2018digital, ali_understanding_2021}, and human moderation \cite{10.1145/3290605.3300390, 10.1145/3491102.3501999, 10.1145/3025453.3025985}. However, the challenge of countering the growing volume of toxic language has yet to be resolved \cite{10.1145/3415179, 10.1145/3415163}. This is perhaps due to the diversification of toxic language \cite{10.1145/3491101.3516502} and online hate speech that increasingly contain neologisms \cite{10.1145/3170427.3188407}, coded expressions, or subtle and indirect phases that mask harmful language \cite{grudin1988cscw}. 

Both microaggressions and hate speech are similar in that they can both stem from underlying biases and perceptions. However, they differ significantly in their manifestations. Compared to hatespeech, microaggressions, are often unintentional and usually emerge as stereotypes, micro-invalidations, and micro-insults \cite{sue_racial_2007, sue_microaggressions_2020} . By contrast, hatespeech is characterized by deliberate use of explicit language, often expressed as profanity, name-calling, bullying, or accusations \cite{10.1145/3479158, 10.1145/2858036.2858548}. Moreover, while microaggressions tend to be subtle and context sensitive, hatespeech is generally less nuanced, context-independent, and less subject to interpretation. As a result, online hatespeech is often regulated by clearer definitions and guidelines \cite{pavlopoulos-etal-2020-toxicity, waseem-etal-2017-understanding}, while there is a noticeable lack of comparable guidelines and definitions for managing online microaggressions. Consequently, detecting online microaggressions is difficult for both humans and machines, making content moderation decisions more challenging \cite{timmer_can_2021, 10.1145/3313831.3376301}. {Recent scholarship in NLP has aimed to uncover linguistic patterns in implicit hate speech by utilizing neural models}\cite{elsherief_latent_2021-1, zannettou2020measuring, breitfeller_finding_2019, 10.1145/3517428.3544801}. However, such studies use data that focus on a broad array of topics. Our work adds to this existing body of research by providing and analyzing a dataset that is more contextually targeted to the topic of race and racism in the U.S.

Furthermore, a growing body of NLP research in language and fairness has highlighted how language models unintentionally capture, reflect, or even amplify various social biases that manifest in the data they are trained on \cite{bolukbasi2016man, sap_risk_2019}. For example, linguistic models that power YouTube’s automatically generated captions tend to identify the language spoken by male and white users more efficiently than they do of female and minority users \cite{doring_male_2019}. Scholars have also demonstrated racial disparities in NLP systems by showing how widely used commercial, off-the-shelf language models fail to recognize  African-American English compared to other dialects \cite{blodgett2016demographic}. Furthermore, state-of-the-art language models pretrained on large amounts of data from the internet collected at specific points in time are susceptible to learning unintended biases towards real-world entities \cite{prabhakaran_perturbation_2019}. For example, even though the phrases, \textit{“I hate Justin Timberlake”} and \textit{“I hate Rihanna”} both express the same semantics based on identical constructions, language models tend to classify the former as significantly more toxic than the latter \cite{prabhakaran_perturbation_2019}, exhibiting gender disparity in toxicity scoring. Such issues may arise from disagreements in annotation labels, especially when it comes to annotating texts associated with gender and race-related content – a task that is immensely difficult for human annotators to reach consensus around what is considered ground truth \cite{10.1145/3411764.3445402, dumitrache2021empirical}. Throughout our analyses, we strive to be aware of such aforementioned biases. Hence, we host a workshop with Black participants through which we iteratively discuss, learn and validate our labels of acts, and use the annotations from the Black participants as the gold standard to resolve any discrepancies between non-Black annotators. 

Finally, another problem with neural classifiers is that the results from such models are difficult to understand due to their lack of interpretability \cite{ettinger2020bert, 10.1145/3313831.3376219}. Therefore, users across online platforms that rely on toxicity classifiers powered by neural networks might question how their posts were evaluated \cite{recast-2, jhaver2019does}. Therefore, as a step towards encouraging linguistic tools that can provide end-users an explanation as to why their post was (mis)classified as toxic, we interpret the classification result from our best performing language model that predicts acts apart from recalls of online microaggressions.

\section{Methods}
\subsection{Data Collection}

\begin{table}[!h]
\resizebox{0.8\columnwidth}{!}{
\begin{tabular}{|c|l|}
\hline
\textbf{Type} & \multicolumn{1}{c|}{\textbf{List of subreddits}}                                                      \\ \hline
Acts &
  \begin{tabular}[c]{@{}l@{}}askblackpeople, askScience, BlackPeopleTwitter, casualconversation, \\ circlejerk, confessions, darkjokes, darkjokeunlocked, explainlikeimfive,\\ Forwardsfromgrandma, gatekeeping, hiphopcirclejerk, insanepeoplefacebook,\\ Jokes, offensivejokes, outoftheloop, pewdiepiesubmissions, relationship\_advice,    \\ shitliberalssays, shitredditsays, shitaskscience, shittylifeprotips,\\ showerthoughts, subredditdrama, Tankiejerk, TrueOffMychest,\\ TrueUnpopularopinion, unpopularOpinion, WhitePeopleTwitter\end{tabular} \\ \hline
Recalls       & \multicolumn{1}{c|}{Blackladies, Cptsd\_bipoc, datingadvice, interracialdating, Mixedrace, TwoXChromosomes} \\ \hline
Both          & askReddit, nostupidquestions, Teenagers, TooAfraidToAsk                                                    \\ \hline
\end{tabular}%
}
\vspace{5pt}
\caption{List of Subreddits Used to Retrieve the Posts and Comments for Acts and Recalls.
}
\label{tab:table3}
\vspace{-7mm}
\end{table}

\begin{table}[!hbt]
\resizebox{0.8\columnwidth}{!}{
\begin{tabular}{|c|l|}
\hline
\textbf{Type} &
  \multicolumn{1}{c|}{\textbf{List of search keywords}} \\ \hline
Acts &
  \begin{tabular}[c]{@{}l@{}}‘African American people’,  ‘African American men’, ‘African American women’,\\ ‘African American individual’, ‘African American individuals’, ‘African American person’,\\ ‘African American man’, ‘African American woman’, ‘African American girl’,\\ ‘African American boy’, ’African American ladies’, ‘African American guy’,\\ ‘African American gal’,  ‘African American lady’, ‘African American dude’,\\ ‘African American kid’,  African American chick’, ‘African American parent’,\\  ‘African American student’, ‘black people’, ‘black boy’, ‘black girl’,\\ ‘black men’, ‘black women’, ‘black man’, ‘black woman’, ‘black person’,\\ ‘black individual’, ‘black individuals’, ‘black dude’, ‘black guy’, ‘black gal’,\\ ‘black lady’,  ‘black ladies’, ‘black chick’, ‘black kid’, ‘black parent’, ‘black student’\end{tabular} \\ \hline
Recalls &
  \begin{tabular}[c]{@{}l@{}}“Microaggressions”;  “Microaggressions, I face as a Black” + man, woman, kid, girl, gal,\\ individual,  person, guy, lady, dude, parent, student; “As a black” + man, woman, kid,\\ girl, gal, individual, person, guy,   lady, dude, parent, student;  “I’m a black” + man, woman,\\ kid, girl, gal, individual, person, guy, lady, dude, parent, student; “I’m an African American”\\ + man, woman, kid, girl, gal, individual, person, guy, lady, dude, parent, student\end{tabular} \\ \hline
\end{tabular}%
}
\vspace{8pt}
\caption{List of Search Terms Used to Collect Acts and Recalls From Reddit.}
\label{tab:table2}
\vspace{-8mm}
\end{table}
We introduce a new corpus called Recalls and Acts of Racial Microaggressions (RAMA) containing 2,000 instances of acts and 1,264 recalls of racial microaggressions from posts and comments from Reddit and Tumblr. Our data consists of acts and recalls of racial microaggressions, specifically against Black people. For Reddit, two authors first manually examined acts and recalls across posts and comments on subreddits known to contain racial microaggressions (e.g., r/showerthoughts, r/TooAfriadToAsk, r/unpopularopinion)\footnote{We identified that a large share of the acts in our data came from comments reacting to recalls on r/BlackPeopleTwitter.}. As subreddit profile pages display a list of other similar subreddits, we used this information and adopted a snowball approach to expand our list of subreddits to manually look for posts and comments containing acts and recalls of racial microaggressions targeting Black people. Upon examining approximately 150 subreddit pages, we finalized our list of subreddits to those shown in Table \ref{tab:table3}. Using these subreddits, three researchers then manually verified 300 acts and 200 recalls of racial microaggression posts and comments across a diverse array of topics. We then used these 500 posts and comments to identify an initial set of common keywords to be used in the API search as shown in Table \ref{tab:table2}. We iteratively expanded on the search keywords through multiple discussions to ensure their relevance and search strength. These keywords are similar to those used in prior work examining posts containing or recalling experiences of microaggressions \cite{breitfeller_finding_2019}. We then used these keywords to collect posts, comments using Reddit's official API PRAW \cite{Liu2021} and pushshift.io\footnote{https://pushshift.io/}. A randomized selection of posts and comments collected from the API search were then inspected by the authors and annotated and verified by workshop participants (Refer to Section 3.2).

Furthermore, we also added posts and comments from a Tumblr website\footnote{https://www.microaggressions.com/} that contains a collection of self-reported accounts of acts of microaggressions across various topics \cite{breitfeller_finding_2019}. We scraped all posts and comments topically pertaining to racial microaggressions and manually verified for acts and recalls related to Black people. Data from Tumblr were also annotated and verified by workshop participants.

\subsection{Participant Workshop: Verifying Labels }

In order to verify and annotate our RAMA corpus, we hosted workshop sessions with a total of 15 participants (6 Black and 9 Non-Black, see Table \ref{tab:table4} for demographic information). The purpose of the workshop was to obtain high-quality human labels as to whether or not a given social media post or comment contained an act of racial microaggression against a Black person/people. 
 
We sent out workshop flyers through campus mailing lists and contacted respondents through email. Participants were then invited to a 90 minute workshop session, which was held over lunch that was provided and paid for. We conducted a total of two separate, but procedurally identical sessions - one with Black and another with non-Black participants. We used the labels from the Black participants as the gold standard to resolve any discrepancies between non-Black annotators.  
We began each workshop by introducing ourselves, the project background, key research motivations behind identifying acts vs. recalls of racial microaggressions, and the take-home annotation task. We aimed to minimize participant fatigue and encourage a safe discussion environment given the difficult nature of the topic and sensitive content associated with racial microaggressions. Hence, participants were informed that they could take a break or leave at any point during the workshop and that there was no time constraints for the take-home task. In addition, participants had the opportunity to receive research credits by obtaining approval from their respective course instructors. This meant that their active participation in the workshop could be recognized and counted towards fulfilling their research credit requirements.
 
Once participants introduced themselves to one another, we used a guided PowerPoint presentation to explain and walk through multiple examples of comments and posts that contained acts and recalls of racial microaggressions across various themes (Table \ref{tab:tablea}), and invited participants to reflect on these examples and to share their perspectives.
 
Subsequently, we collectively annotated 20 carefully chosen examples, representing distinct types of microaggressions as defined by Sue et al. (2009), through in-depth group discussions. During the discussions, participants shared why they did or did not choose to classify a post/comment as an act of racial microaggression. Participants explained what part of the comment or the post/comment specifically contributed to their decision. At the end of the workshop, we invited participants to annotate approximately 300 random samples as a take-home task. There was no requirement to further participate nor a deadline imposed on the take-home tasks. All workshop material (presentation guidelines and annotation examples) were shared with the participants at the conclusion of the workshop.

All workshop sessions were audio-recorded for transcription with participants' consent. Once we received the completed take-home tasks from participants, we checked for inter-annotator agreement. Annotation agreement across all acts was substantial considering the difficulty of understanding the subtle nature of acts of microaggressions (k=0.77)  \cite{landis_application_1977}. Each of our workshop participants annotated approximately 300 posts, such that every post in our dataset received at least 3 annotations to ensure reliability and robustness in annotation verification. Labels from Black participants were used as ground truth values to resolve discrepancies among annotations from non-Black participants. [Refer to Section 3.4]

\begin{table}[!h]
\resizebox{0.7\columnwidth}{!}{
\begin{tabular}{|c|c|c|c|c|}
\hline
\textbf{ID} & \textbf{Racial-Ethnic group} & \textbf{Age} & \textbf{Gender} & \textbf{Ongoing/ Highest Degree} \\ \hline
P1  & Black/ African American     & 19 & M & BA  \\ \hline
P2  & Black/ African American     & 20 & M & BA  \\ \hline
P3  & Black/ African American     & 20 & M & BA  \\ \hline
P4  & Black/  Caribbean American & 20 & F & BA  \\ \hline
P5  & Black/  Caribbean American & 21 & F & MS  \\ \hline
P6  & Black/ Caribbean American    & 25 & M & PhD \\ \hline
P7  & South East Asian                  & 24 & M & PhD \\ \hline
P8  & South Asian                  & 24 & F & MS  \\ \hline
P9  & Middle Eastern                   & 26 & F & PhD \\ \hline
P10 & Middle Eastern                 & 25 & F & PhD \\ \hline
P11 & South Asian                  & 25 & F & MS  \\ \hline
P12 & South Asian                  & 25 & M & MS  \\ \hline
P13 & Asian American     & 26 & F & MS  \\ \hline
P14 & South Asian      & 21 & F & MS  \\ \hline
P15 & South Asian      & 25 & F & BA  \\ \hline
\end{tabular}%
}
\vspace{5pt}
\caption{Participant’s Demographic Data.}
\label{tab:table4}
\vspace{-6mm}
\end{table}

\subsection{Analysis}
\setlength{\parindent}{0pt}
\textit{\textbf{RQ1:} How can we leverage state-of-the-art language models to differentiate acts and recalls?}

To answer RQ1, we initially tested traditional ML-based classifiers known to work well with small amounts of data. To further understand the challenges of implicit hate detection and achieve compositional understanding beyond simple keyword-matching, we fine-tuned large language models (LLMs) such as BERT, RoBERTa and XLNet to distinguish acts and recalls of racial microaggressions. First, using a 80-20 split on the RAMA corpus,  we balanced the distribution of the target class in our data in both train (80\%, N=1680) and test (20\%, N=420) sets. For the traditional ML models, we used the Naive Bayes (NB), Support Vector Machine (SVM) and Logistic Regression (LR) with standard unigrams, Term Frequency-Inverse Document Frequency (TF-IDF), and GloVe embedding (Pennington et al., 2014) features. We used scikit-learn’s \textit{feature\_extraction} attribute to extract features and \textit{CountVectorizer} module with a (default) minimum word frequency $=2, (2,2)$ n-gram range. We used k-fold cross validation with $k=5$ to avoid overfitting of the models. For neural models, we fine-tuned BERT, RoBERTa and XLNet and set the batch size to 16 with 8 training epochs and used AdamW for optimization with 2e-05 learning rate.  All baseline ML models were implemented using sklearn\footnote{https://scikit-learn.org/} and LLMs using PyTorch\footnote{https://pytorch.org/}.

\vspace{3mm}
\textit{\textbf{RQ2:} What are the similarities and differences between acts and recalls in terms of:}
\begin{itemize}
    \item [a)] \textbf{Contexts:} \textit{what contextual words are most predictive of acts vs. recalls?}
    \item [b)] \textbf{Themes:} \textit{what themes best characterize acts vs. recalls?}
\end{itemize}

\paragraph{Identification of Influential Tokens using Integrated Gradients (IG)}
While state-of-the-art neural models are effective at high-level hate speech classification, they are not effective at spelling out more fine-grained categories with detailed explanations of the implied message \cite{10.1145/3403676.3403691}. First, to address this, we employ Integrated Gradients (IG) \cite{Sundararajan2017AxiomaticAF}, a model interpretability technique in deep learning (DL) that helps identify key input features that contribute most to the model’s predictive decision \cite{madsen_post-hoc_2022} and are calculated by computing the gradient of the model’s prediction output to its input features. IG (\ref{equation1}) provides intuitive explanation for output decisions from transformer based models like BERT that often lack interpretive insight \cite{chen_toward_2021, ig_tv}. We used the best performing classification model in RQ1 along with IG to derive words predictive of the model’s decision (RQ2a). In IG, a feature’s contribution to the output of a neural classifier is calculated by considering the gradient of the model prediction with respect to that of the input feature. Integrated Gradients calculate the average of gradients at all points along a straight line from the baseline ${x}'$, which is set to zero vector for text-based models to input $x$ \cite{chen_toward_2021}. Formally, if: $R_n \rightarrow [0, 1]$ represents BERT, then the integrated gradient of the i-th dimension is:

\begin{equation}
\text { Integrated Gradients }(x ; F)=\left(x-x_i^{\prime}\right) x \int_{\Theta=0}^1 \partial F\left(x^{\prime}+\Theta\left(x-x^{\prime}\right)\right) \div \partial x^{\prime} d \Theta
\label{equation1}
\end{equation}

The final attribution score of a particular word is the sum of the integrated gradients for each dimension of that word’s embedded vector \cite{du_towards_2021}.


\paragraph{Uncovering Themes and Contexts using Qualitative Analysis}
Once we identified all the tokens that were most predictive of acts vs. recalls, we manually examined every occurrence of each IG token and its contextual use across the entire dataset of acts (2,000) and recalls (1,264) – this resulted in a total of 490 acts and 626 recalls that contained at least one of the top 40 tokens that were most predictive of their respective class. We then manually examined every single instance of these posts (490 acts, 626 recalls) by examining the themes in which these tokens most frequently occurred in context. This process enabled us to identify key themes and contexts within the posts, offering thematic insights into the nature of acts and recalls of racial microaggressions.

\vspace{3mm}
\textit{\textbf{RQ3:} What are the similarities and differences between the linguistic signature of acts and recalls?}
\setlength{\parindent}{8pt}
The methodological strength and nature of Integrated Gradients (IG) is such that it identifies words that contribute most to why a post is classified as one class versus another, by focusing on factors that differentiate the two classes. Hence, as with most deep learning (DL) interpretation techniques used in explaining binary classification decisions, IG may overlook overlapping characteristics shared between the two classes.  Furthermore, IG as with most NLP methods, involves preprocessing of textual data (e.g., lemmatization) which may bypass detecting verb tenses or other linguistic aspects that might be characteristic of acts and recalls. Hence, to ensure a more robust and comprehensive examination of acts and recalls in RQ3, we overcome this limitation by analyzing a randomly selected set of posts from our entire data– not just those that included IG tokens, but those that did not as well.

To analyze the rhetorical styles of acts and recalls, we followed an iterative open coding procedure. We first approached the transcribed text through axial \cite{Corbin2008BasicsOQ}, thematic coding \cite{Gibbs2007AnalyzingQD} and then performed discourse analysis \cite{jimgee}. Two of the authors independently coded a test sample of 100 randomly selected acts and recalls drawn from the larger racial microaggressions dataset, then discussed each post together with assigned codes to establish a shared set of categories for acts and recalls respectively. The authors then coded another 100 randomly selected recalls and similarly discussed them one by one in detail. Applying the resulting codebook, the authors  coded a final set of randomly selected 100 acts and recalls.

To analyze linguistic patterns across acts and recalls, we then proceeded to perform discourse analysis.  We used discourse analysis given its methodological focus on understanding how people contextually use language to convey underlying intent \cite{jimgee}. Discourse analysis has also been widely used across CSCW and HCI scholarship examining social media comments \cite{10.1093/fampra/cmp038, rho_fostering_2018, rho_political_2020}. Using this method, we identified  differences and similarities in the linguistic patterns between acts and recalls with regards to how posters conveyed underlying intent. Further, we validated our qualitative findings with the help of workshop discussions with the Black participants.

\subsection{Understanding Participant Rationale for Labeling Posts as Acts of Microaggressions}
\setlength{\parindent}{5pt}
In this section, we highlight key instances where Black and non-Black participants differed in labeling posts as racial microaggressions, aiming to elucidate their respective reasoning behind such annotation decisions. Black participants, informed by their personal encounters with racial bias, exhibited heightened sensitivity to language nuances in the posts, often missed by non-Black participants. Black participants adeptly identified posts perpetuating harmful stereotypes, discerned underlying intent, and identified subtle racial microaggressions, elements often missed by non-Black participants. Non-Black participants' labeling errors predominantly originated from a lack of awareness regarding these stereotypes and a mistaken sense of color neutrality, leading them to overlook microaggressions subtly embedded in seemingly innocuous questions or statistical assertions.

\paragraph{\textbf{Implicit Bias}} For example, all Black participants labeled the post, \textit{"Why are all Black people so good at sports? They must be naturally athletic"} as an act. They observed that the language used in such posts subtly reinforced implicit biases about Black people. By contrast, two non-Black participants did not label this post as an act due to the seemingly positive sentiment attached to phrases like \textit{"naturally athletic"}.  According to Black participants, while such expressions might initially present positive stereotypes, they carry harmful implications. They argued that such statements can diminish the individuality and diversity of Black people, attributing their achievements in sports solely to inherent physical abilities rather than hard work, dedication, and skill. Black participants labeled such posts under the racial microaggresion themes, such as \textit{"Assumptions about intelligence, competence, or status" }and \textit{"Pathologizing minority culture or appearance"} as shown in \ref{appendix}.

\paragraph{\textbf{Lack of Familiarity with Stereotypes}}
Black participants displayed more profound knowledge of stereotypes than their non-Black counterparts, as exemplified in several annotation discrepancies. Notably, non-Black participants overlooked the microaggressions embedded in acts like \textit{"Do black people find it hard to swim?"} and \textit{"Why do Black people like watermelon?"} which subtly perpetuate stereotypes about Black individuals that were immediately recognizable to our Black participants. This disparity in awareness underlines the essential role of cultural insight and firsthand lived experiences in identifying and understanding microaggressions associated with one's social group.

\paragraph{\textbf{Failure to recognize perceived intent or broader implications}} 
On certain occasions, non-Black participants fell short in understanding the underlying intent of posts or the broader implications they carried, aspects that Black participants consistently discerned. For example, with the post, \textit{"Do Black people even tan?"}, non-Black participants saw it as an innocent question, likely stemming from curiosity or a lack of knowledge. Conversely, Black participants questioned the poster's underlying intent, recognizing that the question seemed to subtly propagate belittling stereotypes through curiosity and humor.

This pattern was also evident when participants encountered the post - \textit{"90\% of violent outbreaks were caused by Black people"}. Three non-Black participants  interpreted this as a mere statistical observation, failing to consider the broader implications. In contrast, all Black participants pointed out the potential harm of such assertions, underscoring that statistics can be misleading and that the intent behind their use is critically important. As one Black participant  noted, \textit{"Definitely heard that one before, it is trying to use statistics to conclude that black people are dangerous; however, there is a lot more to that statistic because the original intent of the police was to monitor minority groups. If the intent is to monitor minority groups, it is almost obvious that this statistic is going to be there"} -P2. Another, Black participant stated, \textit{"The context matters - who is saying it..."} -P4, arguing that the identity of the speaker—their race, relationship to the listener, etc.—and the context in which the statement is made—whether in jest or during a serious conversation—can significantly sway its interpretation as a microaggression.

\paragraph{\textbf{False Colorblindness}} In some instances, non-Black participants failed to identify acts due to the absence of explicit racial references.  In contrast, Black participants, acknowledging the role of race as an inseparable component of their experiences,  leveraged this understanding to detect even subtly embedded microaggressions in posts.  This difference was particularly evident in labeling posts like \textit{"I don't care if they are black, gay, green, alien, or inanimate objects."} While several non-Black participants perceived this as an attempt to equate all individuals irrespective of their race,  Black participants highlighted the problem of such seemingly neutral statements. For example, one Black participant stated,  \textit{"They are trying to see everyone the same, but race plays a factor in our lives - cut and dry. Just saying you don't see color doesn't change the fact that there is systemic racism. It is more important to acknowledge the oppression that people face."} -P5.

\paragraph{\textbf{Recognizing Hedging Language as a Red Flag}} The use of hedging expressions like \textit{"Not to sound racist"} or \textit{"I don't mean to be racist"} was adeptly identified by Black participants as a potential marker for microaggressions. Such phrases, rather than mitigating the impact of a poster's statements, served as warnings to Black participants, hinting at microaggressions. One illustrative example is the post: \textit{"Why do some Black people have large yellow eyes? Ok, I know this probably sounds really racist, but I don't mean to be one."} Black participants classified this post as an act, recognizing that the hedging phrases did not negate or neutralize the racially charged comments that followed. This understanding allowed Black participants to navigate microaggressions more adeptly than non-Black participants. As one participant stated, \textit{"Just because you say 'I don't mean to be such and such' does not automatically absolve you of what you said before or after"} -P6. This perception, however, was often missed by non-Black participants, emphasizing the importance of sensitivity and awareness in detecting and understanding microaggressions.

\section{Findings}
\setlength{\parindent}{8pt}
In this section, we first provide our findings related to classifier performance for differentiating acts and recalls of racial microaggressions (RQ1). Further, our findings revealed higher level differences between acts and recalls across various themes identified via Integrated Gradients (RQ2), as well as more granular differences in their linguistic signatures via discourse analysis and focus group interactions (RQ3).

\subsection{RQ1: Classifying Acts vs. Recalls}  
To answer RQ1, we first built natural language based classification models to classify acts vs. recalls of racial microaggressions. We experimented with several traditional models such as Support Vector Machine (SVM), Naive Bayes, and Logistic Regression as baseline classifiers along with three different feature extraction methods (GloVe, TF-IDF, n-grams), as well as state-of-the-art neural models such as XLNet, BERT, and RoBERTa, to classify a user post/comment into an act or recall. We observed that RoBERTa and XLNet achieved a high level of performance, with accuracies at $0.954$ and $0.917$, respectively; in contrast, SVM and BERT had far lower accuracy scores at $0.760$ and $0.906$, respectively. The modeling results are presented in Table \ref{tab:table5}.

\begin{table}[]
\begin{tabular}{llcccc}
\hline
\multicolumn{2}{c}{}                          & \multicolumn{4}{c}{Binary Classification Result} \\\hline
\multicolumn{2}{c}{Model}                     & Precision    & Recall    & F-1      & Accuracy   \\\hline
\multicolumn{2}{l}{SVM (n-grams)}             & 0.752        & 0.747     & 0.749    & 0.716      \\
\multicolumn{2}{l}{SVM (TF-IDF)}              & 0.639        & 0.653     & 0.645    & 0.618      \\
\multicolumn{2}{l}{SVM (GloVe)}               & 0.709        & 0.735     & 0.721    & 0.708      \\
\multicolumn{2}{l}{Naive Bayes (n-grams)}     & 0.717        & 0.721     & 0.718    & 0.722      \\
\multicolumn{2}{l}{Naive Bayes (TF-IDF)}      & 0.622        & 0.608     & 0.614    & 0.621      \\
\multicolumn{2}{l}{Naive Bayes (GloVe)}       & 0.667        & 0.525     & 0.587    & 0.545      \\
\multicolumn{2}{l}{Log. Regression (n-grams)} & 0.633        & 0.692     & 0.661    & 0.693      \\
\multicolumn{2}{l}{Log. Regression (TF-IDF)}  & 0.769        & 0.724     & 0.746    & 0.719      \\
\multicolumn{2}{l}{Log. Regression (GloVe)}   & 0.719        & 0.72      & 0.719    & 0.733     \\
\multicolumn{2}{l}{BERT}   & 0.864        & 0.925      & 0.893    & 0.906     \\
\multicolumn{2}{l}{XLNET}   & 0.891        & 0.968      & 0.927    & 0.917     \\
\multicolumn{2}{l}{RoBERTa}   & \textbf{0.921}        & \textbf{0.950}      & \textbf{0.934}    & \textbf{0.954}     \\\hline
\vspace{-5pt}
\end{tabular}
\caption{Classification Metrics for Acts vs. Recalls (Best Performance Is Bolded)}
\label{tab:table5}
\vspace{-4mm}
\end{table}

\subsection{RQ2: Interpreting Acts vs. Recalls}

To explain the  predictive decisions of our best-performing neural language classifier, we leverage Integrated Gradients (IG), a model interpretability technique, to extract key words that the model considered most predictive of acts vs. recalls. We use IG to identify specific tokens that contribute most to the model’s predictive decision by computing attribution scores for each of the tokens. Attribution scores indicate how much a specific token correlates to the model's prediction judgment. We categorized the top 40 tokens with the highest and lowest attribution scores ranging from positive (predictive of recalls) to negative (predictive of acts) values by each class (act/recall), with the magnitude indicating the predictive strength for each class. We then manually examined every occurrence of each IG token and its contextual use across the entire dataset of acts and recalls and identified prevalent themes in which the top 40 tokens most frequently occurred in context. Table \ref{table:tableig} details the themes that emerged in the acts and recalls respectively. In total, we identified three themes unique to acts \textit{("Questions", "Ethnicity", "Evolution" and "Human Race")}, four themes associated with recalls \textit{("Relationships", "Workplace", "Everyday Life", "Geographical Location")}, and five overlapping themes\textit{("Appearance", "Criminality", "Ability", "Personality", "Sexual Exoticism")}. Tokens are listed in the descending order of their predictive strength and grouped by salient themes based on the most common contexts in which these tokens appeared across sentences.


\subsubsection{Themes of Acts}\paragraph{\textbf{Questions}} 

Among posters of acts of racial microaggression, the tokens, \textit{why} and \textit{question,} are often used to pose a question along the lines of why a certain stereotype exists.  Consider this example of an act: \textit{“Why do black men have such dry hands?...I have just realized that while making this question that I’ve never [shook] the hand of a black woman.”} The poster asks a question (\textit{‘Why do black men have such dry hands?’}) and then proceeds to acknowledge that they posed a question.  By questioning the validity of their own question while writing the post, the poster appears to be thinking out loud, thereby not filtering their thoughts.

\paragraph{\textbf{Ethnicity}} 
The tokens \textit{ethnic, indian, hispanic,} and \textit{latino} are often used by posters of acts to compare Black people to other races, especially racial and ethnic minorities, such as Indian and Hispanic people. For example, the poster of the following act creates a hierarchy of intelligence based on race in order to highlight their belief that Black people are less intelligent than other people, using different racial majorities (\textit{‘caucasians’}) as well as racial and ethnic minorities (\textit{‘asians,’, ‘indians,’ ‘latinos/hispanics’}) to elucidate their point: 
\begin{center}
\textit{“Some say that you can order races on their intelligence with asians on top, indians and caucasians a little below, latinos/hispanics about 1/4 std. deviation below whites, and blacks about 2/3-1 standard deviation below whites.”}
\end{center}

\paragraph{\textbf{Evolution and Human Race}}
Tokens such as \textit{earth} and \textit{evolution} are frequently used by posters of acts to discuss where Black people came from as well as what led them to develop their distinct appearance and abilities.  In addition to using the token \textit{earth} to  imply a point of origin, the poster also uses the word to casually place Black people outside of the human race: \textit{“If Adam and Eve are the first people in the Earth and they are white, why are there Black people?”}  Moreover, the following act uses the word \textit{evolution} as a way to justify why Black people are faster than other races by connecting a common stereotype to evolution: \textit{“Black people are faster because of evolution.”}

\subsubsection{Themes of Recalls}
\paragraph{\textbf{Relationships}}
Posters of recalls commonly use the word \textit{friend} to describe the perpetrator of an act of racial microaggression or the person that experiences a racial microaggression alongside them.  The poster of this recall uses the word \textit{friend} to describe the person that they were with when they both experienced a racial microaggression: \textit{“I’m Dominican-American and one day me and my friend who’s Bengali went to the mall. We walked into a MAC store and a White lady approached us and asked us: Where are you guys from? You guys look exotic.”}  Our findings suggest that this \textit{friend} is often times Black or another racial minority, such as Bengali.  On the other hand, the word friend is also often used to describe the perpetrator of an act of racial microaggression: \textit{“I had a white male friend of mine in high school tell me that I’m the darkest he would go.”} Our findings also suggest that the token, \textit{husband}, is often used to describe the person that the victim of the racial microaggression is compared to by the perpetrator of the act: \textit{“After expressing his shock upon seeing my husband’s last name (whose family is German) he said, “Oh, so you’ve got a good Jewish boy, huh?” You must feel lucky!”}

\paragraph{\textbf{Workplace}}
Recalls often contain tokens such as \textit{job} and \textit{interviews} to describe acts of racial microaggression that occur in the workplace: \textit{“At job interviews, I tell them where I’m from, born and raised in the Dominican Republic, and they say “Oooh” with a tone of disappointment.”}  Here, the words \textit{job} and \textit{interviews} serve to specify where in particular the poster has experienced a racial microaggression.
  \begin{table}[!h]
    \begin{minipage}{.49\textwidth}
      \centering
\resizebox{1.00\columnwidth}{!}{
\begin{tabular}{|cclcc|}
\hline
\multicolumn{5}{|c|}{\textbf{Attributed to Acts}}                                                                                                                                           \\ \hline
\multicolumn{1}{|c|}{\textbf{Themes}}                                                                  & \multicolumn{2}{c|}{\textbf{Token}}        & \multicolumn{1}{c|}{\textbf{Attribution Score}} & \textbf{Freq.} \\ \hline
\multicolumn{1}{|c|}{\multirow{2}{*}{\begin{tabular}[c]{@{}c@{}}Questions (247)\end{tabular}}} & \multicolumn{2}{c|}{question}     & \multicolumn{1}{c|}{-0.077}            & 29    \\ \cline{2-5} 
\multicolumn{1}{|c|}{}                                                                        & \multicolumn{2}{c|}{why}          & \multicolumn{1}{c|}{-0.169}            & 218   \\ \hline
\multicolumn{1}{|c|}{\multirow{4}{*}{Ethnicity (23)}}                                              & \multicolumn{2}{c|}{ethnic}       & \multicolumn{1}{c|}{-0.0147}           & 9     \\ \cline{2-5} 
\multicolumn{1}{|c|}{}                                                                        & \multicolumn{2}{c|}{indian}       & \multicolumn{1}{c|}{-0.020}            & 11    \\ \cline{2-5} 
\multicolumn{1}{|c|}{}                                                                        & \multicolumn{2}{c|}{hispanic}     & \multicolumn{1}{c|}{-0.029}            & 6     \\ \cline{2-5} 
\multicolumn{1}{|c|}{}                                                                        & \multicolumn{2}{c|}{latino}       & \multicolumn{1}{c|}{-0.044}            & 3     \\ \hline
\multicolumn{1}{|c|}{\multirow{7}{*}{Evolution and Human Race (22)}}                               & \multicolumn{2}{c|}{earth}        & \multicolumn{1}{c|}{-0.009}            & 2     \\ \cline{2-5} 
\multicolumn{1}{|c|}{}                                                                        & \multicolumn{2}{c|}{existence}    & \multicolumn{1}{c|}{-0.017}            & 5     \\ \cline{2-5} 
\multicolumn{1}{|c|}{}                                                                        & \multicolumn{2}{c|}{evolution}    & \multicolumn{1}{c|}{-0.017}            & 2     \\ \cline{2-5} 
\multicolumn{1}{|c|}{}                                                                        & \multicolumn{2}{c|}{population}   & \multicolumn{1}{c|}{-0.017}            & 6     \\ \cline{2-5} 
\multicolumn{1}{|c|}{}                                                                        & \multicolumn{2}{c|}{civilization} & \multicolumn{1}{c|}{-0.038}            & 3     \\ \cline{2-5} 
\multicolumn{1}{|c|}{}                                                                        & \multicolumn{2}{c|}{humans}       & \multicolumn{1}{c|}{-0.066}            & 2     \\ \cline{2-5} 
\multicolumn{1}{|c|}{}                                                                        & \multicolumn{2}{c|}{primates}     & \multicolumn{1}{c|}{-0.015}            & 2     \\ \hline
\multicolumn{1}{|c|}{\multirow{5}{*}{Appearance (42)}}                                             & \multicolumn{2}{c|}{fat}          & \multicolumn{1}{c|}{-0.004}            & 2     \\ \cline{2-5} 
\multicolumn{1}{|c|}{}                                                                        & \multicolumn{2}{c|}{monkeys}      & \multicolumn{1}{c|}{-0.006}            & 2     \\ \cline{2-5} 
\multicolumn{1}{|c|}{}                                                                        & \multicolumn{2}{c|}{palms}        & \multicolumn{1}{c|}{-0.007}            & 6     \\ \cline{2-5} 
\multicolumn{1}{|c|}{}                                                                        & \multicolumn{2}{c|}{gorilla}      & \multicolumn{1}{c|}{-0.009}            & 4     \\ \cline{2-5} 
\multicolumn{1}{|c|}{}                                                                        & \multicolumn{2}{c|}{skinned}      & \multicolumn{1}{c|}{-0.024}            & 28    \\ \hline
\multicolumn{1}{|c|}{\multirow{7}{*}{Criminality (52)}}                                            & \multicolumn{2}{c|}{police}       & \multicolumn{1}{c|}{-0.003}            & 6     \\ \cline{2-5} 
\multicolumn{1}{|c|}{}                                                                        & \multicolumn{2}{c|}{robbed}       & \multicolumn{1}{c|}{-0.014}            & 2     \\ \cline{2-5} 
\multicolumn{1}{|c|}{}                                                                        & \multicolumn{2}{c|}{dangerous}    & \multicolumn{1}{c|}{-0.014}            & 11    \\ \cline{2-5} 
\multicolumn{1}{|c|}{}                                                                        & \multicolumn{2}{c|}{commit}       & \multicolumn{1}{c|}{-0.016}            & 14    \\ \cline{2-5} 
\multicolumn{1}{|c|}{}                                                                        & \multicolumn{2}{c|}{violent}      & \multicolumn{1}{c|}{-0.023}            & 11    \\ \cline{2-5} 
\multicolumn{1}{|c|}{}                                                                        & \multicolumn{2}{c|}{streets}      & \multicolumn{1}{c|}{-0.056}            & 5     \\ \cline{2-5} 
\multicolumn{1}{|c|}{}                                                                        & \multicolumn{2}{c|}{attacking}    & \multicolumn{1}{c|}{-0.058}            & 3     \\ \hline
\multicolumn{1}{|c|}{\multirow{3}{*}{Ability (17)}}                                                & \multicolumn{2}{c|}{sports}       & \multicolumn{1}{c|}{-0.003}            & 7     \\ \cline{2-5} 
\multicolumn{1}{|c|}{}                                                                        & \multicolumn{2}{c|}{intelligent}  & \multicolumn{1}{c|}{-0.020}            & 3     \\ \cline{2-5} 
\multicolumn{1}{|c|}{}                                                                        & \multicolumn{2}{c|}{IQ}           & \multicolumn{1}{c|}{-0.048}            & 7     \\ \hline
\multicolumn{1}{|c|}{\multirow{8}{*}{Personality (54)}}                                            & \multicolumn{2}{c|}{ignorance}    & \multicolumn{1}{c|}{-1.300}            & 5     \\ \cline{2-5} 
\multicolumn{1}{|c|}{}                                                                        & \multicolumn{2}{c|}{scary}        & \multicolumn{1}{c|}{-0.001}            & 3     \\ \cline{2-5} 
\multicolumn{1}{|c|}{}                                                                        & \multicolumn{2}{c|}{names}        & \multicolumn{1}{c|}{-0.004}            & 15    \\ \cline{2-5} 
\multicolumn{1}{|c|}{}                                                                        & \multicolumn{2}{c|}{dumb}         & \multicolumn{1}{c|}{-0.004}            & 5     \\ \cline{2-5} 
\multicolumn{1}{|c|}{}                                                                        & \multicolumn{2}{c|}{disgusting}   & \multicolumn{1}{c|}{-0.004}            & 2     \\ \cline{2-5} 
\multicolumn{1}{|c|}{}                                                                        & \multicolumn{2}{c|}{funny}        & \multicolumn{1}{c|}{-0.007}            & 6     \\ \cline{2-5} 
\multicolumn{1}{|c|}{}                                                                        & \multicolumn{2}{c|}{ghetto}       & \multicolumn{1}{c|}{-0.009}            & 7     \\ \cline{2-5} 
\multicolumn{1}{|c|}{}                                                                        & \multicolumn{2}{c|}{loud}         & \multicolumn{1}{c|}{-0.018}            & 11    \\ \hline
\multicolumn{1}{|c|}{\multirow{4}{*}{Sexual Exoticism (33)}}                                       & \multicolumn{2}{c|}{sex}          & \multicolumn{1}{c|}{-0.017}            & 10    \\ \cline{2-5} 
\multicolumn{1}{|c|}{}                                                                        & \multicolumn{2}{c|}{attracted}    & \multicolumn{1}{c|}{-0.037}            & 12    \\ \cline{2-5} 
\multicolumn{1}{|c|}{}                                                                        & \multicolumn{2}{c|}{hotter}       & \multicolumn{1}{c|}{-0.005}            & 7     \\ \cline{2-5} 
\multicolumn{1}{|c|}{}                                                                        & \multicolumn{2}{c|}{sexual}       & \multicolumn{1}{c|}{-0.065}            & 4     \\ \hline
\end{tabular}}
    \end{minipage}
    \begin{minipage}{.49\textwidth}
      \centering
\resizebox{0.975\columnwidth}{!}{
\begin{tabular}{|cclcc|}
\hline
\multicolumn{5}{|c|}{\textbf{Attributed to Recalls}}                                                                                                  \\ \hline
\multicolumn{1}{|c|}{\textbf{Themes}}                            & \multicolumn{2}{c|}{\textbf{Token}}        & \multicolumn{1}{c|}{\textbf{Attribution Score}} & \textbf{Freq.} \\ \hline
\multicolumn{1}{|c|}{\multirow{7}{*}{Relationships (176)}}    & \multicolumn{2}{c|}{friend}       & \multicolumn{1}{c|}{0.1820}            & 76    \\ \cline{2-5} 
\multicolumn{1}{|c|}{}                                  & \multicolumn{2}{c|}{husband}      & \multicolumn{1}{c|}{0.1597}            & 12    \\ \cline{2-5} 
\multicolumn{1}{|c|}{}                                  & \multicolumn{2}{c|}{mother}       & \multicolumn{1}{c|}{0.1419}            & 13    \\ \cline{2-5} 
\multicolumn{1}{|c|}{}                                  & \multicolumn{2}{c|}{partner}      & \multicolumn{1}{c|}{0.1004}            & 7     \\ \cline{2-5} 
\multicolumn{1}{|c|}{}                                  & \multicolumn{2}{c|}{father}       & \multicolumn{1}{c|}{0.0987}            & 14    \\ \cline{2-5} 
\multicolumn{1}{|c|}{}                                  & \multicolumn{2}{c|}{family}       & \multicolumn{1}{c|}{0.0848}            & 37    \\ \cline{2-5} 
\multicolumn{1}{|c|}{}                                  & \multicolumn{2}{c|}{boyfriend}    & \multicolumn{1}{c|}{0.0697}            & 17    \\ \hline
\multicolumn{1}{|c|}{\multirow{8}{*}{Workplace (217)}}        & \multicolumn{2}{c|}{job}          & \multicolumn{1}{c|}{0.1751}            & 5     \\ \cline{2-5} 
\multicolumn{1}{|c|}{}                                  & \multicolumn{2}{c|}{manager}      & \multicolumn{1}{c|}{0.1665}            & 31    \\ \cline{2-5} 
\multicolumn{1}{|c|}{}                                  & \multicolumn{2}{c|}{company}      & \multicolumn{1}{c|}{0.1596}            & 6     \\ \cline{2-5} 
\multicolumn{1}{|c|}{}                                  & \multicolumn{2}{c|}{office}       & \multicolumn{1}{c|}{0.1508}            & 75    \\ \cline{2-5} 
\multicolumn{1}{|c|}{}                                  & \multicolumn{2}{c|}{teacher}      & \multicolumn{1}{c|}{0.1221}            & 13    \\ \cline{2-5} 
\multicolumn{1}{|c|}{}                                  & \multicolumn{2}{c|}{interview}    & \multicolumn{1}{c|}{0.1122}            & 14    \\ \cline{2-5} 
\multicolumn{1}{|c|}{}                                  & \multicolumn{2}{c|}{employment}   & \multicolumn{1}{c|}{0.104}             & 54    \\ \cline{2-5} 
\multicolumn{1}{|c|}{}                                  & \multicolumn{2}{c|}{internship}   & \multicolumn{1}{c|}{0.1025}            & 19    \\ \hline
\multicolumn{1}{|c|}{\multirow{6}{*}{Everyday Life (15)}}    & \multicolumn{2}{c|}{cafe}         & \multicolumn{1}{c|}{0.1577}            & 1     \\ \cline{2-5} 
\multicolumn{1}{|c|}{}                                  & \multicolumn{2}{c|}{salon}        & \multicolumn{1}{c|}{0.1464}            & 1     \\ \cline{2-5} 
\multicolumn{1}{|c|}{}                                  & \multicolumn{2}{c|}{shopping}     & \multicolumn{1}{c|}{0.1340}            & 3     \\ \cline{2-5} 
\multicolumn{1}{|c|}{}                                  & \multicolumn{2}{c|}{church}       & \multicolumn{1}{c|}{0.1195}            & 4     \\ \cline{2-5} 
\multicolumn{1}{|c|}{}                                  & \multicolumn{2}{c|}{grocery}      & \multicolumn{1}{c|}{0.0941}            & 4     \\ \cline{2-5} 
\multicolumn{1}{|c|}{}                                  & \multicolumn{2}{c|}{gym}          & \multicolumn{1}{c|}{0.0776}            & 2     \\ \hline
\multicolumn{1}{|c|}{\multirow{5}{*}{Geographical Location (11)}}         & \multicolumn{2}{c|}{california}   & \multicolumn{1}{c|}{0.2480}            & 1     \\ \cline{2-5} 
\multicolumn{1}{|c|}{}                                  & \multicolumn{2}{c|}{london}       & \multicolumn{1}{c|}{0.153584}          & 1     \\ \cline{2-5} 
\multicolumn{1}{|c|}{}                                  & \multicolumn{2}{c|}{midwest}      & \multicolumn{1}{c|}{0.2480}            & 4     \\ \cline{2-5} 
\multicolumn{1}{|c|}{}                                  & \multicolumn{2}{c|}{europe}       & \multicolumn{1}{c|}{0.1228}            & 3     \\ \cline{2-5} 
\multicolumn{1}{|c|}{}                                  & \multicolumn{2}{c|}{chicago}      & \multicolumn{1}{c|}{0.1185}            & 2     \\ \hline
\multicolumn{1}{|c|}{\multirow{3}{*}{Appearance (151)}}       & \multicolumn{2}{c|}{hair}         & \multicolumn{1}{c|}{0.0843}            & 143   \\ \cline{2-5} 
\multicolumn{1}{|c|}{}                                  & \multicolumn{2}{c|}{straightened} & \multicolumn{1}{c|}{0.0113}            & 6     \\ \cline{2-5} 
\multicolumn{1}{|c|}{}                                  & \multicolumn{2}{c|}{ape}          & \multicolumn{1}{c|}{0.0003}            & 2     \\ \hline
\multicolumn{1}{|c|}{\multirow{2}{*}{Criminality (12)}}      & \multicolumn{2}{c|}{criminals}    & \multicolumn{1}{c|}{0.0078}            & 3     \\ \cline{2-5} 
\multicolumn{1}{|c|}{}                                  & \multicolumn{2}{c|}{neighborhood} & \multicolumn{1}{c|}{0.0017}            & 9     \\ \hline
\multicolumn{1}{|c|}{\multirow{2}{*}{Ability (8)}}          & \multicolumn{2}{c|}{basketball}   & \multicolumn{1}{c|}{0.0040}            & 4     \\ \cline{2-5} 
\multicolumn{1}{|c|}{}                                  & \multicolumn{2}{c|}{athletic}     & \multicolumn{1}{c|}{0.0283}            & 4     \\ \hline
\multicolumn{1}{|c|}{\multirow{5}{*}{Personality (18)}}      & \multicolumn{2}{c|}{smelled}      & \multicolumn{1}{c|}{0.0115}            & 2     \\ \cline{2-5} 
\multicolumn{1}{|c|}{}                                  & \multicolumn{2}{c|}{lazy}         & \multicolumn{1}{c|}{0.0110}            & 3     \\ \cline{2-5} 
\multicolumn{1}{|c|}{}                                  & \multicolumn{2}{c|}{stronger}     & \multicolumn{1}{c|}{0.0034}            & 4     \\ \cline{2-5} 
\multicolumn{1}{|c|}{}                                  & \multicolumn{2}{c|}{ignorant}     & \multicolumn{1}{c|}{0.0029}            & 6     \\ \cline{2-5} 
\multicolumn{1}{|c|}{}                                  & \multicolumn{2}{c|}{stupid}       & \multicolumn{1}{c|}{0.0004}            & 3     \\ \hline
\multicolumn{1}{|c|}{\multirow{2}{*}{Sexual Exoticism (18)}} & \multicolumn{2}{c|}{exotic}       & \multicolumn{1}{c|}{0.0466}            & 6     \\ \cline{2-5} 
\multicolumn{1}{|c|}{}                                  & \multicolumn{2}{c|}{attractive}   & \multicolumn{1}{c|}{0.0322}            & 12    \\ \hline

\end{tabular}}
    \end{minipage}
\vspace{5pt}
\caption{Themes Emerged From Attributive Words of Acts and Recalls With Scores in Descending Order. In addition to the attribution scores, we provide the number of times each token occurred in the corpus (Freq.) }
\vspace{-4mm}
\label{table:tableig}

  \end{table}
\paragraph{\textbf{Everyday Life}}
Recalls often contain tokens relating to everyday activities or places that people frequent on a day to day basis such as a \textit{café} in order to convey what they were doing when experiencing an act of racial microaggression: \textit{“On Friday morning, as I walked to the café between classes at my predominantly white university, the school appointed photographer offered me a free coffee if I agreed to play the role of the cheerful token black woman in a group of strangers.”}

\paragraph{\textbf{Geographical Location}}
Posters of recalls often use tokens relating to geographical location (countries, cities, regions, etc.), such as \textit{California}, to communicate where they experienced a racial microaggression: \textit{“I walk into a gas station market in California with about ten of my Latina/o and black high school students to buy snacks for our college road trip, and within five minutes, we hear, “SECURITY CHECK ON ALL AISLES.”}

\subsubsection{Overlapping Themes Across Acts and Recalls}
\paragraph{\textbf{Appearance}}
Both acts and recalls contain words such as \textit{gorilla} and \textit{ape}, respectively.  While posters of acts commonly use the word gorilla in order to imply that Black people do not deserve to be treated as human beings: \textit{“Black people appear more closely related to gorillas than human,”}  posters of recalls commonly use the word \textit{ape} in order to describe instances in which they have been compared to such an animal: \textit{“He was also the worst of the people making these jokes in high-school and shortly after, making hundreds of  ``all black people are criminals" jokes and *comparing me to an ape* one time.”}  As it relates to the topic of appearance, given that ape is an umbrella term that includes several species, one of which is a gorilla, it follows that acts within this theme are often more specific than recalls.  Similarly, while acts utilize a variety of words such as \textit{fat, palms, and skinned} with the purpose of negatively stereotyping Black peoples’ appearances: \textit{“Black people are fat because McDonalds is all they can afford”}, recalls frequently use words relating to a single feature, hair (\textit{hair, straightened}) to describe their experience of being the victim of microaggressions about the texture of their hair: \textit{“This is why I don’t ever straighten my hair anymore, even though it’s something I used to like to do for fun on occasion, because the compliments always seem to insinuate that my normal hair is unprofessional, unruly, or otherwise socially unacceptable.”}

\paragraph{\textbf{Criminality}}
Posters of acts often use anecdotal evidence and a variety of words related to \textit{criminality}, such as \textit{police, robbed, and dangerous} in order to make stereotypical statements about Black people: \textit{“Today I almost I got my car robbed from me. I’ve gotten robbed twice by a black person and this is the 3rd time but this time I was able to get away.”}  In contrast, posters of recalls frequently use a single word, \textit{criminals}, in order to describe being associated with criminals: \textit{“He was also the worst of the people making these jokes in high-school and shortly after, making hundreds of ``all black people are criminals" jokes and *comparing me to an ape* one time.”}

\paragraph{\textbf{Ability}}
While acts commonly contain tokens such as \textit{sports} and \textit{IQ}, recalls frequently contain tokens related to sports, such as \textit{basketball} and \textit{athletic}.  For example, in this act the poster uses the word sports to highlight that Black people are only good at rap and sports: \textit{“Black people of Reddit, How does it feel to be inferior and only good at rap and sports?”}. In addition, the acronym \textit{IQ} is used in order to demonstrate that Black people are naturally less intelligent than other races, as IQ is seen as more of an inherent intelligence as opposed to intelligence derived from hard work: \textit{``Black people have a way lower IQ across the board compared to their white counterparts."}  In contrast, recalls focus solely on sports: \textit{``The normal stereotypical things like us liking fried chicken and watermelon, every black person knows how to dance (my brothers are proof that's a lie) we're all good at basketball (I'm proof that's a lie)."}  Finally, similar to criminality and appearance, recalls within this theme contain a fewer variety of words, indicating that the content of acts is more dispersed as compared to recalls.

\paragraph{\textbf{Personality}}
Posters of both acts and recalls commonly make use of tokens related to negative personality characteristics such as \textit{dumb} and \textit{stupid}.  While acts frequently use the word dumb to criticize Black people: \textit{``They never know what I’m talking about because they are dumb,"} recalls frequently make use of the word stupid to describe how they are perceived by others: \textit{``So that’s why I’m not only likely to be a thief, I’m likely to be a stupid thief!"}

\paragraph{\textbf{Sexual Exoticism}}
Posters of acts frequently use the tokens, \textit{attracted} and \textit{sex}, to highlight their perception of Black people as highly sexually desirable just because of their race: \textit{``I’m extremely sexually attracted to Black men and women."}  In addition, this act uses the word sex to underscore that being Black is a condition that must be satisfied for them to agree to have sex with someone: \textit{``I will only have sex with Black men."}  In contrast, posters of recalls frequently use the word attractive to describe that other races typically find their race unattractive: \textit{``Like when people tell you aren’t black because you’re attractive."}  

While Integrated Gradient (IG) offers valuable insights into the contexts and themes of racial microaggressions, as with most deep learning interpretation techniques used to explain binary classification decisions \cite{ribeiro-etal-2016-trust,NIPS2017_8a20a862}, IG too has limitations in capturing complex linguistic patterns. For instance, it may struggle to capture the \textit{overlapping} characteristics between acts and recalls, which are essential in understanding the nuances in posts that may result in FPs and FNs. Additionally, IG relies on preprocessing steps like lemmatization and stop word removal, which may not fully capture linguistic cues such as verb tense or the use of absolute terminology in acts and recalls. To overcome these limitations, we adopted a more comprehensive approach in RQ3 by analyzing a randomly selected set of posts from our entire data, including those that did not contain the top 40 IG tokens. This allowed us to be more exhaustive in our examination of linguistic patterns that characterize acts and recalls of microaggressions.

\subsection{RQ3: Characterizing Acts vs. Recalls}
To further understand the nature of acts and recalls of racial microaggressions in Reddit posts and comments, we examined the linguistic attributes manifest in their content using discourse analysis to better understand the social purpose underlying the linguistic patterns observed in acts and recalls of racial microaggressions.  Tables \ref{tab:table7}, \ref{tab:table8}, \ref{tab:table9} represent the linguistic pattern that emerged from the analysis and examples for each linguistic pattern unique to acts and recalls respectively. Our findings suggest three linguistic patterns in both acts and recalls of racial microaggression from our data.  We utilize linguistic analysis as well as data gathered from our workshop participants to better understand the functional purposes of the similarities and differences we observed in the linguistic patterns of acts and recalls.  

\subsubsection{Linguistic Signature of Acts}

\paragraph{\textbf{Questions}}
Our findings revealed questions to be a key linguistic pattern in acts of racial microaggression.  Consider this example of an act: \textit{“Why are black people so athletic? Not to sound racist, but I have recently noticed that black students excel at all the sports in my school. I am White and most people in my school are white, However the black players are the best in every single sport for our school. Why is that? I don’t mean to be racist or offend anyone. Just wondering.”} This statement serves to create a broad generalization of all Black people as being athletic, falling under several themes of racial microaggressions such as racial categorization and sameness, assumptions about intelligence, competence, or status, and connecting via stereotypes \cite{williams_after_2021-1}. Consider the statement below from one of our workshop participants, P4:

\begin{table}[]
\resizebox{0.9\columnwidth}{!}{%
\begin{tabular}{|c|l|l}
\cline{1-2}
\textbf{Linguistic Pattern} &
  \multicolumn{1}{c|}{\textbf{Examples of acts of Racial Microaggression}} &
   \\ \cline{1-2}
Questions &
  \textit{\begin{tabular}[c]{@{}l@{}} {}\\ $\bullet$  \textbf{Why} are black people so bad at swimming? \\ {}\\ $\bullet$  \textbf{Why} do so many black people litter? I don’t think a day\\    goes by that I don’t see someone throw trash out of their\\    car, 9 times out of 10 that person is black. \textbf{Why}?\\ {}\\ $\bullet$  \textbf{Black people of Reddit}, which one of you stole my bike?\\ {}\\$\bullet$  \textbf{Black people of Reddit}, can I touch your hair? \\{}\end{tabular}} &
   \\ \cline{1-2}
\begin{tabular}[c]{@{}c@{}}Use of absolute terminology \\ (‘all’, ‘never’, ‘ever’, ‘should’, ‘absolutely’, ‘only’)\end{tabular} &
  \textit{\begin{tabular}[c]{@{}l@{}}{}\\ $\bullet$ {[}Black people{]} are \textbf{all} mentally handicapped and physically\\    incapable of supporting themselves.\\ {}\\ $\bullet$  I would \textbf{never, ever} hire someone with a ‘black’ name on\\   their resume. I wouldn’t even interview them.\\{}\end{tabular}} &
   \\ \cline{1-2}
Use of statistics &
  \textit{\begin{tabular}[c]{@{}l@{}}{}\\ $\bullet$  Black people are not oppressed and if they want to be in \\ prison less, they should not be committing \textbf{53\%} of all \\ homicides while  only being \textbf{12\%} of the population.\\{}\\ $\bullet$ 
 \textbf{12\%} of the population is black people, yet they commit so \\ much  more crimes.\\{}\end{tabular}} &
   \\ \cline{1-2}
\begin{tabular}[c]{@{}c@{}}Use of modifying adverbs or \\ adjectives (‘most’, ‘usually’, ‘consistently’)\end{tabular} &
  \textit{\begin{tabular}[c]{@{}l@{}} {}\\ $\bullet$  Black people \textbf{usually} name their kids after stuff they can’t \\afford.  Like Mercedes, Diamond, Hope, and Insurance.\\ {}\\ $\bullet$ Black people are \textbf{consistently} the most rude, demanding, \\ignorant, of what want and shady.\\{}\\ $\bullet$  I fail black students way more often because, objectively, \\they make the \textbf{most} mistakes on driving test.\\{}\end{tabular}} &
   \\ \cline{1-2}
\end{tabular}%
}
\vspace{5pt}
\caption{Linguistic Patterns Observed Using Discourse Analysis for Acts With Examples.}
\label{tab:table7}
\vspace{-4mm}
\end{table}

\medskip
\begin{center}
\textit{“It is still a microaggression in the form of a question. You are still generalizing black people and you kind of believe that statement which is why you are curious about it” —P4}
\end{center}
\medskip
This statement is a racial microaggression masked in the form of a question, which can give the statement more of a tone of curiosity than aggressiveness [Refer to Table 8].  Nevertheless, just like P4 highlighted, this curiosity is still a form of generalization.  The poster of this statement seems to believe that all Black people are athletic, which is why they are curious about why that is the case. Consider the statement below from another one of our workshop participants, P5:
\medskip
\begin{center}
\textit{“There are only a few black players that are the best but when they are asking this question, they clearly generalize. If the star athletes weren’t black, you wouldn’t be wondering, “why are these people so athletic?” It would just be obvious that they probably practice a lot”-P5} 
\end{center}
\medskip
Despite the subtlety of this statement, P4 further highlights how this statement is a generalization, emphasizing that if the star players were white, the posters would likely attribute their talent to practice and hard work instead of race.  On the other hand, our findings also suggest that posters commonly disguise microaggressions using a combination of curiosity and humor.  Consider this example of an act: \textit{``Black people of Reddit, which one of you stole my bike?"}  This statement uses a common stereotype of Black people, criminality, to make a joke in the form of a question.  By directly addressing members of the Black community on Reddit \textit{(“Black people of Reddit”)} and assuming one of these members stole their bike \textit{(``which one of you")}, the poster of this act utilizes humor to soften a harsh stereotype.  

\paragraph{\textbf{Use of Absolute terminology and statistics}} 
In addition to questions, our findings show that acts utilize absolute terminology and statistics in order to justify making a racial microaggression.  Consider this example of an act: \textit{“Black people are not oppressed and if they want to be in prison less they should not be committing 53\% of all homicides while only being 12\% of the population.”} This statement uses the problematic ``13/50" argument \cite{kelis_2021}, which is commonly used to stereotype Black crime in order to make statements appear factual as opposed to stereotypical \cite{13/50-article}. The 13/50 argument is an overused and often misleading talking point that poses that black people make up only 13\% of the population but commit 50\% of all known crimes \cite{doi:10.1177/1043986207306870}. 
Consider the statement below from one of our workshop participants, P3:
\smallskip
\begin{center}
\textit{“There is a lot more to that statistic because the original intent of the police was to monitor Black people”-P3}
\end{center}
\medskip
According to P4, such statistical reference is biased because the United States history of systemic racism has resulted in Black people often being the target of police \cite{Moore2017, 10.2307/43670469}. Moreover, another participant highlights how the use of  statistics matters in determining whether a statement should be considered a racial microaggression:
\smallskip
\begin{center}
\textit{“There is a difference between using a statistic to prove that there is a problem with the prison system versus using it to say something about Black people.”-P4}
\end{center}
\medskip
In this statement, P4 highlights that she believes that using a statistic to make a statement about an institution, such as the prison system, is different from using a statistic to make a statement about a certain race, such as Black people.  Given that institutions such as the prison system are not human, she implies that using a statistic to make a seemingly \textit{“factual”} statement that is negative about a particular group of humans is bound to be hurtful to people.  The poster of this act uses the statistic in order to justify making a racial microaggression on the grounds that they are unbiased and are just disseminating a fact.

\paragraph{\textbf{Use of modifying adverbs or adjectives}}
The last key linguistic pattern distinct to acts is the use of modifying adverbs/adjectives.  Consider this example of an act: \textit{``Black people usually name their kids after things they can’t afford. Like Mercedes, Diamond, Hope, or Insurance."} In response to this statement, one of our workshop participants highlights why this statement is a stereotypical/generalized statement, stating that she doesn’t know anyone with those names:
\smallskip
\begin{center}
\textit{“This statement is too generalizing.  I don’t know a single person with those names. It is not as common as you think.”-P5} 
\end{center}
\medskip
The poster of this act uses the modifying adverb, \textit{`usually'}, in order to characterize this behavior of Black people as frequent. Unlike the posters of acts that use statistics to justify making a racist remark, the poster of this act uses a colloquial term, \textit{`usually'}.  By using the modifying adverb, \textit{`usually'}, the poster of this act seeks to give the impression that their personal knowledge is sufficient to justify such a claim.  Thus, similar to posters of acts that utilize statistics, this poster tries to justify making a stereotypical racist remark.

\subsubsection{Linguistic signature of Recalls}
\begin{table}[]
\resizebox{0.9 \columnwidth}{!}{%
\begin{tabular}{|c|l|l}
\cline{1-2}
\textbf{Lingusitic Pattern} &
  \multicolumn{1}{c|}{\textbf{Examples of Recalls of Racial Micraoggression}} &
   \\ \cline{1-2}
`White' and `White people' &
  \textit{\begin{tabular}[c]{@{}l@{}} {} \\ $\bullet$ I wish \textbf{white people} in general would stop commenting on my appearance\\ unless it’s to compliment me or to tell me that I have something stuck in \\ my teeth.\\ {} \\ $\bullet$ \textbf{White people} singling me out at social events to make small talk with me\\  about race/politics.\\ {}\end{tabular}} &
   \\ \cline{1-2}
Paste Tense Verbs &
  \textit{\begin{tabular}[c]{@{}l@{}} {} \\ $\bullet$ I \textbf{was} at work and the topics of racism came up with my boss who is Italian...\\   {} \\$\bullet$ I \textbf{felt} irritated at having to explain that yes, I am a REAL programmer.\\ {}\end{tabular}} &
   \\ \cline{1-2}
'Only Black' &
  \textit{\begin{tabular}[c]{@{}l@{}} {} \\ $\bullet$ I had an English teacher who loved to talk, and whenever she’d say anything\\  about race or Black culture, she’s turn to me(the \textbf{only black kid} in the room)\\  as if to validate/confirm the statement.\\ {} \\$\bullet$ I was the \textbf{only black girl} in the room with him and ten other coworkers\\ {}\end{tabular}} &
   \\ \cline{1-2}
\end{tabular}%
}
\vspace{5pt}
\caption{ Linguistic Patterns Observed Using Discourse Analysis for Recalls With Examples.}
\label{tab:table8}
\vspace{-4mm}
\end{table}

\paragraph{\textbf{`White' and `White people'}}
One notable linguistic feature of recalls is the use of the phrases \textit{‘White’} and \textit{‘White people.’} Consider this example of a recall: \textit{“White people singling me out at social events to make small talk with me about race/politics.  I think they want to see me get impassioned or educate them. No I’m tired, I came out to have fun.”} In this recall, the poster is expressing his/her discomfort of being \textit{‘singled out’} at social events by \textit{‘white people’}. Clearly stating the subject \textit{(‘white people)’} that is causing his/her discomfort, the poster uses this phrase to highlight the source of their tiredness. By making the source of their tiredness very clear, the poster seeks to have his/her experiences validated. 

\paragraph{\textbf{Past Tense Verbs}}
Another common linguistic feature we noticed in recalls was the use of past tense verbs.  Since recalls are recounts of acts of racial microaggressions, it follows that most recalls describe events that took place in the past.  Consider this example of a recall: \textit{“I was at work when the topic of racism came up with my boss who is Italian...”}.  The poster of this recall is describing the setting of when he/she experienced an act of racial microaggression. The use of past tense verbs, such as \textit{‘was,’} and \textit{‘came’} aids in disclosing the setting in which the act took place. This is a key element in the poster’s recount of their experience.  Prior work in human communications research suggests that lying individuals use fewer words and fewer past tense verb forms \cite{10.1111/j.1468-2958.1982.tb00684.x}. These verbs help provide readers with a confidence that the poster must be telling the truth because they seem to be recounting their past experience very clearly.  This type of disclosure helps create an overall tone of honesty and authenticity in the statement, as the poster hopes to have their experience validated by others.
	
\paragraph{\textbf{‘Only Black'}}
The use of the phrase \textit{‘only black’} was the last prominent linguistic pattern distinct to recalls that our findings revealed.  Consider this example of a recall: \textit{“I was the only black girl in the room with him and ten other coworkers.”}   The poster’s use of the phrase \textit{‘only Black’} in contrast to \textit{‘him and ten other coworkers’} underscores her discomfort at being the only Black girl in the room. 

\subsubsection{Linguistic Similarities between Acts and Recalls}

\paragraph{\textbf{Use of First-Person}}
One notable difference between acts and recalls of racial microaggression is the role that first person voice serves in context.  Consider the example: \textit{"I am not a racist, but it seems whenever I sit near a group of black people I can't hear the movie over all the noise they make."}  Here, the poster uses first person to preemptively defend himself/herself from being called a racist.  One of our workshop participants, P3, states that this defensiveness does nothing to absolve the poster of what he/she said:
\medskip
\begin{center}
\textit{
“Just because you say I don’t be mean to be such and such does not automatically resolve you of what you said before or after. Maybe they don’t mean to be racist but words are words and the implication is still going to be there regardless of intent. ”-P3}
\end{center}
\medskip
While this type of linguistic pattern appears to mask a racial microaggression, P1 points out that while the poster may not have bad intentions, the statement is still an act of racial microaggression, and therefore, the hedging does nothing to reduce the severity of the statement. Unlike posters of acts that seek anonymity, posters of recalls typically use first person voice to thoroughly describe themselves-\textit{“I am a 22 year old, brown skinned African American girl. In school in Maryland. I felt out of place and isolated."}  Here, the poster utilizes the first person voice twice to describe her age and feelings. Based on LIWC analyses \cite{doi:10.1177/0261927X09351676} and manual inspection, our results indicate that roughly 40\% of recalls that utilize first person do so to describe themselves. This self-disclosure serves to create an overall tone of authenticity and honesty, which the poster hopes will allow his/her experiences to be validated. 
\medskip
\begin{justify}
    Moreover, our findings suggest that posters of acts utilize first person voice in even subtler ways to hedge the aggressiveness of their comments.  Consider this example of an act from our dataset:
    \begin{center}
    \textit{“Speaking as a capital “C” Conservative. I totally agree. I enjoy good, well developed characters, I don’t care if they are black, gay, green, alien, or inanimate objects. Scandal is a great show, orange is the new black was great for the first couple seasons, and Jesus is a great character in the walking dead. (That’s coming from an orthodox catholic conservative). Now to be honest, I hated black panther, just didn’t think it was a good movie.”}
    \end{center}
\end{justify}
\medskip
Workshop participants agreed that the last sentence of the post contains the microaggression; nevertheless, the poster makes several statements before in order to take attention away from it.  The repetitive use of first person voice (\textit{“I”}) prior to the last statement serves to highlight the poster’s desire to portray themselves as an objective critic of entertainment.  Similar to our previous findings, one of our workshop participants highlights that this superfluous build up is just another method of hedging a racist remark:
\medskip
\begin{center}
\textit{“They try to use color blindness to remove themselves from what they are saying”-P6}
\end{center}
\medskip
\begin{justify}
The poster seeks to show that they see everyone as equal by saying, \textit{“I don’t care if they are black, gay, green, alien, or inanimate objects.”}  By making a somewhat extreme statement, that they don’t see color, the poster seeks to curb the aggressiveness of his/her statement.  Moreover, one of our workshop participants, P3, highlights this type of hedging has consequences beyond just being a way to remove oneself from the consequences of making a racist remark:
\end{justify}
\medskip
\begin{center}
\textit{“I definitely can take offensive to this since you are not validating who people are-you are not acknowledging where they came from or what they experienced”-P2}
\end{center}
\medskip
\begin{justify}
This type of false color blindness fails to acknowledge the systemic racism that has existed in American culture for many decades \cite{williams_after_2021-1}.  Nevertheless, like P3 points out, these types of statements are often invalidating for Black people if they are proud of their identity or have suffered because of it [for more details on the definition and examples of color blindness, refer to Table \ref{tab:tablea}, Appendix]. While acts utilize superfluous build up and false color blindness to hedge the severity of microaggressions, our findings suggest that posters of recalls also incorporate build up prior to the point they are trying to make.
\end{justify}
\medskip
\begin{center}
\textit{“I will ALWAYS be one to want to expand my viewpoints, appreciate history, and under the social climates, but I'm fucking tired... I 10000\% care about race and gender issues within our communities, but I'm tired of including "others" in the conversation. How can we navigate things like institutionalized racism without begging “others” to see us as human? Unfortunately, it's never going to change. Maybe I'm just being a pessimist, but...”}
\end{center}
\medskip

By emphasizing the word \textit{always} and using the first two sentences to prove that he/she is an open minded individual, the poster of this recall is trying to establish themselves as someone that is legitimate and trustworthy in the Black community.  Therefore, the poster uses superfluous build up to gain peoples’ trust and have his/her thoughts and feelings heard.

\begin{table}[]\resizebox{0.9\columnwidth}{!}{
\begin{tabular}{|c|l|l|}
\hline
\textbf{Linguistic Pattern} & \textbf{Examples of acts of Racial Microaggression} & \textbf{Examples of recalls of Racial Microaggression} \\ \hline
\multirow{4}{*}{\begin{tabular}[c]{@{}c@{}}{}\\ {}\\ {}\\ {}\\ Use of First Person\end{tabular}}     & \textit{\begin{tabular}[c]{@{}l@{}} {}\\ $\bullet$ \textbf{I’m} not a racist, but it seems \\ whenever \textbf{I} see sit near a group of Black \\ people, \textbf{I} can’t hear the movie over all the \\ noise they make.\\{} \end{tabular}} & \textit{\begin{tabular}[c]{@{}l@{}}$\bullet$ \textbf{I’m} an African American graduate \\ student, and \textbf{I} teach at a large \\ university.\\{}\end{tabular}}                          \\
& \textit{\begin{tabular}[c]{@{}l@{}}$\bullet$ \textbf{I} don’t hate them, \textbf{I} don’t bully them, \\ but \textbf{I’m} careful with them, as if they \\ were criminals.\\{}\end{tabular}}                                           & \textit{\begin{tabular}[c]{@{}l@{}}$\bullet$ It angers me when people measure \\ my race by the way \textbf{I talk,} \\ \textbf{dress and carry myself.}\end{tabular}}                   \\ 
& \textit{\begin{tabular}[c]{@{}l@{}} $\bullet$ \textbf{I know I’ll be downvoted}, but anyone\\  in the U.S. who works for tips knows that \\ Black people are far less likely to tip…\\{}\end{tabular}}    &     \\ \cline{1-3}
\multirow{2}{*}{“Us” vs. “Them” Language} & \textit{\begin{tabular}[c]{@{}l@{}} {}\\ $\bullet$ Black redditors, what is your take on\\  having a white friend? What do you see \\ \textbf{us} as? Or any other race? \\{}\end{tabular}}                                  

&\textit{\begin{tabular}[c]{@{}l@{}} {}\\ $\bullet$ How can \textbf{we} navigate things like\\  institutionalized racism without \\ begging “others” to see \textbf{us} as human? \\{}\end{tabular}} \\

& \textit{\begin{tabular}[c]{@{}l@{}}$\bullet$ Maybe we shouldn’t let \textbf{them} (Black \\  people) vote.\\{}\end{tabular}}                                       

& \textit{\begin{tabular}[c]{@{}l@{}}$\bullet$ \textbf{They} always said it was a joke but \\  \textbf{they} kept doing it over and over. \\{}\end{tabular}}                                     \\ \cline{1-3}

\end{tabular}}

\vspace{5pt}
\caption{Linguistic Patterns Observed for Acts and Recalls Using Discourse Analysis With Examples.}
\label{tab:table9}
\vspace{-4mm}
\end{table}

\paragraph{\textbf{Use of “Us” vs. “Them” Language}}
Another notable characteristic of acts and recalls of racial microaggression is the use of Us vs. Them language.  Consider the example from our discourse analysis: \textit{“Maybe we shouldn’t let them (Black people) vote”.} By utilizing both \textit{we} and \textit{them}, this statement is characteristic of Us vs. Them language.  The juxtaposition of these words serves to highlight the presence of two different groups, \textit{we} or \textit{us} and \textit{them} or \textit{“Black people.”}  This creation of an in-group and out-group serves to portray Black people as second class citizens.
On the other hand, recalls utilize \textit{Us} vs. \textit{Them} language in order or to emphasize their feelings of being discriminated against.  Consider this recall from our dataset: \textit{‘They always said it was a joke but they kept doing it over and over.”}  The use of the word \textit{“they”} in this statement is characteristic of \textit{“them”} language in \textit{Us} vs. \textit{Them} language.  The sentence highlights that the out-group is hurting the in-group's feelings by making what they considered to be a joke about the victim’s race. In context, the use of \textit{Us} vs. \textit{Them} language helps underscore the victim’s discomfort due to the “joke” by highlighting the actions of the out-group.


 As we conclude our findings for RQ3, it's important to reflect on how they differ from our findings from RQ2. While both research questions aimed to strengthen our understanding of racial microaggressions in online content, they each focused on different aspects of these interactions. RQ2 centered on \textbf{`what'} was being discussed in the posts, identifying themes and contexts of racial microaggressions through the analysis of influential tokens and their contextual use. This analysis revealed themes such as \textit{`Questions'},\textit{ `Evolution'}, and \textit{`Criminality' } for acts (among others), and \textit{`Relationships'}, \textit{`Ability'}, and \textit{`Workplace'} for recalls (among others), offering thematic insights into the content of racial microaggressions. On the other hand, RQ3 delved into \textbf{`how'} these discussions were being framed and expressed. We manually examined a randomized selection of posts from the entire dataset, identifying linguistic patterns and discourse structures (such as the "Use of modifying adverbs or adjectives", "Past tense verbs", and "Use of first-person" etc) that characterize acts versus recalls.  

To this end, our findings from RQ2 and RQ3 complement each other, with RQ2 providing thematic insights into the content of racial microaggressions, and RQ3 offering linguistic insights into their expression. Together, they contribute to a more holistic understanding of racial microaggressions in online content, each illuminating different aspects of these complex interactions.

\section{Discussion}

\subsection{Enhancing Content Moderation Classifiers with Context Awareness of FPs and FNs}

Current hate speech classifiers often rely on predetermined thresholds, such as  classifying a post as "offensive" if its toxicity score exceeds 0.8 (a benchmark recommended by Perspective API for determining harmful posts). As a result, these classifiers may struggle to adequately capture implicit microaggressions, leading to  false negatives (FNs) \cite{elsherief_latent_2021-1}. This raises the question around whether toxicity thresholds alone are  reliable measures for evaluating implicit forms of harmful content.

To improve the precision of these classification systems, our research findings highlight the necessity for a deeper examination into the specific linguistic patterns that trigger FPs and FNs in the first place.  For example, our study reveals that FPs and FNs often exhibit overlapping, yet distinct linguistic patterns and themes, which can be overlooked by  threshold-based classifiers. For instance, posters of acts often use a variety of words (e.g., `police', `robbed', and `dangerous') thematically associated with \textit{criminality} to stereotype Black people as \textit{"criminals"} in their posts, while posters of recalls tend to use the word \textit{`criminal'} as a singular word to recount their experience of being wrongfully perceived as a potential lawbreaker. This strong thematic overlap in words that characterize acts and recalls can make it extremely difficult for classifiers, from simple keyword-matching algorithms (which often overlook the context of the flagged words) to even the more context-aware ones (e.g., BERT), to distinguish the subtle nuances shared between FPs and FNs. Similarly, the presence of common semantic characteristics between recalls and acts can also potentially mislead classifiers. For instance, both posters of acts and recalls use the first person voice  for different purposes.  While the former tend to use the first person alongside hedging language to defend themselves (e.g., \textit{"I am not a racist"}), the latter use it to depict or narrate  past experiences (e.g., \textit{"As a Black woman, I've always been ignored by my coworkers."}). This highlights that classifiers, despite the different contexts in which the language is  used, can fail to discern FPs apart from FNs. 

Hence, we foresee the need for future content moderation systems to incorporate the nuanced contexts around acts and recalls into toxicity classifiers, as decision thresholds alone cannot fully capture the complex linguistic subtleties  shared by recalls and acts, leading them to be misclassified as FPs and FNs, respectively. To address this issue, we suggest that threshold-based classifiers could be improved by incorporating \textit{contextual features} that incorporate the thematic and linguistic patterns of acts and recalls observed in our study. These features can capture the semantic meaning of each linguistic pattern within its specific context. For instance, the word \textit{"I"} would generate different features when used in varying contexts, thereby capturing the nuances of its use in acts versus recalls. By extracting such contextual features, we can improve the accuracy of toxicity classifiers and overcome the limitations of simple decision thresholds, thereby reducing the likelihood of misclassifying posts as FPs or FNs. 

Furthermore, insights from our workshop demonstrate that capturing the true essence of context when distinguishing acts from recalls depends on one's familiarity with stereotypes or being able to discern the role of hedging language in an argument. This involves understanding underlying implications that extend beyond the mere text---a nuance most classifiers overlook. Encapsulating such context more holistically, as shown in this work, is crucial for fostering more accurate and equitable content moderation systems.

\subsection{Increasing Awareness of Acts and Recalls Across Moderators and Users}
\paragraph{\textit{Improving Explanations Around Moderation Decisions}} Prior work on transparency in content moderation calls attention to key challenges in the moderation process \cite{removal_explanations_jhaver}: moderators often fail to articulate \textit{\textbf{what}} aspect of the content prompted moderation or \textbf{\textit{why}} such moderation was necessary. This begs the question: how can we inform users about why their posts were flagged or removed, and which aspects of their posts led to such moderation outcomes? What kind of strategies can moderators undertake to effectively communicate such moderation decisions to users?  Our study provides insights to help address these questions in the context of improving moderation explanations. Our analysis reveals key themes and linguistic patterns that could potentially equip moderators with a deeper understanding around acts versus recalls, thereby enabling more informed decisions.  For instance, posters of acts often pose questions that are seemingly driven by curiosity, or statements incorporating statistics about Black people.  It is possible that individuals committing acts of microaggressions are not fully cognizant of their behavior in their discussions of certain racial groups \cite{sue_microaggressions_2020}. For example, posters frequently pose questions, such as \textit{"Why are black people so bad at swimming?"} or \textit{"Why did evolution turn us white people white when dark skinned Africans have no problem surviving in places like Northern Europe?}". These questions, while appearing innocent and curiosity-driven at first glance, contain offensive undertones and implications that the person posing the question might overlook. As our findings show, acts such as these examples, frequently feature themes around Black people's \textit{ability} or their status in the \textit{evolution} or \textit{human race}. Moderators could employ the themes and linguistic patterns (e.g., `Questions', etc.) identified in our research that commonly characterize acts to articulate their content moderation decisions to users. They can explain that posts like the examples above, containing seemingly harmless questions, are in fact associated with common themes of racial microaggressions. By leveraging these insights, moderators could bring a new level of transparency and understanding to their moderation process, potentially minimizing confusion and dispute within the community. Similarly, moderators can leverage the detailed characterizations of recalls from our research to  identify potential misclassifications of recalls in their content moderation decisions. This would enable moderators to effectively communicate to posters of recalls who are mistakenly caught up in false positives, elaborating on reasons why their post was erroneously flagged as harmful.

\paragraph{\textit{Improving Moderation and Community Guidelines}} Prior work in content moderation has highlighted the issue of vague and unclear guidelines on social media platforms \cite{modsandbox,looking_glass_tanu, 10.1145/3359294}. These rules often lack explicit and specific wording, making their operationalization and enforcement processes opaque and non-transparent \cite{looking_glass_tanu}. As a result, it becomes challenging for both users and moderators to understand how these rules are applied and why certain content is flagged or removed \cite{10.1145/3359294}. To solve this, we can incorporate clear illustrative examples of posts analyzed from our study into the moderation and community guidelines.  Specifically, we can update moderation guidelines to include illustrative posts (e.g., \textit{"Why are all black people so loud?"}) highlighted with keywords derived from IG (e.g., \textit{`Why'},  \textit{`loud'}) as as well as clear descriptions of the  themes (e.g., \textit{`Questions'}, \textit{`Personality'}) and linguistic patterns (e.g., \textit{`Questions'}, \textit{`Use of absolute terminology'}) associated with acts and recalls  in our study.  Such comprehensive guidelines can empower moderators to make more informed and consistent decisions when evaluating user-generated content.  Moreover, prior research suggests that “explicit rules and guidelines increase the ability for community members to know the norms” \cite{Kiesler2010RegulatingBI, 10.1145/3359294}. Our findings can be used to establish clear and explicit community guidelines that informs users what type of content is acceptable to post on the community while minimizing the risk of unintentionally perpetuating harmful narratives (e.g., \textit{acts}) or disregarding the experiences of marginalized groups (e.g., \textit{recalls}).

\subsection{Language Mimicry and Relational Dynamics in Online Discussion Communities}
 
Online discussion communities, such as Reddit, naturally embody relational dynamics across users based on community roles (e.g., moderators/ admins vs. regular users) and membership statuses (e.g., old vs. new members). Research shows that such social structures and relational hierarchies within online discussion groups can potentially play a role in language mimicry and adoption. Language coordination is a phenomenon in which people tend to unconsciously mimic the language of others by responding with similar words or phrases \cite{ismsi}. Research in computational linguistics has demonstrated how language coordination persists across conversations in ways that reflect power differentials between people. For example, in Supreme Court case settings, lawyers tend to linguistically mimic the language of the Supreme Court justices rather than vice versa \cite{danescu-niculescu-mizil_echoes_2012}. Such language coordination also occurs online: Wikipedeans tend to echo the linguistic style of admins significantly more than that of non-admins who are perceived to have a lower status within the community \cite{danescu-niculescu-mizil_echoes_2012}. Our findings show that acts of racial microaggressions on social media embody persistent linguistic patterns, such as absolutist expressions (e.g., \textit{never, ever hire someone with a Black name}) and modifying adverbs (e.g., \textit{Black people are consistently the rudest}) that generalize or racially discriminate against Black people. Acts are also frequently masked in the form of questions disguised as genuine curiosity, or conveyed with statistics that tend to factualize selective information as broader truths. Given the presence of relational dynamics in online communities on top of platform affordances (e.g., up/down votes, likes, volume of comments) that interplay with such dynamics, linguistic patterns of acts can be mimicked and adopted across users, potentially amplifying racial biases, and endorsing harmful assumptions that underly acts of racial microaggressions. For future work, we intend to empirically capture how membership statuses and power dynamics within online discussion groups are associated with the adoption and spread of linguistic patterns of acts and recalls.

\subsection{Critical Race Theory in Language and the Importance of Counter-Storytelling}

According to Critical Race Theory, social conceptions of race and racism shape, and are shaped by laws, social movements, politics, and the media \cite{ogbonnaya-ogburu_critical_2020}. Such an argument is well-reflected across the theoretical premise of several anthropological research studies on race and language. Anthropological linguists have long recognized the importance of treating racial categories and concepts "not as objective facts about the world, but as the outcome of discursive processes that operate across intersecting scales of space and time” \cite{chun2015language}. That is because, the process through which language itself is racialized, or the way language racializes certain groups of people over time, inevitably involves linking certain objects, ideas, and themes to a racial group \cite{briggs_stories_2003, hoffman_review_2012, inoue_word_2018}, thereby concretizing stereotypes about a particular race, as shown in our findings. For example, the dominant themes that emerge across acts tend to link Blackness with \textbf{crime, sexual exoticism, and questionable belonging to the human race}, which falsely perpetuate racial tropes about Black people’s \textbf{personality} (e.g., \textit{funny, loud, dumb, creepy, ghetto, etc.}), \textbf{ability} (\textit{sports, intelligent, IQ, etc.}), and \textbf{appearance} (\textit{fat, hair,} etc.). Interestingly, many of these identical themes appear in recalls, wherein Black users engage in what critical race scholars describe as counter-storytelling. Counter-storytelling is the act of recounting an individual’s experience with racism, typically through language that operates as a discursive tool for challenging majoritarian perspectives in culturally dominant discourses on race \cite{10.2307/1073418}. Stereotypes perpetuated through racial attitudes, and conceptions of race and racism that have persisted across centuries, tend to become normalized into culturally dominant narratives \cite{delgado1991images}. As a result, implicit racism as observed through online acts of racial microaggressions in our data can falsely appear as race neutral. Black users, as shown in our findings, in essence, call-out such biases through counter-storytelling, through which they directly challenge racial stereotypes and attitudes by conversing on the same topics and themes present in the acts through autobiographical language. Sociotechnical systems that fail to distinguish acts and recalls risk suppressing these counter-stories shared by Black users. Both critical race scholars and historians argue that sharing personal stories has always been essential to the survival and liberation of racially oppressed groups \cite{delgado1989we}. Ensuring sociotechnical systems that safeguard rather than impede important conversations and experiential knowledge shared through counter-stories, such as the ones shown in this work, are critical to establishing more inclusive and enriching environments for online discourse. 

\section{Conclusion}
Through this work, we call for a deeper understanding of the semantic differences between acts and recalls of racial microaggressions on social media, and a re-evaluation of how users and current socio-technical systems differentiate the two. As an initial step towards this effort, we manually curated a corpus of acts and recalls, which were discussed, hand-annotated, and verified by Black participants through a workshop session. We then used this data to classify, interpret and characterize the language underlying acts vs. recalls of  racial microaggressions associated with Black people. By doing so, we provide an empirical characterization of the underlying themes, contexts, and the linguistic signature between acts and recalls. 

\section{Limitations}
While our research is the first to systematically investigate acts and recalls of racial microaggressions comprehensively, our work is not without limitations. First, our analysis is limited to the context of racism in the U.S. Hence, implications around our findings may not be generalized to foreign contexts of racism against Blacks in other countries. Second, since Reddit is a global site, we have limited understanding of whether all the posters are from the U.S or not, which skews the earlier assumption of our analysis being limited to the context of racism in the U.S. Third, given the limited Black population in our area, we were only able to recruit Black college students for our workshop discussions. We plan to extend this study to a broader group in the future. Further, obtaining “ground-truth” labels for discursive data such as ours, are often fraught with subjective interpretations of race and social values linked with race-related matters, which are subject to the annotator’s own perspectives on racism, personal experiences, identity, and social background. Hence, while we endeavored towards obtaining “ground-truth” labels by discerning insights from discussions with Black participants through our workshop, we acknowledge that this process too, can be subject to biases. Finally, in our data collection, we excluded posts with neutral references to race to maintain focus on acts and recalls of microaggressions. This decision, while necessary for our study, may limit the generalizability of our machine learning models to broader or different types of contexts.

\section{Acknowledgments}

We thank our workshop participants for providing helpful discussions and sharing their personal experiences with us.

\bibliographystyle{ACM-Reference-Format}
\bibliography{sample-base}


\begin{thebibliography}{163}


\ifx \showCODEN    \undefined \def \showCODEN     #1{\unskip}     \fi
\ifx \showDOI      \undefined \def \showDOI       #1{#1}\fi
\ifx \showISBNx    \undefined \def \showISBNx     #1{\unskip}     \fi
\ifx \showISBNxiii \undefined \def \showISBNxiii  #1{\unskip}     \fi
\ifx \showISSN     \undefined \def \showISSN      #1{\unskip}     \fi
\ifx \showLCCN     \undefined \def \showLCCN      #1{\unskip}     \fi
\ifx \shownote     \undefined \def \shownote      #1{#1}          \fi
\ifx \showarticletitle \undefined \def \showarticletitle #1{#1}   \fi
\ifx \showURL      \undefined \def \showURL       {\relax}        \fi
\providecommand\bibfield[2]{#2}
\providecommand\bibinfo[2]{#2}
\providecommand\natexlab[1]{#1}
\providecommand\showeprint[2][]{arXiv:#2}

\bibitem[pap(9 10)]%
        {papacharissi_social_2010}
 \bibinfo{year}{2010-09-10}\natexlab{}.
\newblock \showarticletitle{Social Network Sites as Networked Publics: Affordances, Dynamics, and Implications}.
\newblock In \bibinfo{booktitle}{\emph{A Networked Self} (\bibinfo{edition}{0} ed.)}, \bibfield{editor}{\bibinfo{person}{Zizi Papacharissi}} (Ed.). \bibinfo{publisher}{Routledge}, \bibinfo{pages}{47--66}.
\newblock
\showISBNx{978-0-203-87652-7}
\urldef\tempurl%
\url{https://doi.org/10.4324/9780203876527-8}
\showDOI{\tempurl}


\bibitem[Ackerman-Barger et~al\mbox{.}(2020)]%
        {AckermanBarger2020}
\bibfield{author}{\bibinfo{person}{Kupiri Ackerman-Barger}, \bibinfo{person}{Dowin Boatright}, \bibinfo{person}{Rosana Gonzalez-Colaso}, \bibinfo{person}{Regina Orozco}, {and} \bibinfo{person}{Darin Latimore}.} \bibinfo{year}{2020}\natexlab{}.
\newblock \showarticletitle{Seeking Inclusion Excellence}.
\newblock \bibinfo{journal}{\emph{Academic Medicine}} \bibinfo{volume}{95}, \bibinfo{number}{5} (\bibinfo{date}{May} \bibinfo{year}{2020}), \bibinfo{pages}{758--763}.
\newblock
\urldef\tempurl%
\url{https://doi.org/10.1097/acm.0000000000003077}
\showDOI{\tempurl}


\bibitem[Ali et~al\mbox{.}(6 21)]%
        {ali_understanding_2021}
\bibfield{author}{\bibinfo{person}{Shiza Ali}, \bibinfo{person}{Mohammad~Hammas Saeed}, \bibinfo{person}{Esraa Aldreabi}, \bibinfo{person}{Jeremy Blackburn}, \bibinfo{person}{Emiliano De~Cristofaro}, \bibinfo{person}{Savvas Zannettou}, {and} \bibinfo{person}{Gianluca Stringhini}.} \bibinfo{year}{2021-06-21}\natexlab{}.
\newblock \showarticletitle{Understanding the Effect of Deplatforming on Social Networks}. In \bibinfo{booktitle}{\emph{13th {ACM} Web Science Conference 2021}} (New York, {NY}, {USA}) \emph{(\bibinfo{series}{{WebSci} '21})}. \bibinfo{publisher}{Association for Computing Machinery}, \bibinfo{pages}{187--195}.
\newblock
\showISBNx{978-1-4503-8330-1}
\urldef\tempurl%
\url{https://doi.org/10.1145/3447535.3462637}
\showDOI{\tempurl}


\bibitem[Allyn(2020)]%
        {allyn_2020}
\bibfield{author}{\bibinfo{person}{Bobby Allyn}.} \bibinfo{year}{2020}\natexlab{}.
\newblock \bibinfo{title}{{It's `our fault': Nextdoor CEO takes blame for deleting of black lives matter posts}}.
\newblock
\newblock
\urldef\tempurl%
\url{https://tinyurl.com/3353yeb4}
\showURL{%
\tempurl}


\bibitem[Andalibi et~al\mbox{.}(5 07)]%
        {andalibi_understanding_2016}
\bibfield{author}{\bibinfo{person}{Nazanin Andalibi}, \bibinfo{person}{Oliver~L. Haimson}, \bibinfo{person}{Munmun De~Choudhury}, {and} \bibinfo{person}{Andrea Forte}.} \bibinfo{year}{2016-05-07}\natexlab{}.
\newblock \showarticletitle{Understanding Social Media Disclosures of Sexual Abuse Through the Lenses of Support Seeking and Anonymity}. In \bibinfo{booktitle}{\emph{Proceedings of the 2016 {CHI} Conference on Human Factors in Computing Systems}} (San Jose California {USA}). \bibinfo{publisher}{{ACM}}, \bibinfo{pages}{3906--3918}.
\newblock
\showISBNx{978-1-4503-3362-7}
\urldef\tempurl%
\url{https://doi.org/10.1145/2858036.2858096}
\showDOI{\tempurl}


\bibitem[Ashktorab and Vitak(2016)]%
        {10.1145/2858036.2858548}
\bibfield{author}{\bibinfo{person}{Zahra Ashktorab} {and} \bibinfo{person}{Jessica Vitak}.} \bibinfo{year}{2016}\natexlab{}.
\newblock \showarticletitle{Designing Cyberbullying Mitigation and Prevention Solutions through Participatory Design With Teenagers}. In \bibinfo{booktitle}{\emph{Proceedings of the 2016 CHI Conference on Human Factors in Computing Systems}} (San Jose, California, USA) \emph{(\bibinfo{series}{CHI '16})}. \bibinfo{publisher}{Association for Computing Machinery}, \bibinfo{address}{New York, NY, USA}, \bibinfo{pages}{3895–3905}.
\newblock
\showISBNx{9781450333627}
\urldef\tempurl%
\url{https://doi.org/10.1145/2858036.2858548}
\showDOI{\tempurl}


\bibitem[Assari and Moghani~Lankarani(2018)]%
        {assari_depressive_2018}
\bibfield{author}{\bibinfo{person}{Shervin Assari} {and} \bibinfo{person}{Maryam Moghani~Lankarani}.} \bibinfo{year}{2018}\natexlab{}.
\newblock \showarticletitle{Depressive Symptoms and Self-Esteem in White and Black Older Adults in the United States}.
\newblock \bibinfo{journal}{\emph{Brain sciences}} \bibinfo{volume}{8}, \bibinfo{number}{6} (\bibinfo{year}{2018}), \bibinfo{pages}{105}.
\newblock


\bibitem[Balayn et~al\mbox{.}(2021)]%
        {10.1145/3479158}
\bibfield{author}{\bibinfo{person}{Agathe Balayn}, \bibinfo{person}{Jie Yang}, \bibinfo{person}{Zoltan Szlavik}, {and} \bibinfo{person}{Alessandro Bozzon}.} \bibinfo{year}{2021}\natexlab{}.
\newblock \showarticletitle{Automatic Identification of Harmful, Aggressive, Abusive, and Offensive Language on the Web: A Survey of Technical Biases Informed by Psychology Literature}.
\newblock \bibinfo{journal}{\emph{Trans. Soc. Comput.}} \bibinfo{volume}{4}, \bibinfo{number}{3}, Article \bibinfo{articleno}{11} (\bibinfo{date}{oct} \bibinfo{year}{2021}), \bibinfo{numpages}{56}~pages.
\newblock
\showISSN{2469-7818}
\urldef\tempurl%
\url{https://doi.org/10.1145/3479158}
\showDOI{\tempurl}


\bibitem[Banks(2010)]%
        {banks2010regulating}
\bibfield{author}{\bibinfo{person}{James Banks}.} \bibinfo{year}{2010}\natexlab{}.
\newblock \showarticletitle{Regulating hate speech online}.
\newblock \bibinfo{journal}{\emph{International Review of Law, Computers \& Technology}} \bibinfo{volume}{24}, \bibinfo{number}{3} (\bibinfo{year}{2010}), \bibinfo{pages}{233--239}.
\newblock


\bibitem[Bar-Tal(2017a)]%
        {bar-tal_self-censorship_2017-1}
\bibfield{author}{\bibinfo{person}{Daniel Bar-Tal}.} \bibinfo{year}{2017}\natexlab{a}.
\newblock \showarticletitle{Self-censorship as a socio-political-psychological phenomenon: Conception and research}.
\newblock \bibinfo{journal}{\emph{Political Psychology}}  \bibinfo{volume}{38} (\bibinfo{year}{2017}), \bibinfo{pages}{37--65}.
\newblock


\bibitem[Bar-Tal(2017b)]%
        {bar-tal_self-censorship_2017}
\bibfield{author}{\bibinfo{person}{Daniel Bar-Tal}.} \bibinfo{year}{2017}\natexlab{b}.
\newblock \showarticletitle{Self-censorship: The conceptual framework}.
\newblock \bibinfo{journal}{\emph{Self-censorship in contexts of conflict: Theory and research}} (\bibinfo{year}{2017}), \bibinfo{pages}{1--18}.
\newblock


\bibitem[Barnes(2015)]%
        {10.2307/43670469}
\bibfield{author}{\bibinfo{person}{Mario~L. Barnes}.} \bibinfo{year}{2015}\natexlab{}.
\newblock \bibinfo{journal}{\emph{Law \& Society Review}} \bibinfo{volume}{49}, \bibinfo{number}{1} (\bibinfo{year}{2015}), \bibinfo{pages}{279--282}.
\newblock
\showISSN{00239216, 15405893}
\urldef\tempurl%
\url{http://www.jstor.org/stable/43670469}
\showURL{%
\tempurl}


\bibitem[Baumeister et~al\mbox{.}(2001)]%
        {baumeister_bad_2001}
\bibfield{author}{\bibinfo{person}{Roy~F Baumeister}, \bibinfo{person}{Ellen Bratslavsky}, \bibinfo{person}{Catrin Finkenauer}, {and} \bibinfo{person}{Kathleen~D Vohs}.} \bibinfo{year}{2001}\natexlab{}.
\newblock \showarticletitle{Bad is stronger than good}.
\newblock \bibinfo{journal}{\emph{Review of general psychology}} \bibinfo{volume}{5}, \bibinfo{number}{4} (\bibinfo{year}{2001}), \bibinfo{pages}{323--370}.
\newblock


\bibitem[Blodgett et~al\mbox{.}(2016)]%
        {blodgett2016demographic}
\bibfield{author}{\bibinfo{person}{Su~Lin Blodgett}, \bibinfo{person}{Lisa Green}, {and} \bibinfo{person}{Brendan O'Connor}.} \bibinfo{year}{2016}\natexlab{}.
\newblock \showarticletitle{Demographic dialectal variation in social media: A case study of African-American English}.
\newblock \bibinfo{journal}{\emph{arXiv preprint arXiv:1608.08868}} (\bibinfo{year}{2016}).
\newblock


\bibitem[Blodgett et~al\mbox{.}(2022)]%
        {10.1145/3491101.3516502}
\bibfield{author}{\bibinfo{person}{Su~Lin Blodgett}, \bibinfo{person}{Q.~Vera Liao}, \bibinfo{person}{Alexandra Olteanu}, \bibinfo{person}{Rada Mihalcea}, \bibinfo{person}{Michael Muller}, \bibinfo{person}{Morgan~Klaus Scheuerman}, \bibinfo{person}{Chenhao Tan}, {and} \bibinfo{person}{Qian Yang}.} \bibinfo{year}{2022}\natexlab{}.
\newblock \showarticletitle{{Responsible Language Technologies: Foreseeing and Mitigating Harms}}. In \bibinfo{booktitle}{\emph{Extended Abstracts of the 2022 CHI Conference on Human Factors in Computing Systems}} (New Orleans, LA, USA) \emph{(\bibinfo{series}{CHI EA '22})}. \bibinfo{publisher}{Association for Computing Machinery}, \bibinfo{address}{New York, NY, USA}, Article \bibinfo{articleno}{152}, \bibinfo{numpages}{3}~pages.
\newblock
\showISBNx{9781450391566}
\urldef\tempurl%
\url{https://doi.org/10.1145/3491101.3516502}
\showDOI{\tempurl}


\bibitem[Blume et~al\mbox{.}(2012)]%
        {aw_relationship_2012}
\bibfield{author}{\bibinfo{person}{Arthur~W Blume}, \bibinfo{person}{Laura~V Lovato}, \bibinfo{person}{Bryan~N Thyken}, {and} \bibinfo{person}{Natasha Denny}.} \bibinfo{year}{2012}\natexlab{}.
\newblock \showarticletitle{The relationship of microaggressions with alcohol use and anxiety among ethnic minority college students in a historically White institution.}
\newblock \bibinfo{journal}{\emph{Cultural diversity and ethnic minority psychology}} \bibinfo{volume}{18}, \bibinfo{number}{1} (\bibinfo{year}{2012}), \bibinfo{pages}{45}.
\newblock


\bibitem[Bolukbasi et~al\mbox{.}(2016)]%
        {bolukbasi2016man}
\bibfield{author}{\bibinfo{person}{Tolga Bolukbasi}, \bibinfo{person}{Kai-Wei Chang}, \bibinfo{person}{James~Y Zou}, \bibinfo{person}{Venkatesh Saligrama}, {and} \bibinfo{person}{Adam~T Kalai}.} \bibinfo{year}{2016}\natexlab{}.
\newblock \showarticletitle{{Man is to computer programmer as woman is to homemaker? debiasing word embeddings}}.
\newblock \bibinfo{journal}{\emph{Advances in neural information processing systems}}  \bibinfo{volume}{29} (\bibinfo{year}{2016}).
\newblock


\bibitem[Bonam et~al\mbox{.}(2016)]%
        {bonam_polluting_2016}
\bibfield{author}{\bibinfo{person}{Courtney~M. Bonam}, \bibinfo{person}{Hilary~B. Bergsieker}, {and} \bibinfo{person}{Jennifer~L. Eberhardt}.} \bibinfo{year}{2016}\natexlab{}.
\newblock \showarticletitle{Polluting Black space}.
\newblock   \bibinfo{volume}{145} (\bibinfo{year}{2016}), \bibinfo{pages}{1561--1582}.
\newblock
\showISSN{1939-2222}
\urldef\tempurl%
\url{https://doi.org/10.1037/xge0000226}
\showDOI{\tempurl}
\newblock
\shownote{Place: {US} Publisher: American Psychological Association}.


\bibitem[Borsotti and Bjorn(2022)]%
        {cscw-humour}
\bibfield{author}{\bibinfo{person}{Valeria Borsotti} {and} \bibinfo{person}{Pernille Bjorn}.} \bibinfo{year}{2022}\natexlab{}.
\newblock \showarticletitle{Humor and Stereotypes in Computing: An Equity-focused Approach to Institutional Accountability}.
\newblock \bibinfo{journal}{\emph{Computer Supported Cooperative Work (CSCW)}}  \bibinfo{volume}{31} (\bibinfo{date}{07} \bibinfo{year}{2022}).
\newblock
\urldef\tempurl%
\url{https://doi.org/10.1007/s10606-022-09440-9}
\showDOI{\tempurl}


\bibitem[Boyd(2010)]%
        {boyd2010social}
\bibfield{author}{\bibinfo{person}{Danah Boyd}.} \bibinfo{year}{2010}\natexlab{}.
\newblock \showarticletitle{Social Network Sites as Networked Publics: Affordances, Dynamics, and Implications}.
\newblock
\urldef\tempurl%
\url{https://api.semanticscholar.org/CorpusID:158379198}
\showURL{%
\tempurl}


\bibitem[Breitfeller et~al\mbox{.}(9 11)]%
        {breitfeller_finding_2019}
\bibfield{author}{\bibinfo{person}{Luke Breitfeller}, \bibinfo{person}{Emily Ahn}, \bibinfo{person}{David Jurgens}, {and} \bibinfo{person}{Yulia Tsvetkov}.} \bibinfo{year}{2019-11}\natexlab{}.
\newblock \showarticletitle{{Finding Microaggressions in the Wild: A Case for Locating Elusive Phenomena in Social Media Posts}}. In \bibinfo{booktitle}{\emph{Proceedings of the 2019 Conference on Empirical Methods in Natural Language Processing and the 9th International Joint Conference on Natural Language Processing ({EMNLP}-{IJCNLP})}} (Hong Kong, China). \bibinfo{publisher}{Association for Computational Linguistics}, \bibinfo{pages}{1664--1674}.
\newblock
\urldef\tempurl%
\url{https://doi.org/10.18653/v1/D19-1176}
\showDOI{\tempurl}


\bibitem[Bruner(1957)]%
        {bruner_perceptual_1957}
\bibfield{author}{\bibinfo{person}{Jerome~S Bruner}.} \bibinfo{year}{1957}\natexlab{}.
\newblock \showarticletitle{On perceptual readiness.}
\newblock \bibinfo{journal}{\emph{Psychological review}} \bibinfo{volume}{64}, \bibinfo{number}{2} (\bibinfo{year}{1957}), \bibinfo{pages}{123}.
\newblock


\bibitem[Caselli et~al\mbox{.}(2020)]%
        {caselli-etal-2020-feel}
\bibfield{author}{\bibinfo{person}{Tommaso Caselli}, \bibinfo{person}{Valerio Basile}, \bibinfo{person}{Jelena Mitrovi{\'c}}, \bibinfo{person}{Inga Kartoziya}, {and} \bibinfo{person}{Michael Granitzer}.} \bibinfo{year}{2020}\natexlab{}.
\newblock \showarticletitle{{I} Feel Offended, Don{'}t Be Abusive! Implicit/Explicit Messages in Offensive and Abusive Language}. In \bibinfo{booktitle}{\emph{Proceedings of the Twelfth Language Resources and Evaluation Conference}}. \bibinfo{publisher}{European Language Resources Association}, \bibinfo{address}{Marseille, France}, \bibinfo{pages}{6193--6202}.
\newblock
\showISBNx{979-10-95546-34-4}
\urldef\tempurl%
\url{https://aclanthology.org/2020.lrec-1.760}
\showURL{%
\tempurl}


\bibitem[Cavazza and Guidetti(2014)]%
        {cavazza_swearing_2014}
\bibfield{author}{\bibinfo{person}{Nicoletta Cavazza} {and} \bibinfo{person}{Margherita Guidetti}.} \bibinfo{year}{2014}\natexlab{}.
\newblock \showarticletitle{Swearing in political discourse: Why vulgarity works}.
\newblock \bibinfo{journal}{\emph{Journal of Language and Social Psychology}} \bibinfo{volume}{33}, \bibinfo{number}{5} (\bibinfo{year}{2014}), \bibinfo{pages}{537--547}.
\newblock


\bibitem[Chancellor et~al\mbox{.}(2017)]%
        {10.1145/3025453.3025985}
\bibfield{author}{\bibinfo{person}{Stevie Chancellor}, \bibinfo{person}{Yannis Kalantidis}, \bibinfo{person}{Jessica~A. Pater}, \bibinfo{person}{Munmun De~Choudhury}, {and} \bibinfo{person}{David~A. Shamma}.} \bibinfo{year}{2017}\natexlab{}.
\newblock \showarticletitle{{Multimodal Classification of Moderated Online Pro-Eating Disorder Content}}. In \bibinfo{booktitle}{\emph{Proceedings of the 2017 CHI Conference on Human Factors in Computing Systems}} (Denver, Colorado, USA) \emph{(\bibinfo{series}{CHI '17})}. \bibinfo{publisher}{Association for Computing Machinery}, \bibinfo{address}{New York, NY, USA}, \bibinfo{pages}{3213–3226}.
\newblock
\showISBNx{9781450346559}
\urldef\tempurl%
\url{https://doi.org/10.1145/3025453.3025985}
\showDOI{\tempurl}


\bibitem[Chancellor et~al\mbox{.}(2016)]%
        {10.1145/2858036.2858248}
\bibfield{author}{\bibinfo{person}{Stevie Chancellor}, \bibinfo{person}{Zhiyuan~(Jerry) Lin}, {and} \bibinfo{person}{Munmun De~Choudhury}.} \bibinfo{year}{2016}\natexlab{}.
\newblock \showarticletitle{``{This Post Will Just Get Taken Down}": {Characterizing Removed Pro-Eating Disorder Social Media Content}}. In \bibinfo{booktitle}{\emph{Proceedings of the 2016 CHI Conference on Human Factors in Computing Systems}} (San Jose, California, USA) \emph{(\bibinfo{series}{CHI '16})}. \bibinfo{publisher}{Association for Computing Machinery}, \bibinfo{address}{New York, NY, USA}, \bibinfo{pages}{1157–1162}.
\newblock
\showISBNx{9781450333627}
\urldef\tempurl%
\url{https://doi.org/10.1145/2858036.2858248}
\showDOI{\tempurl}


\bibitem[Chandrasekharan et~al\mbox{.}(2019)]%
        {chandrasekharan2019crossmod}
\bibfield{author}{\bibinfo{person}{Eshwar Chandrasekharan}, \bibinfo{person}{Chaitrali Gandhi}, \bibinfo{person}{Matthew~Wortley Mustelier}, {and} \bibinfo{person}{Eric Gilbert}.} \bibinfo{year}{2019}\natexlab{}.
\newblock \showarticletitle{{Crossmod: A Cross-Community Learning-based System to Assist Reddit Moderators}}.
\newblock \bibinfo{journal}{\emph{Proceedings of the ACM on human-computer interaction}} \bibinfo{volume}{3}, \bibinfo{number}{CSCW} (\bibinfo{year}{2019}), \bibinfo{pages}{1--30}.
\newblock


\bibitem[Chandrasekharan et~al\mbox{.}(2018)]%
        {chandrasekharan2018internet}
\bibfield{author}{\bibinfo{person}{Eshwar Chandrasekharan}, \bibinfo{person}{Mattia Samory}, \bibinfo{person}{Shagun Jhaver}, \bibinfo{person}{Hunter Charvat}, \bibinfo{person}{Amy Bruckman}, \bibinfo{person}{Cliff Lampe}, \bibinfo{person}{Jacob Eisenstein}, {and} \bibinfo{person}{Eric Gilbert}.} \bibinfo{year}{2018}\natexlab{}.
\newblock \showarticletitle{The {Internet's} hidden rules: An empirical study of {Reddit} norm violations at micro, meso, and macro scales}.
\newblock \bibinfo{journal}{\emph{Proceedings of the ACM on Human-Computer Interaction}} \bibinfo{volume}{2}, \bibinfo{number}{CSCW} (\bibinfo{year}{2018}), \bibinfo{pages}{1--25}.
\newblock


\bibitem[Chen et~al\mbox{.}(2021)]%
        {chen_toward_2021}
\bibfield{author}{\bibinfo{person}{Lijuan Chen}, \bibinfo{person}{Yanyan Lan}, \bibinfo{person}{Liang Pang}, \bibinfo{person}{Jiafeng Guo}, {and} \bibinfo{person}{Xueqi Cheng}.} \bibinfo{year}{2021}\natexlab{}.
\newblock \showarticletitle{{Toward the Understanding of Deep Text Matching Models for Information Retrieval}}.
\newblock \bibinfo{journal}{\emph{arXiv preprint arXiv:2108.07081}} (\bibinfo{year}{2021}).
\newblock


\bibitem[Chun and Lo(2015)]%
        {chun2015language}
\bibfield{author}{\bibinfo{person}{Elaine~W Chun} {and} \bibinfo{person}{Adrienne Lo}.} \bibinfo{year}{2015}\natexlab{}.
\newblock \showarticletitle{{Language and racialization}}.
\newblock In \bibinfo{booktitle}{\emph{The Routledge handbook of linguistic anthropology}}. \bibinfo{publisher}{Routledge}, \bibinfo{pages}{220--233}.
\newblock


\bibitem[Cook et~al\mbox{.}(2022)]%
        {Cook2022AweVA}
\bibfield{author}{\bibinfo{person}{Chrissy Cook}, \bibinfo{person}{Jie Cai}, {and} \bibinfo{person}{Donghee~Yvette Wohn}.} \bibinfo{year}{2022}\natexlab{}.
\newblock \showarticletitle{{Awe Versus Aww: The Effectiveness of Two Kinds of Positive Emotional Stimulation on Stress Reduction for Online Content Moderators}}.
\newblock \bibinfo{journal}{\emph{Proceedings of the ACM on Human-Computer Interaction}}  \bibinfo{volume}{6} (\bibinfo{year}{2022}), \bibinfo{pages}{1 -- 19}.
\newblock


\bibitem[Corbin and Strauss(2008)]%
        {Corbin2008BasicsOQ}
\bibfield{author}{\bibinfo{person}{Juliet~M. Corbin} {and} \bibinfo{person}{Anselm Strauss}.} \bibinfo{year}{2008}\natexlab{}.
\newblock \showarticletitle{Basics of Qualitative Research (3rd ed.): Techniques and Procedures for Developing Grounded Theory}.
\newblock


\bibitem[Danescu-Niculescu-Mizil et~al\mbox{.}(2012)]%
        {danescu-niculescu-mizil_echoes_2012}
\bibfield{author}{\bibinfo{person}{Cristian Danescu-Niculescu-Mizil}, \bibinfo{person}{Lillian Lee}, \bibinfo{person}{Bo Pang}, {and} \bibinfo{person}{Jon Kleinberg}.} \bibinfo{year}{2012}\natexlab{}.
\newblock \showarticletitle{Echoes of Power: Language Effects and Power Differences in Social Interaction}. In \bibinfo{booktitle}{\emph{Proceedings of the 21st International Conference on World Wide Web}} (Lyon, France) \emph{(\bibinfo{series}{WWW '12})}. \bibinfo{publisher}{Association for Computing Machinery}, \bibinfo{address}{New York, NY, USA}, \bibinfo{pages}{699–708}.
\newblock
\showISBNx{9781450312295}
\urldef\tempurl%
\url{https://doi.org/10.1145/2187836.2187931}
\showDOI{\tempurl}


\bibitem[Das et~al\mbox{.}(2020)]%
        {Das2020FastAA}
\bibfield{author}{\bibinfo{person}{Anubrata Das}, \bibinfo{person}{Brandon Dang}, {and} \bibinfo{person}{Matthew Lease}.} \bibinfo{year}{2020}\natexlab{}.
\newblock \showarticletitle{Fast, Accurate, and Healthier: Interactive Blurring Helps Moderators Reduce Exposure to Harmful Content}. In \bibinfo{booktitle}{\emph{AAAI Conference on Human Computation \& Crowdsourcing}}.
\newblock


\bibitem[Delgado and Stefancic(1989)]%
        {delgado1989we}
\bibfield{author}{\bibinfo{person}{Richard Delgado} {and} \bibinfo{person}{Jean Stefancic}.} \bibinfo{year}{1989}\natexlab{}.
\newblock \showarticletitle{Why do we tell the same stories?: Law reform, critical librarianship, and the triple helix dilemma}.
\newblock \bibinfo{journal}{\emph{Stanford Law Review}} (\bibinfo{year}{1989}), \bibinfo{pages}{207--225}.
\newblock


\bibitem[Delgado and Stefancic(1991)]%
        {delgado1991images}
\bibfield{author}{\bibinfo{person}{Richard Delgado} {and} \bibinfo{person}{Jean Stefancic}.} \bibinfo{year}{1991}\natexlab{}.
\newblock \showarticletitle{{Images of the outsider in American law and culture: Can free expression remedy systemic social ills}}.
\newblock \bibinfo{journal}{\emph{Cornell L. Rev.}}  \bibinfo{volume}{77} (\bibinfo{year}{1991}), \bibinfo{pages}{1258}.
\newblock


\bibitem[Delgado and Stefancic(1993)]%
        {10.2307/1073418}
\bibfield{author}{\bibinfo{person}{Richard Delgado} {and} \bibinfo{person}{Jean Stefancic}.} \bibinfo{year}{1993}\natexlab{}.
\newblock \showarticletitle{Critical Race Theory: An Annotated Bibliography}.
\newblock \bibinfo{journal}{\emph{Virginia Law Review}} \bibinfo{volume}{79}, \bibinfo{number}{2} (\bibinfo{year}{1993}), \bibinfo{pages}{461--516}.
\newblock
\showISSN{00426601}
\urldef\tempurl%
\url{http://www.jstor.org/stable/1073418}
\showURL{%
\tempurl}


\bibitem[Ding et~al\mbox{.}(2023)]%
        {ig_tv}
\bibfield{author}{\bibinfo{person}{Xiaohan Ding}, \bibinfo{person}{Michael Horning}, {and} \bibinfo{person}{Eugenia~H. Rho}.} \bibinfo{year}{2023}\natexlab{}.
\newblock \showarticletitle{Same Words, Different Meanings: Semantic Polarization in Broadcast Media Language Forecasts Polarity in Online Public Discourse}.
\newblock \bibinfo{journal}{\emph{Proceedings of the International AAAI Conference on Web and Social Media}} \bibinfo{volume}{17}, \bibinfo{number}{1} (\bibinfo{date}{Jun.} \bibinfo{year}{2023}), \bibinfo{pages}{161--172}.
\newblock
\urldef\tempurl%
\url{https://doi.org/10.1609/icwsm.v17i1.22135}
\showDOI{\tempurl}


\bibitem[D{\"o}ring and Mohseni(2019)]%
        {doring_male_2019}
\bibfield{author}{\bibinfo{person}{Nicola D{\"o}ring} {and} \bibinfo{person}{M~Rohangis Mohseni}.} \bibinfo{year}{2019}\natexlab{}.
\newblock \showarticletitle{Male dominance and sexism on YouTube: results of three content analyses}.
\newblock \bibinfo{journal}{\emph{Feminist Media Studies}} \bibinfo{volume}{19}, \bibinfo{number}{4} (\bibinfo{year}{2019}), \bibinfo{pages}{512--524}.
\newblock


\bibitem[Dosono(2018)]%
        {10.1145/3148330.3152697}
\bibfield{author}{\bibinfo{person}{Bryan Dosono}.} \bibinfo{year}{2018}\natexlab{}.
\newblock \showarticletitle{{AAPI Identity Work on Reddit: Toward Social Support and Collective Action}}. In \bibinfo{booktitle}{\emph{Proceedings of the 2018 ACM Conference on Supporting Groupwork}} (Sanibel Island, Florida, USA) \emph{(\bibinfo{series}{GROUP '18})}. \bibinfo{publisher}{Association for Computing Machinery}, \bibinfo{address}{New York, NY, USA}, \bibinfo{pages}{373–378}.
\newblock
\showISBNx{9781450355629}
\urldef\tempurl%
\url{https://doi.org/10.1145/3148330.3152697}
\showDOI{\tempurl}


\bibitem[Dosono and Semaan(2019)]%
        {10.1145/3290605.3300372}
\bibfield{author}{\bibinfo{person}{Bryan Dosono} {and} \bibinfo{person}{Bryan Semaan}.} \bibinfo{year}{2019}\natexlab{}.
\newblock \showarticletitle{{Moderation Practices as Emotional Labor in Sustaining Online Communities: The Case of AAPI Identity Work on Reddit}}. In \bibinfo{booktitle}{\emph{Proceedings of the 2019 CHI Conference on Human Factors in Computing Systems}} (Glasgow, Scotland Uk) \emph{(\bibinfo{series}{CHI '19})}. \bibinfo{publisher}{Association for Computing Machinery}, \bibinfo{address}{New York, NY, USA}, \bibinfo{pages}{1–13}.
\newblock
\showISBNx{9781450359702}
\urldef\tempurl%
\url{https://doi.org/10.1145/3290605.3300372}
\showDOI{\tempurl}


\bibitem[Du et~al\mbox{.}(2021)]%
        {du_towards_2021}
\bibfield{author}{\bibinfo{person}{Mengnan Du}, \bibinfo{person}{Varun Manjunatha}, \bibinfo{person}{Rajiv Jain}, \bibinfo{person}{Ruchi Deshpande}, \bibinfo{person}{Franck Dernoncourt}, \bibinfo{person}{Jiuxiang Gu}, \bibinfo{person}{Tong Sun}, {and} \bibinfo{person}{Xia Hu}.} \bibinfo{year}{2021}\natexlab{}.
\newblock \showarticletitle{Towards interpreting and mitigating shortcut learning behavior of {NLU} models}.
\newblock \bibinfo{journal}{\emph{arXiv preprint arXiv:2103.06922}} (\bibinfo{year}{2021}).
\newblock


\bibitem[Duggan(2017)]%
        {duggan_2020}
\bibfield{author}{\bibinfo{person}{Maeve Duggan}.} \bibinfo{year}{2017}\natexlab{}.
\newblock \bibinfo{title}{{1 in 4 Black Americans have faced online harassment because of their race or ethnicity}}.
\newblock
\newblock
\urldef\tempurl%
\url{https://tinyurl.com/5n8n845e}
\showURL{%
\tempurl}


\bibitem[Dulaney(2006)]%
        {10.1111/j.1468-2958.1982.tb00684.x}
\bibfield{author}{\bibinfo{person}{Jr. Dulaney, Earl~F.}} \bibinfo{year}{2006}\natexlab{}.
\newblock \showarticletitle{{Changes in Language Behavior as a Function of Veracity}}.
\newblock \bibinfo{journal}{\emph{Human Communication Research}} \bibinfo{volume}{9}, \bibinfo{number}{1} (\bibinfo{date}{03} \bibinfo{year}{2006}), \bibinfo{pages}{75--82}.
\newblock
\showISSN{0360-3989}
\urldef\tempurl%
\url{https://doi.org/10.1111/j.1468-2958.1982.tb00684.x}
\showDOI{\tempurl}
\showeprint{https://academic.oup.com/hcr/article-PDF/9/1/75/22343175/jhumcom0075.PDF}


\bibitem[Dumitrache et~al\mbox{.}(2021)]%
        {dumitrache2021empirical}
\bibfield{author}{\bibinfo{person}{Anca Dumitrache}, \bibinfo{person}{Oana Inel}, \bibinfo{person}{Benjamin Timmermans}, \bibinfo{person}{Carlos Ortiz}, \bibinfo{person}{Robert-Jan Sips}, \bibinfo{person}{Lora Aroyo}, {and} \bibinfo{person}{Chris Welty}.} \bibinfo{year}{2021}\natexlab{}.
\newblock \showarticletitle{Empirical methodology for crowdsourcing ground truth}.
\newblock \bibinfo{journal}{\emph{Semantic Web}} \bibinfo{volume}{12}, \bibinfo{number}{3} (\bibinfo{year}{2021}), \bibinfo{pages}{403--421}.
\newblock


\bibitem[Dwoskin et~al\mbox{.}(2021)]%
        {dwoskin_tiku_timberg_2021}
\bibfield{author}{\bibinfo{person}{Elizabeth Dwoskin}, \bibinfo{person}{Nitasha Tiku}, {and} \bibinfo{person}{Craig Timberg}.} \bibinfo{year}{2021}\natexlab{}.
\newblock \bibinfo{title}{{Facebook's race-blind practices around hate speech came at the expense of black users, new documents show}}.
\newblock
\newblock
\urldef\tempurl%
\url{https://www.washingtonpost.com/technology/2021/11/21/facebook-algorithm-biased-race/}
\showURL{%
\tempurl}


\bibitem[E~Shaw and Bailey(2009)]%
        {10.1093/fampra/cmp038}
\bibfield{author}{\bibinfo{person}{Sara E~Shaw} {and} \bibinfo{person}{Julia Bailey}.} \bibinfo{year}{2009}\natexlab{}.
\newblock \showarticletitle{{Discourse analysis: what is it and why is it relevant to family practice?}}
\newblock \bibinfo{journal}{\emph{Family Practice}} \bibinfo{volume}{26}, \bibinfo{number}{5} (\bibinfo{date}{06} \bibinfo{year}{2009}), \bibinfo{pages}{413--419}.
\newblock
\showISSN{0263-2136}
\urldef\tempurl%
\url{https://doi.org/10.1093/fampra/cmp038}
\showDOI{\tempurl}
\showeprint{https://academic.oup.com/fampra/article-PDF/26/5/413/1779242/cmp038.PDF}


\bibitem[Ehsan et~al\mbox{.}(2021)]%
        {10.1145/3411764.3445188}
\bibfield{author}{\bibinfo{person}{Upol Ehsan}, \bibinfo{person}{Q.~Vera Liao}, \bibinfo{person}{Michael Muller}, \bibinfo{person}{Mark~O. Riedl}, {and} \bibinfo{person}{Justin~D. Weisz}.} \bibinfo{year}{2021}\natexlab{}.
\newblock \showarticletitle{Expanding Explainability: Towards Social Transparency in {AI} Systems}. In \bibinfo{booktitle}{\emph{Proceedings of the 2021 CHI Conference on Human Factors in Computing Systems}} (Yokohama, Japan) \emph{(\bibinfo{series}{CHI '21})}. \bibinfo{publisher}{Association for Computing Machinery}, \bibinfo{address}{New York, NY, USA}, Article \bibinfo{articleno}{82}, \bibinfo{numpages}{19}~pages.
\newblock
\showISBNx{9781450380966}
\urldef\tempurl%
\url{https://doi.org/10.1145/3411764.3445188}
\showDOI{\tempurl}


\bibitem[{ElSherief} et~al\mbox{.}(1 11)]%
        {elsherief_latent_2021-1}
\bibfield{author}{\bibinfo{person}{Mai {ElSherief}}, \bibinfo{person}{Caleb Ziems}, \bibinfo{person}{David Muchlinski}, \bibinfo{person}{Vaishnavi Anupindi}, \bibinfo{person}{Jordyn Seybolt}, \bibinfo{person}{Munmun De~Choudhury}, {and} \bibinfo{person}{Diyi Yang}.} \bibinfo{year}{2021-11}\natexlab{}.
\newblock \showarticletitle{Latent Hatred: A Benchmark for Understanding Implicit Hate Speech}. In \bibinfo{booktitle}{\emph{Proceedings of the 2021 Conference on Empirical Methods in Natural Language Processing}} (Online and Punta Cana, Dominican Republic). \bibinfo{publisher}{Association for Computational Linguistics}, \bibinfo{pages}{345--363}.
\newblock
\urldef\tempurl%
\url{https://doi.org/10.18653/v1/2021.emnlp-main.29}
\showDOI{\tempurl}


\bibitem[Eschmann(2021)]%
        {doi:10.1177/2332649220933307}
\bibfield{author}{\bibinfo{person}{Rob Eschmann}.} \bibinfo{year}{2021}\natexlab{}.
\newblock \showarticletitle{Digital Resistance: How Online Communication Facilitates Responses to Racial Microaggressions}.
\newblock \bibinfo{journal}{\emph{Sociology of Race and Ethnicity}} \bibinfo{volume}{7}, \bibinfo{number}{2} (\bibinfo{year}{2021}), \bibinfo{pages}{264--277}.
\newblock
\urldef\tempurl%
\url{https://doi.org/10.1177/2332649220933307}
\showDOI{\tempurl}
\showeprint{https://doi.org/10.1177/2332649220933307}


\bibitem[Eschmann et~al\mbox{.}(2020)]%
        {doi:10.1177/2056305120975716}
\bibfield{author}{\bibinfo{person}{Rob Eschmann}, \bibinfo{person}{Jacob Groshek}, \bibinfo{person}{Rachel Chanderdatt}, \bibinfo{person}{Khea Chang}, {and} \bibinfo{person}{Maysa Whyte}.} \bibinfo{year}{2020}\natexlab{}.
\newblock \showarticletitle{Making a Microaggression: Using Big Data and Qualitative Analysis to Map the Reproduction and Disruption of Microaggressions through Social Media}.
\newblock \bibinfo{journal}{\emph{Social Media + Society}} \bibinfo{volume}{6}, \bibinfo{number}{4} (\bibinfo{year}{2020}), \bibinfo{pages}{2056305120975716}.
\newblock
\urldef\tempurl%
\url{https://doi.org/10.1177/2056305120975716}
\showDOI{\tempurl}
\showeprint{https://doi.org/10.1177/2056305120975716}


\bibitem[Espaillat et~al\mbox{.}(2019)]%
        {Espaillat2019AnES}
\bibfield{author}{\bibinfo{person}{Andre Espaillat}, \bibinfo{person}{Danielle Panna}, \bibinfo{person}{Dianne~L. Goede}, \bibinfo{person}{Matthew~J. Gurka}, \bibinfo{person}{Maureen~A Novak}, {and} \bibinfo{person}{Zareen Zaidi}.} \bibinfo{year}{2019}\natexlab{}.
\newblock \showarticletitle{An exploratory study on microaggressions in medical school: What are they and why should we care?}
\newblock \bibinfo{journal}{\emph{Perspectives on Medical Education}}  \bibinfo{volume}{8} (\bibinfo{year}{2019}), \bibinfo{pages}{143 -- 151}.
\newblock


\bibitem[Essed(1991)]%
        {essed_understanding_1991}
\bibfield{author}{\bibinfo{person}{Philomena Essed}.} \bibinfo{year}{1991}\natexlab{}.
\newblock \bibinfo{booktitle}{\emph{Understanding everyday racism: An interdisciplinary theory}}.
\newblock \bibinfo{publisher}{Sage Publications, Inc}.
\newblock
\showISBNx{978-0-8039-4255-4 978-0-8039-4256-1}
\newblock
\shownote{Pages: x, 322}.


\bibitem[Ettinger(2020)]%
        {ettinger2020bert}
\bibfield{author}{\bibinfo{person}{Allyson Ettinger}.} \bibinfo{year}{2020}\natexlab{}.
\newblock \showarticletitle{What BERT is not: Lessons from a new suite of psycholinguistic diagnostics for language models}.
\newblock \bibinfo{journal}{\emph{Transactions of the Association for Computational Linguistics}}  \bibinfo{volume}{8} (\bibinfo{year}{2020}), \bibinfo{pages}{34--48}.
\newblock


\bibitem[Feuston et~al\mbox{.}(2020)]%
        {feuston2020conformity}
\bibfield{author}{\bibinfo{person}{Jessica~L. Feuston}, \bibinfo{person}{Alex~S. Taylor}, {and} \bibinfo{person}{Anne~Marie Piper}.} \bibinfo{year}{2020}\natexlab{}.
\newblock \showarticletitle{Conformity of Eating Disorders through Content Moderation}.
\newblock \bibinfo{journal}{\emph{Proc. ACM Hum.-Comput. Interact.}} \bibinfo{volume}{4}, \bibinfo{number}{CSCW1}, Article \bibinfo{articleno}{40} (\bibinfo{date}{may} \bibinfo{year}{2020}), \bibinfo{numpages}{28}~pages.
\newblock
\urldef\tempurl%
\url{https://doi.org/10.1145/3392845}
\showDOI{\tempurl}


\bibitem[Gebel(2020)]%
        {gebel_2020}
\bibfield{author}{\bibinfo{person}{Meira Gebel}.} \bibinfo{year}{2020}\natexlab{}.
\newblock \bibinfo{title}{{Black creators say {Tiktok} still secretly hides their content}}.
\newblock
\newblock
\urldef\tempurl%
\url{https://www.digitaltrends.com/social-media/black-creators-claim-tiktok-still-secretly-blocking-content/}
\showURL{%
\tempurl}


\bibitem[Gee(2014)]%
        {jimgee}
\bibfield{author}{\bibinfo{person}{James~Paul Gee}.} \bibinfo{year}{2014}\natexlab{}.
\newblock \bibinfo{booktitle}{\emph{An {Introduction} to {Discourse} {Analysis}: Theory and Method}}.
\newblock \bibinfo{publisher}{routledge}.
\newblock


\bibitem[Ghosh et~al\mbox{.}(2011)]%
        {10.1145/1993574.1993599}
\bibfield{author}{\bibinfo{person}{Arpita Ghosh}, \bibinfo{person}{Satyen Kale}, {and} \bibinfo{person}{Preston McAfee}.} \bibinfo{year}{2011}\natexlab{}.
\newblock \showarticletitle{Who Moderates the Moderators? Crowdsourcing Abuse Detection in User-Generated Content}. In \bibinfo{booktitle}{\emph{Proceedings of the 12th ACM Conference on Electronic Commerce}} (San Jose, California, USA) \emph{(\bibinfo{series}{EC '11})}. \bibinfo{publisher}{Association for Computing Machinery}, \bibinfo{address}{New York, NY, USA}, \bibinfo{pages}{167–176}.
\newblock
\showISBNx{9781450302616}
\urldef\tempurl%
\url{https://doi.org/10.1145/1993574.1993599}
\showDOI{\tempurl}


\bibitem[Gibbs(2018)]%
        {Gibbs2007AnalyzingQD}
\bibfield{author}{\bibinfo{person}{Graham~R Gibbs}.} \bibinfo{year}{2018}\natexlab{}.
\newblock \bibinfo{booktitle}{\emph{Analyzing qualitative data}}. Vol.~\bibinfo{volume}{6}.
\newblock \bibinfo{publisher}{Sage}.
\newblock


\bibitem[Gillespie(2018)]%
        {gillespie2018custodians}
\bibfield{author}{\bibinfo{person}{Tarleton Gillespie}.} \bibinfo{year}{2018}\natexlab{}.
\newblock \bibinfo{booktitle}{\emph{Custodians of the Internet: Platforms, content moderation, and the hidden decisions that shape social media}}.
\newblock \bibinfo{publisher}{Yale University Press}.
\newblock


\bibitem[Gillespie(2022)]%
        {fitzsimmons_2021}
\bibfield{author}{\bibinfo{person}{Tarleton Gillespie}.} \bibinfo{year}{2022}\natexlab{}.
\newblock \showarticletitle{{Do Not Recommend? Reduction as a Form of Content Moderation}}.
\newblock \bibinfo{journal}{\emph{Social Media + Society}} \bibinfo{volume}{8}, \bibinfo{number}{3} (\bibinfo{year}{2022}), \bibinfo{pages}{20563051221117552}.
\newblock


\bibitem[Goff et~al\mbox{.}(2008)]%
        {goff_not_2008}
\bibfield{author}{\bibinfo{person}{Phillip~Atiba Goff}, \bibinfo{person}{Jennifer~L Eberhardt}, \bibinfo{person}{Melissa~J Williams}, {and} \bibinfo{person}{Matthew~Christian Jackson}.} \bibinfo{year}{2008}\natexlab{}.
\newblock \showarticletitle{Not yet human: implicit knowledge, historical dehumanization, and contemporary consequences.}
\newblock \bibinfo{journal}{\emph{Journal of personality and social psychology}} \bibinfo{volume}{94}, \bibinfo{number}{2} (\bibinfo{year}{2008}), \bibinfo{pages}{292}.
\newblock


\bibitem[Goodwin(2002)]%
        {briggs_stories_2003}
\bibfield{author}{\bibinfo{person}{Joseph~P Goodwin}.} \bibinfo{year}{2002}\natexlab{}.
\newblock \bibinfo{booktitle}{\emph{Stories in the time of cholera: Racial profiling during a medical nightmare}}.
\newblock \bibinfo{publisher}{JSTOR}.
\newblock


\bibitem[Gorwa et~al\mbox{.}(2020)]%
        {doi:10.1177/2053951719897945}
\bibfield{author}{\bibinfo{person}{Robert Gorwa}, \bibinfo{person}{Reuben Binns}, {and} \bibinfo{person}{Christian Katzenbach}.} \bibinfo{year}{2020}\natexlab{}.
\newblock \showarticletitle{Algorithmic content moderation: Technical and political challenges in the automation of platform governance}.
\newblock \bibinfo{journal}{\emph{Big Data \& Society}} \bibinfo{volume}{7}, \bibinfo{number}{1} (\bibinfo{year}{2020}), \bibinfo{pages}{2053951719897945}.
\newblock
\urldef\tempurl%
\url{https://doi.org/10.1177/2053951719897945}
\showDOI{\tempurl}
\showeprint{https://doi.org/10.1177/2053951719897945}


\bibitem[Grudin(1988)]%
        {grudin1988cscw}
\bibfield{author}{\bibinfo{person}{Jonathan Grudin}.} \bibinfo{year}{1988}\natexlab{}.
\newblock \showarticletitle{{Why CSCW applications fail: problems in the design and evaluationof organizational interfaces}}. In \bibinfo{booktitle}{\emph{Proceedings of the 1988 ACM conference on Computer-supported cooperative work}}. \bibinfo{pages}{85--93}.
\newblock


\bibitem[Guynn(2019)]%
        {guynn_2020}
\bibfield{author}{\bibinfo{person}{Jessica Guynn}.} \bibinfo{year}{July 2019}\natexlab{}.
\newblock \showarticletitle{{Facebook while black: Users call it getting ‘Zucked’ say talking about racism is censored as hate speech}}.
\newblock \bibinfo{journal}{\emph{Usa today}}  \bibinfo{volume}{24} (\bibinfo{year}{July 2019}).
\newblock


\bibitem[Haimson et~al\mbox{.}(2021)]%
        {10.1145/3479610}
\bibfield{author}{\bibinfo{person}{Oliver~L. Haimson}, \bibinfo{person}{Daniel Delmonaco}, \bibinfo{person}{Peipei Nie}, {and} \bibinfo{person}{Andrea Wegner}.} \bibinfo{year}{2021}\natexlab{}.
\newblock \showarticletitle{{Disproportionate Removals and Differing Content Moderation Experiences for Conservative, Transgender, and Black Social Media Users: Marginalization and Moderation Gray Areas}}.
\newblock \bibinfo{journal}{\emph{Proc. ACM Hum.-Comput. Interact.}} \bibinfo{volume}{5}, \bibinfo{number}{CSCW2}, Article \bibinfo{articleno}{466} (\bibinfo{date}{oct} \bibinfo{year}{2021}), \bibinfo{numpages}{35}~pages.
\newblock
\urldef\tempurl%
\url{https://doi.org/10.1145/3479610}
\showDOI{\tempurl}


\bibitem[Hall and Fields(2015)]%
        {hall_its_2015}
\bibfield{author}{\bibinfo{person}{Joanne~M Hall} {and} \bibinfo{person}{Becky Fields}.} \bibinfo{year}{2015}\natexlab{}.
\newblock \showarticletitle{{“It’s} killing us!” {Narratives} of {Black} adults about microaggression experiences and related health stress}.
\newblock \bibinfo{journal}{\emph{Global qualitative nursing research}}  \bibinfo{volume}{2} (\bibinfo{year}{2015}), \bibinfo{pages}{2333393615591569}.
\newblock


\bibitem[Haynes et~al\mbox{.}(2016)]%
        {haynes_three_2016}
\bibfield{author}{\bibinfo{person}{Chayla Haynes}, \bibinfo{person}{Saran Stewart}, {and} \bibinfo{person}{Evette Allen}.} \bibinfo{year}{2016}\natexlab{}.
\newblock \showarticletitle{Three paths, one struggle: Black women and girls battling invisibility in {U.S} classrooms}.
\newblock \bibinfo{journal}{\emph{Journal of Negro Education}} \bibinfo{volume}{85}, \bibinfo{number}{3} (\bibinfo{year}{2016}), \bibinfo{pages}{380--391}.
\newblock


\bibitem[Hettiachchi and Goncalves(2020)]%
        {10.1145/3369457.3369491}
\bibfield{author}{\bibinfo{person}{Danula Hettiachchi} {and} \bibinfo{person}{Jorge Goncalves}.} \bibinfo{year}{2020}\natexlab{}.
\newblock \showarticletitle{Towards Effective Crowd-Powered Online Content Moderation}. In \bibinfo{booktitle}{\emph{Proceedings of the 31st Australian Conference on Human-Computer-Interaction}} (Fremantle, WA, Australia) \emph{(\bibinfo{series}{OZCHI'19})}. \bibinfo{publisher}{Association for Computing Machinery}, \bibinfo{address}{New York, NY, USA}, \bibinfo{pages}{342–346}.
\newblock
\showISBNx{9781450376969}
\urldef\tempurl%
\url{https://doi.org/10.1145/3369457.3369491}
\showDOI{\tempurl}


\bibitem[Heung et~al\mbox{.}(2022)]%
        {10.1145/3517428.3544801}
\bibfield{author}{\bibinfo{person}{Sharon Heung}, \bibinfo{person}{Mahika Phutane}, \bibinfo{person}{Shiri Azenkot}, \bibinfo{person}{Megh Marathe}, {and} \bibinfo{person}{Aditya Vashistha}.} \bibinfo{year}{2022}\natexlab{}.
\newblock \showarticletitle{Nothing Micro About It: Examining Ableist Microaggressions on Social Media}. In \bibinfo{booktitle}{\emph{Proceedings of the 24th International ACM SIGACCESS Conference on Computers and Accessibility}} (Athens, Greece) \emph{(\bibinfo{series}{ASSETS '22})}. \bibinfo{publisher}{Association for Computing Machinery}, \bibinfo{address}{New York, NY, USA}, Article \bibinfo{articleno}{27}, \bibinfo{numpages}{14}~pages.
\newblock
\showISBNx{9781450392587}
\urldef\tempurl%
\url{https://doi.org/10.1145/3517428.3544801}
\showDOI{\tempurl}


\bibitem[Hurwitz and Peffley(1997)]%
        {13/50-article}
\bibfield{author}{\bibinfo{person}{Jon Hurwitz} {and} \bibinfo{person}{Mark Peffley}.} \bibinfo{year}{1997}\natexlab{}.
\newblock \showarticletitle{Public Perceptions of Race and Crime: The Role of Racial Stereotypes}.
\newblock \bibinfo{journal}{\emph{American Journal of Political Science}}  \bibinfo{volume}{41} (\bibinfo{date}{04} \bibinfo{year}{1997}), \bibinfo{pages}{375}.
\newblock
\urldef\tempurl%
\url{https://doi.org/10.2307/2111769}
\showDOI{\tempurl}


\bibitem[Hutson et~al\mbox{.}(2018)]%
        {10.1145/3274342}
\bibfield{author}{\bibinfo{person}{Jevan~A. Hutson}, \bibinfo{person}{Jessie~G. Taft}, \bibinfo{person}{Solon Barocas}, {and} \bibinfo{person}{Karen Levy}.} \bibinfo{year}{2018}\natexlab{}.
\newblock \showarticletitle{Debiasing Desire: Addressing Bias \& Discrimination on Intimate Platforms}.
\newblock \bibinfo{journal}{\emph{Proc. ACM Hum.-Comput. Interact.}} \bibinfo{volume}{2}, \bibinfo{number}{CSCW}, Article \bibinfo{articleno}{73} (\bibinfo{date}{Nov} \bibinfo{year}{2018}), \bibinfo{numpages}{18}~pages.
\newblock
\urldef\tempurl%
\url{https://doi.org/10.1145/3274342}
\showDOI{\tempurl}


\bibitem[Inoue(2018)]%
        {inoue_word_2018}
\bibfield{author}{\bibinfo{person}{Miyako Inoue}.} \bibinfo{year}{2018}\natexlab{}.
\newblock \showarticletitle{Word for word: Verbatim as political technologies}.
\newblock \bibinfo{journal}{\emph{Annual Review of Anthropology}}  \bibinfo{volume}{47} (\bibinfo{year}{2018}), \bibinfo{pages}{217--232}.
\newblock


\bibitem[Jhaver et~al\mbox{.}(2019a)]%
        {10.1145/3359294}
\bibfield{author}{\bibinfo{person}{Shagun Jhaver}, \bibinfo{person}{Darren~Scott Appling}, \bibinfo{person}{Eric Gilbert}, {and} \bibinfo{person}{Amy Bruckman}.} \bibinfo{year}{2019}\natexlab{a}.
\newblock \showarticletitle{"Did You Suspect the Post Would Be Removed?": Understanding User Reactions to Content Removals on Reddit}.
\newblock \bibinfo{journal}{\emph{Proc. ACM Hum.-Comput. Interact.}} \bibinfo{volume}{3}, \bibinfo{number}{CSCW}, Article \bibinfo{articleno}{192} (\bibinfo{date}{nov} \bibinfo{year}{2019}), \bibinfo{numpages}{33}~pages.
\newblock
\urldef\tempurl%
\url{https://doi.org/10.1145/3359294}
\showDOI{\tempurl}


\bibitem[Jhaver et~al\mbox{.}(2021)]%
        {10.1145/3479525}
\bibfield{author}{\bibinfo{person}{Shagun Jhaver}, \bibinfo{person}{Christian Boylston}, \bibinfo{person}{Diyi Yang}, {and} \bibinfo{person}{Amy Bruckman}.} \bibinfo{year}{2021}\natexlab{}.
\newblock \showarticletitle{Evaluating the effectiveness of deplatforming as a moderation strategy on {Twitter}}.
\newblock \bibinfo{journal}{\emph{Proceedings of the ACM on Human-Computer Interaction}} \bibinfo{volume}{5}, \bibinfo{number}{CSCW2} (\bibinfo{year}{2021}), \bibinfo{pages}{1--30}.
\newblock


\bibitem[Jhaver et~al\mbox{.}(2019b)]%
        {jhaver2019does}
\bibfield{author}{\bibinfo{person}{Shagun Jhaver}, \bibinfo{person}{Amy Bruckman}, {and} \bibinfo{person}{Eric Gilbert}.} \bibinfo{year}{2019}\natexlab{b}.
\newblock \showarticletitle{Does transparency in moderation really matter? User behavior after content removal explanations on reddit}.
\newblock \bibinfo{journal}{\emph{Proceedings of the ACM on Human-Computer Interaction}} \bibinfo{volume}{3}, \bibinfo{number}{CSCW} (\bibinfo{year}{2019}), \bibinfo{pages}{1--27}.
\newblock


\bibitem[Jhaver et~al\mbox{.}(2019c)]%
        {removal_explanations_jhaver}
\bibfield{author}{\bibinfo{person}{Shagun Jhaver}, \bibinfo{person}{Amy Bruckman}, {and} \bibinfo{person}{Eric Gilbert}.} \bibinfo{year}{2019}\natexlab{c}.
\newblock \showarticletitle{Does Transparency in Moderation Really Matter? User Behavior After Content Removal Explanations on Reddit}.
\newblock \bibinfo{journal}{\emph{Proc. ACM Hum.-Comput. Interact.}} \bibinfo{volume}{3}, \bibinfo{number}{CSCW}, Article \bibinfo{articleno}{150} (\bibinfo{date}{nov} \bibinfo{year}{2019}), \bibinfo{numpages}{27}~pages.
\newblock
\urldef\tempurl%
\url{https://doi.org/10.1145/3359252}
\showDOI{\tempurl}


\bibitem[Jhaver et~al\mbox{.}(2018)]%
        {10.1145/3185593}
\bibfield{author}{\bibinfo{person}{Shagun Jhaver}, \bibinfo{person}{Sucheta Ghoshal}, \bibinfo{person}{Amy Bruckman}, {and} \bibinfo{person}{Eric Gilbert}.} \bibinfo{year}{2018}\natexlab{}.
\newblock \showarticletitle{Online Harassment and Content Moderation: The Case of Blocklists}.
\newblock \bibinfo{journal}{\emph{ACM Trans. Comput.-Hum. Interact.}} \bibinfo{volume}{25}, \bibinfo{number}{2}, Article \bibinfo{articleno}{12} (\bibinfo{date}{mar} \bibinfo{year}{2018}), \bibinfo{numpages}{33}~pages.
\newblock
\showISSN{1073-0516}
\urldef\tempurl%
\url{https://doi.org/10.1145/3185593}
\showDOI{\tempurl}


\bibitem[Juneja et~al\mbox{.}(2020)]%
        {looking_glass_tanu}
\bibfield{author}{\bibinfo{person}{Prerna Juneja}, \bibinfo{person}{Deepika Rama~Subramanian}, {and} \bibinfo{person}{Tanushree Mitra}.} \bibinfo{year}{2020}\natexlab{}.
\newblock \showarticletitle{Through the Looking Glass: Study of Transparency in Reddit's Moderation Practices}.
\newblock \bibinfo{journal}{\emph{Proc. ACM Hum.-Comput. Interact.}} \bibinfo{volume}{4}, \bibinfo{number}{GROUP}, Article \bibinfo{articleno}{17} (\bibinfo{date}{jan} \bibinfo{year}{2020}), \bibinfo{numpages}{35}~pages.
\newblock
\urldef\tempurl%
\url{https://doi.org/10.1145/3375197}
\showDOI{\tempurl}


\bibitem[Kaur et~al\mbox{.}(2020)]%
        {10.1145/3313831.3376219}
\bibfield{author}{\bibinfo{person}{Harmanpreet Kaur}, \bibinfo{person}{Harsha Nori}, \bibinfo{person}{Samuel Jenkins}, \bibinfo{person}{Rich Caruana}, \bibinfo{person}{Hanna Wallach}, {and} \bibinfo{person}{Jennifer Wortman~Vaughan}.} \bibinfo{year}{2020}\natexlab{}.
\newblock \showarticletitle{Interpreting Interpretability: Understanding Data Scientists' Use of Interpretability Tools for Machine Learning}. In \bibinfo{booktitle}{\emph{Proceedings of the 2020 CHI Conference on Human Factors in Computing Systems}} (Honolulu, HI, USA) \emph{(\bibinfo{series}{CHI '20})}. \bibinfo{publisher}{Association for Computing Machinery}, \bibinfo{address}{New York, NY, USA}, \bibinfo{pages}{1–14}.
\newblock
\showISBNx{9781450367080}
\urldef\tempurl%
\url{https://doi.org/10.1145/3313831.3376219}
\showDOI{\tempurl}


\bibitem[Keels et~al\mbox{.}(2017)]%
        {keels_psychological_2017}
\bibfield{author}{\bibinfo{person}{Micere Keels}, \bibinfo{person}{Myles Durkee}, {and} \bibinfo{person}{Elan Hope}.} \bibinfo{year}{2017}\natexlab{}.
\newblock \showarticletitle{The psychological and academic costs of school-based racial and ethnic microaggressions}.
\newblock \bibinfo{journal}{\emph{American Educational Research Journal}} \bibinfo{volume}{54}, \bibinfo{number}{6} (\bibinfo{year}{2017}), \bibinfo{pages}{1316--1344}.
\newblock


\bibitem[Kelis(2021)]%
        {kelis_2021}
\bibfield{author}{\bibinfo{person}{Kennedy Kelis}.} \bibinfo{year}{2021}\natexlab{}.
\newblock \bibinfo{title}{Deconstructing The ``13/50" Argument}.
\newblock
\newblock
\urldef\tempurl%
\url{https://detester.org/publications/162kennedy}
\showURL{%
\tempurl}


\bibitem[Kew(2018)]%
        {kew_2018}
\bibfield{author}{\bibinfo{person}{Ben Kew}.} \bibinfo{year}{2018}\natexlab{}.
\newblock \showarticletitle{{Poll: Two-thirds of Conservatives don't trust Facebook, believe social media censors conservatives}}.
\newblock \bibinfo{journal}{\emph{Breitbart}} (\bibinfo{date}{29 Aug} \bibinfo{year}{2018}).
\newblock
\urldef\tempurl%
\url{https://tinyurl.com/pzavznpb}
\showURL{%
\tempurl}


\bibitem[Khan et~al\mbox{.}(0 01)]%
        {khan_social_2014}
\bibfield{author}{\bibinfo{person}{Gohar~Feroz Khan}, \bibinfo{person}{Bobby Swar}, {and} \bibinfo{person}{Sang~Kon Lee}.} \bibinfo{year}{2014-10-01}\natexlab{}.
\newblock \showarticletitle{Social Media Risks and Benefits: A Public Sector Perspective}.
\newblock  \bibinfo{volume}{32}, \bibinfo{number}{5} (\bibinfo{year}{2014-10-01}), \bibinfo{pages}{606--627}.
\newblock
\showISSN{0894-4393}
\urldef\tempurl%
\url{https://doi.org/10.1177/0894439314524701}
\showDOI{\tempurl}
\newblock
\shownote{Publisher: {SAGE} Publications Inc}.


\bibitem[Kiesler et~al\mbox{.}(2010)]%
        {Kiesler2010RegulatingBI}
\bibfield{author}{\bibinfo{person}{Sara~B. Kiesler}, \bibinfo{person}{Robert~E. Kraut}, \bibinfo{person}{Paul Resnick}, {and} \bibinfo{person}{Aniket Kittur}.} \bibinfo{year}{2010}\natexlab{}.
\newblock \showarticletitle{Regulating Behavior in Online Communities}.
\newblock


\bibitem[Kou and Gui(2020)]%
        {10.1145/3415173}
\bibfield{author}{\bibinfo{person}{Yubo Kou} {and} \bibinfo{person}{Xinning Gui}.} \bibinfo{year}{2020}\natexlab{}.
\newblock \showarticletitle{{Mediating Community-AI Interaction through Situated Explanation: The Case of AI-Led Moderation}}.
\newblock \bibinfo{journal}{\emph{Proc. ACM Hum.-Comput. Interact.}} \bibinfo{volume}{4}, \bibinfo{number}{CSCW2}, Article \bibinfo{articleno}{102} (\bibinfo{date}{oct} \bibinfo{year}{2020}), \bibinfo{numpages}{27}~pages.
\newblock
\urldef\tempurl%
\url{https://doi.org/10.1145/3415173}
\showDOI{\tempurl}


\bibitem[Kou and Gui(2021)]%
        {10.1145/3411764.3445279}
\bibfield{author}{\bibinfo{person}{Yubo Kou} {and} \bibinfo{person}{Xinning Gui}.} \bibinfo{year}{2021}\natexlab{}.
\newblock \showarticletitle{Flag and Flaggability in Automated Moderation: The Case of Reporting Toxic Behavior in an Online Game Community}. In \bibinfo{booktitle}{\emph{Proceedings of the 2021 CHI Conference on Human Factors in Computing Systems}} (Yokohama, Japan) \emph{(\bibinfo{series}{CHI '21})}. \bibinfo{publisher}{Association for Computing Machinery}, \bibinfo{address}{New York, NY, USA}, Article \bibinfo{articleno}{437}, \bibinfo{numpages}{12}~pages.
\newblock
\showISBNx{9781450380966}
\urldef\tempurl%
\url{https://doi.org/10.1145/3411764.3445279}
\showDOI{\tempurl}


\bibitem[Lai et~al\mbox{.}(2022)]%
        {10.1145/3491102.3501999}
\bibfield{author}{\bibinfo{person}{Vivian Lai}, \bibinfo{person}{Samuel Carton}, \bibinfo{person}{Rajat Bhatnagar}, \bibinfo{person}{Q.~Vera Liao}, \bibinfo{person}{Yunfeng Zhang}, {and} \bibinfo{person}{Chenhao Tan}.} \bibinfo{year}{2022}\natexlab{}.
\newblock \showarticletitle{{Human-AI Collaboration via Conditional Delegation: A Case Study of Content Moderation}}. In \bibinfo{booktitle}{\emph{{Proceedings of the 2022 CHI Conference on Human Factors in Computing Systems}}} (New Orleans, LA, USA) \emph{(\bibinfo{series}{CHI '22})}. \bibinfo{publisher}{Association for Computing Machinery}, \bibinfo{address}{New York, NY, USA}, Article \bibinfo{articleno}{54}, \bibinfo{numpages}{18}~pages.
\newblock
\showISBNx{9781450391573}
\urldef\tempurl%
\url{https://doi.org/10.1145/3491102.3501999}
\showDOI{\tempurl}


\bibitem[Landis and Koch(1977)]%
        {landis_application_1977}
\bibfield{author}{\bibinfo{person}{J~Richard Landis} {and} \bibinfo{person}{Gary~G Koch}.} \bibinfo{year}{1977}\natexlab{}.
\newblock \showarticletitle{{An Application of Hierarchical Kappa-type Statistics in the Assessment of Majority Agreement among Multiple Observers}}.
\newblock \bibinfo{journal}{\emph{Biometrics}} (\bibinfo{year}{1977}), \bibinfo{pages}{363--374}.
\newblock


\bibitem[Lees et~al\mbox{.}(1 04)]%
        {lees_capturing_2021}
\bibfield{author}{\bibinfo{person}{Alyssa Lees}, \bibinfo{person}{Daniel Borkan}, \bibinfo{person}{Ian Kivlichan}, \bibinfo{person}{Jorge Nario}, {and} \bibinfo{person}{Tesh Goyal}.} \bibinfo{year}{2021-04}\natexlab{}.
\newblock \showarticletitle{Capturing Covertly Toxic Speech via Crowdsourcing}. In \bibinfo{booktitle}{\emph{Proceedings of the First Workshop on Bridging Human–Computer Interaction and Natural Language Processing}} (Online). \bibinfo{publisher}{Association for Computational Linguistics}, \bibinfo{pages}{14--20}.
\newblock


\bibitem[Lepri et~al\mbox{.}(2018)]%
        {lepri-paper}
\bibfield{author}{\bibinfo{person}{Bruno Lepri}, \bibinfo{person}{Nuria Oliver}, \bibinfo{person}{Emmanuel Letouzé}, \bibinfo{person}{Alex Pentland}, {and} \bibinfo{person}{Patrick Vinck}.} \bibinfo{year}{2018}\natexlab{}.
\newblock \showarticletitle{{Fair, Transparent, and Accountable Algorithmic Decision-making Processes: The Premise, the Proposed Solutions, and the Open Challenges}}.
\newblock \bibinfo{journal}{\emph{Philosophy \& Technology}}  \bibinfo{volume}{31} (\bibinfo{date}{12} \bibinfo{year}{2018}).
\newblock
\urldef\tempurl%
\url{https://doi.org/10.1007/s13347-017-0279-x}
\showDOI{\tempurl}


\bibitem[Leung et~al\mbox{.}(2020)]%
        {10.1145/3313831.3376874}
\bibfield{author}{\bibinfo{person}{Weiwen Leung}, \bibinfo{person}{Zheng Zhang}, \bibinfo{person}{Daviti Jibuti}, \bibinfo{person}{Jinhao Zhao}, \bibinfo{person}{Maximilian Klein}, \bibinfo{person}{Casey Pierce}, \bibinfo{person}{Lionel Robert}, {and} \bibinfo{person}{Haiyi Zhu}.} \bibinfo{year}{2020}\natexlab{}.
\newblock \showarticletitle{Race, Gender and Beauty: The Effect of Information Provision on Online Hiring Biases}. In \bibinfo{booktitle}{\emph{Proceedings of the 2020 CHI Conference on Human Factors in Computing Systems}} (Honolulu, HI, USA) \emph{(\bibinfo{series}{CHI '20})}. \bibinfo{publisher}{Association for Computing Machinery}, \bibinfo{address}{New York, NY, USA}, \bibinfo{pages}{1–11}.
\newblock
\showISBNx{9781450367080}
\urldef\tempurl%
\url{https://doi.org/10.1145/3313831.3376874}
\showDOI{\tempurl}


\bibitem[Lingel(2019)]%
        {lingel_2019}
\bibfield{author}{\bibinfo{person}{Jessa Lingel}.} \bibinfo{year}{2019}\natexlab{}.
\newblock \bibinfo{title}{The gentrifcation of the internet.}
\newblock
\newblock
\urldef\tempurl%
\url{https://culturedigitally.org/2019/03/the-gentrification-of-the-internet/}
\showURL{%
\tempurl}


\bibitem[Link et~al\mbox{.}(2016)]%
        {Link2016AHA}
\bibfield{author}{\bibinfo{person}{Daniel Link}, \bibinfo{person}{Bernd Hellingrath}, {and} \bibinfo{person}{Jie Ling}.} \bibinfo{year}{2016}\natexlab{}.
\newblock \showarticletitle{A Human-in-the-Loop Approach for Semi-Automated Content Moderation}. In \bibinfo{booktitle}{\emph{International Conference on Information Systems for Crisis Response and Management}}.
\newblock


\bibitem[Liu et~al\mbox{.}(2021)]%
        {Liu2021}
\bibfield{author}{\bibinfo{person}{Yang Liu}, \bibinfo{person}{Christopher Whitfield}, \bibinfo{person}{Tianyang Zhang}, \bibinfo{person}{Amanda Hauser}, \bibinfo{person}{Taeyonn Reynolds}, {and} \bibinfo{person}{Mohd Anwar}.} \bibinfo{year}{2021}\natexlab{}.
\newblock \showarticletitle{Monitoring {COVID}-19 pandemic through the lens of social media using natural language processing and machine learning}.
\newblock \bibinfo{journal}{\emph{Health Information Science and Systems}} \bibinfo{volume}{9}, \bibinfo{number}{1} (\bibinfo{date}{June} \bibinfo{year}{2021}).
\newblock
\urldef\tempurl%
\url{https://doi.org/10.1007/s13755-021-00158-4}
\showDOI{\tempurl}


\bibitem[Lundberg and Lee(2017)]%
        {NIPS2017_8a20a862}
\bibfield{author}{\bibinfo{person}{Scott~M Lundberg} {and} \bibinfo{person}{Su-In Lee}.} \bibinfo{year}{2017}\natexlab{}.
\newblock \showarticletitle{A Unified Approach to Interpreting Model Predictions}. In \bibinfo{booktitle}{\emph{Advances in Neural Information Processing Systems}}, \bibfield{editor}{\bibinfo{person}{I.~Guyon}, \bibinfo{person}{U.~Von Luxburg}, \bibinfo{person}{S.~Bengio}, \bibinfo{person}{H.~Wallach}, \bibinfo{person}{R.~Fergus}, \bibinfo{person}{S.~Vishwanathan}, {and} \bibinfo{person}{R.~Garnett}} (Eds.), Vol.~\bibinfo{volume}{30}. \bibinfo{publisher}{Curran Associates, Inc.}
\newblock
\urldef\tempurl%
\url{https://proceedings.neurips.cc/paper_files/paper/2017/file/8a20a8621978632d76c43dfd28b67767-Paper.pdf}
\showURL{%
\tempurl}


\bibitem[Madsen et~al\mbox{.}(4 29)]%
        {madsen_post-hoc_2022}
\bibfield{author}{\bibinfo{person}{Andreas Madsen}, \bibinfo{person}{Siva Reddy}, {and} \bibinfo{person}{Sarath Chandar}.} \bibinfo{year}{2022-04-29}\natexlab{}.
\newblock \bibinfo{title}{Post-hoc Interpretability for Neural {NLP}: A Survey}.
\newblock
\newblock
\showeprint[arxiv]{2108.04840 [cs]}


\bibitem[Marshall et~al\mbox{.}(2021)]%
        {Marshall2021RespondingAN}
\bibfield{author}{\bibinfo{person}{Andrea Marshall}, \bibinfo{person}{Angela~D. Pack}, \bibinfo{person}{Sarah~A. Owusu}, \bibinfo{person}{Rainbo Hultman}, \bibinfo{person}{David Drake}, \bibinfo{person}{Florentine U.~N. Rutaganira}, \bibinfo{person}{Maria Namwanje}, \bibinfo{person}{Chantell~S. Evans}, \bibinfo{person}{Edgar Garza-Lopez}, \bibinfo{person}{Samantha~C. Lewis}, \bibinfo{person}{Cristina Termini}, \bibinfo{person}{Salma AshShareef}, \bibinfo{person}{Innes Hicsasmaz}, \bibinfo{person}{Brittany~L. Taylor}, \bibinfo{person}{Melanie~R. McReynolds}, \bibinfo{person}{Haysetta~D. Shuler}, {and} \bibinfo{person}{Antentor~O. Hinton}.} \bibinfo{year}{2021}\natexlab{}.
\newblock \showarticletitle{Responding and navigating racialized microaggressions in STEM}.
\newblock \bibinfo{journal}{\emph{Pathogens and Disease}}  \bibinfo{volume}{79} (\bibinfo{year}{2021}).
\newblock


\bibitem[Mathew et~al\mbox{.}(2020)]%
        {10.1145/3415163}
\bibfield{author}{\bibinfo{person}{Binny Mathew}, \bibinfo{person}{Anurag Illendula}, \bibinfo{person}{Punyajoy Saha}, \bibinfo{person}{Soumya Sarkar}, \bibinfo{person}{Pawan Goyal}, {and} \bibinfo{person}{Animesh Mukherjee}.} \bibinfo{year}{2020}\natexlab{}.
\newblock \showarticletitle{{Hate Begets Hate: A Temporal Study of Hate Speech}}.
\newblock \bibinfo{journal}{\emph{Proc. ACM Hum.-Comput. Interact.}} \bibinfo{volume}{4}, \bibinfo{number}{CSCW2}, Article \bibinfo{articleno}{92} (\bibinfo{date}{oct} \bibinfo{year}{2020}), \bibinfo{numpages}{24}~pages.
\newblock
\urldef\tempurl%
\url{https://doi.org/10.1145/3415163}
\showDOI{\tempurl}


\bibitem[Moore(2017)]%
        {Moore2017}
\bibfield{author}{\bibinfo{person}{Ryan Moore}.} \bibinfo{year}{2017}\natexlab{}.
\newblock \bibinfo{booktitle}{\emph{The New Jim Crow}}.
\newblock \bibinfo{publisher}{Macat Library}.
\newblock
\urldef\tempurl%
\url{https://doi.org/10.4324/9781912282586}
\showDOI{\tempurl}


\bibitem[Moore-Berg et~al\mbox{.}(2022)]%
        {moore-berg_empathy_2022}
\bibfield{author}{\bibinfo{person}{Samantha~L Moore-Berg}, \bibinfo{person}{Boaz Hameiri}, {and} \bibinfo{person}{Emile~G Bruneau}.} \bibinfo{year}{2022}\natexlab{}.
\newblock \showarticletitle{Empathy, dehumanization, and misperceptions: A media intervention humanizes migrants and increases empathy for their plight but only if misinformation about migrants is also corrected}.
\newblock \bibinfo{journal}{\emph{Social Psychological and Personality Science}} \bibinfo{volume}{13}, \bibinfo{number}{2} (\bibinfo{year}{2022}), \bibinfo{pages}{645--655}.
\newblock


\bibitem[Muller and Erickson(2018)]%
        {10.1145/3170427.3188407}
\bibfield{author}{\bibinfo{person}{Michael Muller} {and} \bibinfo{person}{Thomas Erickson}.} \bibinfo{year}{2018}\natexlab{}.
\newblock \showarticletitle{{In the Data Kitchen: A Review (a Design Fiction on Data Science)}}. In \bibinfo{booktitle}{\emph{Extended Abstracts of the 2018 CHI Conference on Human Factors in Computing Systems}} (Montreal QC, Canada) \emph{(\bibinfo{series}{CHI EA '18})}. \bibinfo{publisher}{Association for Computing Machinery}, \bibinfo{address}{New York, NY, USA}, \bibinfo{pages}{1–10}.
\newblock
\showISBNx{9781450356213}
\urldef\tempurl%
\url{https://doi.org/10.1145/3170427.3188407}
\showDOI{\tempurl}


\bibitem[Muller et~al\mbox{.}(2021)]%
        {10.1145/3411764.3445402}
\bibfield{author}{\bibinfo{person}{Michael Muller}, \bibinfo{person}{Christine~T. Wolf}, \bibinfo{person}{Josh Andres}, \bibinfo{person}{Michael Desmond}, \bibinfo{person}{Narendra~Nath Joshi}, \bibinfo{person}{Zahra Ashktorab}, \bibinfo{person}{Aabhas Sharma}, \bibinfo{person}{Kristina Brimijoin}, \bibinfo{person}{Qian Pan}, \bibinfo{person}{Evelyn Duesterwald}, {and} \bibinfo{person}{Casey Dugan}.} \bibinfo{year}{2021}\natexlab{}.
\newblock \showarticletitle{Designing Ground Truth and the Social Life of Labels}. In \bibinfo{booktitle}{\emph{Proceedings of the 2021 CHI Conference on Human Factors in Computing Systems}} (Yokohama, Japan) \emph{(\bibinfo{series}{CHI '21})}. \bibinfo{publisher}{Association for Computing Machinery}, \bibinfo{address}{New York, NY, USA}, Article \bibinfo{articleno}{94}, \bibinfo{numpages}{16}~pages.
\newblock
\showISBNx{9781450380966}
\urldef\tempurl%
\url{https://doi.org/10.1145/3411764.3445402}
\showDOI{\tempurl}


\bibitem[Nadal et~al\mbox{.}(2014)]%
        {nadal_impact_2014}
\bibfield{author}{\bibinfo{person}{Kevin~L Nadal}, \bibinfo{person}{Katie~E Griffin}, \bibinfo{person}{Yinglee Wong}, \bibinfo{person}{Sahran Hamit}, {and} \bibinfo{person}{Morgan Rasmus}.} \bibinfo{year}{2014}\natexlab{}.
\newblock \showarticletitle{The impact of racial microaggressions on mental health: Counseling implications for clients of color}.
\newblock \bibinfo{journal}{\emph{Journal of Counseling \& Development}} \bibinfo{volume}{92}, \bibinfo{number}{1} (\bibinfo{year}{2014}), \bibinfo{pages}{57--66}.
\newblock


\bibitem[Niederhoffer and Pennebaker(2002)]%
        {ismsi}
\bibfield{author}{\bibinfo{person}{Kate~G Niederhoffer} {and} \bibinfo{person}{James~W Pennebaker}.} \bibinfo{year}{2002}\natexlab{}.
\newblock \showarticletitle{Linguistic style matching in social interaction}.
\newblock \bibinfo{journal}{\emph{Journal of Language and Social Psychology}} \bibinfo{volume}{21}, \bibinfo{number}{4} (\bibinfo{year}{2002}), \bibinfo{pages}{337--360}.
\newblock


\bibitem[Noor(2020)]%
        {noor_2020}
\bibfield{author}{\bibinfo{person}{Poppy Noor}.} \bibinfo{year}{2020}\natexlab{}.
\newblock \bibinfo{title}{{The celebrities who are doing anti-racism right}}.
\newblock
\newblock


\bibitem[Nova et~al\mbox{.}(2021)]%
        {nova2021facebook}
\bibfield{author}{\bibinfo{person}{Fayika~Farhat Nova}, \bibinfo{person}{Michael~Ann DeVito}, \bibinfo{person}{Pratyasha Saha}, \bibinfo{person}{Kazi~Shohanur Rashid}, \bibinfo{person}{Shashwata Roy~Turzo}, \bibinfo{person}{Sadia Afrin}, {and} \bibinfo{person}{Shion Guha}.} \bibinfo{year}{2021}\natexlab{}.
\newblock \showarticletitle{{``Facebook Promotes More Harassment" Social Media Ecosystem, Skill and Marginalized Hijra Identity in Bangladesh}}.
\newblock \bibinfo{journal}{\emph{Proceedings of the ACM on Human-Computer Interaction}} \bibinfo{volume}{5}, \bibinfo{number}{CSCW1} (\bibinfo{year}{2021}), \bibinfo{pages}{1--35}.
\newblock


\bibitem[Nova et~al\mbox{.}(2019)]%
        {10.1145/3287098.3287107}
\bibfield{author}{\bibinfo{person}{Fayika~Farhat Nova}, \bibinfo{person}{MD.~Rashidujjaman Rifat}, \bibinfo{person}{Pratyasha Saha}, \bibinfo{person}{Syed~Ishtiaque Ahmed}, {and} \bibinfo{person}{Shion Guha}.} \bibinfo{year}{2019}\natexlab{}.
\newblock \showarticletitle{{Online Sexual Harassment over Anonymous Social Media in Bangladesh}}. In \bibinfo{booktitle}{\emph{Proceedings of the Tenth International Conference on Information and Communication Technologies and Development}} (Ahmedabad, India) \emph{(\bibinfo{series}{ICTD '19})}. \bibinfo{publisher}{Association for Computing Machinery}, \bibinfo{address}{New York, NY, USA}, Article \bibinfo{articleno}{1}, \bibinfo{numpages}{12}~pages.
\newblock
\showISBNx{9781450361224}
\urldef\tempurl%
\url{https://doi.org/10.1145/3287098.3287107}
\showDOI{\tempurl}


\bibitem[Ogbonnaya-Ogburu et~al\mbox{.}(4 21)]%
        {ogbonnaya-ogburu_critical_2020}
\bibfield{author}{\bibinfo{person}{Ihudiya~Finda Ogbonnaya-Ogburu}, \bibinfo{person}{Angela~D.R. Smith}, \bibinfo{person}{Alexandra To}, {and} \bibinfo{person}{Kentaro Toyama}.} \bibinfo{year}{2020-04-21}\natexlab{}.
\newblock \showarticletitle{Critical Race Theory for {HCI}}. In \bibinfo{booktitle}{\emph{Proceedings of the 2020 {CHI} Conference on Human Factors in Computing Systems}} (Honolulu {HI} {USA}). \bibinfo{publisher}{{ACM}}, \bibinfo{pages}{1--16}.
\newblock
\showISBNx{978-1-4503-6708-0}
\urldef\tempurl%
\url{https://doi.org/10.1145/3313831.3376392}
\showDOI{\tempurl}


\bibitem[Okonofua et~al\mbox{.}(2016)]%
        {okonofua_vicious_2016}
\bibfield{author}{\bibinfo{person}{Jason~A Okonofua}, \bibinfo{person}{Gregory~M Walton}, {and} \bibinfo{person}{Jennifer~L Eberhardt}.} \bibinfo{year}{2016}\natexlab{}.
\newblock \showarticletitle{A vicious cycle: A social--psychological account of extreme racial disparities in school discipline}.
\newblock \bibinfo{journal}{\emph{Perspectives on Psychological Science}} \bibinfo{volume}{11}, \bibinfo{number}{3} (\bibinfo{year}{2016}), \bibinfo{pages}{381--398}.
\newblock


\bibitem[Park et~al\mbox{.}(2022a)]%
        {10.1145/3526113.3545616}
\bibfield{author}{\bibinfo{person}{Joon~Sung Park}, \bibinfo{person}{Lindsay Popowski}, \bibinfo{person}{Carrie Cai}, \bibinfo{person}{Meredith~Ringel Morris}, \bibinfo{person}{Percy Liang}, {and} \bibinfo{person}{Michael~S. Bernstein}.} \bibinfo{year}{2022}\natexlab{a}.
\newblock \showarticletitle{Social Simulacra: Creating Populated Prototypes for Social Computing Systems}. In \bibinfo{booktitle}{\emph{Proceedings of the 35th Annual ACM Symposium on User Interface Software and Technology}} (Bend, OR, USA) \emph{(\bibinfo{series}{UIST '22})}. \bibinfo{publisher}{Association for Computing Machinery}, \bibinfo{address}{New York, NY, USA}, Article \bibinfo{articleno}{74}, \bibinfo{numpages}{18}~pages.
\newblock
\showISBNx{9781450393201}
\urldef\tempurl%
\url{https://doi.org/10.1145/3526113.3545616}
\showDOI{\tempurl}


\bibitem[Park et~al\mbox{.}(2022b)]%
        {Park2022MeasuringTP}
\bibfield{author}{\bibinfo{person}{Joon~Sung Park}, \bibinfo{person}{Joseph Seering}, {and} \bibinfo{person}{Michael~S. Bernstein}.} \bibinfo{year}{2022}\natexlab{b}.
\newblock \showarticletitle{Measuring the Prevalence of Anti-Social Behavior in Online Communities}.
\newblock \bibinfo{journal}{\emph{Proceedings of the ACM on Human-Computer Interaction}}  \bibinfo{volume}{6} (\bibinfo{year}{2022}), \bibinfo{pages}{1 -- 29}.
\newblock


\bibitem[Pavlopoulos et~al\mbox{.}(2020)]%
        {pavlopoulos-etal-2020-toxicity}
\bibfield{author}{\bibinfo{person}{John Pavlopoulos}, \bibinfo{person}{Jeffrey Sorensen}, \bibinfo{person}{Lucas Dixon}, \bibinfo{person}{Nithum Thain}, {and} \bibinfo{person}{Ion Androutsopoulos}.} \bibinfo{year}{2020}\natexlab{}.
\newblock \showarticletitle{Toxicity Detection: Does Context Really Matter?}. In \bibinfo{booktitle}{\emph{Proceedings of the 58th Annual Meeting of the Association for Computational Linguistics}}. \bibinfo{publisher}{Association for Computational Linguistics}, \bibinfo{address}{Online}, \bibinfo{pages}{4296--4305}.
\newblock
\urldef\tempurl%
\url{https://doi.org/10.18653/v1/2020.acl-main.396}
\showDOI{\tempurl}


\bibitem[{PhD}(3 03)]%
        {phd_biased_2020}
\bibfield{author}{\bibinfo{person}{Jennifer L.~Eberhardt {PhD}}.} \bibinfo{year}{2020-03-03}\natexlab{}.
\newblock \bibinfo{booktitle}{\emph{Biased: Uncovering the Hidden Prejudice That Shapes What We See, Think, and Do}}.
\newblock \bibinfo{publisher}{Penguin}.
\newblock
\showISBNx{978-0-7352-2495-7}
\newblock
\shownote{Google-Books-{ID}: {hpDODwAAQBAJ}}.


\bibitem[Prabhakaran et~al\mbox{.}(9 11)]%
        {prabhakaran_perturbation_2019}
\bibfield{author}{\bibinfo{person}{Vinodkumar Prabhakaran}, \bibinfo{person}{Ben Hutchinson}, {and} \bibinfo{person}{Margaret Mitchell}.} \bibinfo{year}{2019-11}\natexlab{}.
\newblock \showarticletitle{Perturbation Sensitivity Analysis to Detect Unintended Model Biases}. In \bibinfo{booktitle}{\emph{Proceedings of the 2019 Conference on Empirical Methods in Natural Language Processing and the 9th International Joint Conference on Natural Language Processing ({EMNLP}-{IJCNLP})}} (Hong Kong, China). \bibinfo{publisher}{Association for Computational Linguistics}, \bibinfo{pages}{5740--5745}.
\newblock
\urldef\tempurl%
\url{https://doi.org/10.18653/v1/D19-1578}
\showDOI{\tempurl}


\bibitem[Rader et~al\mbox{.}(2018)]%
        {10.1145/3173574.3173677}
\bibfield{author}{\bibinfo{person}{Emilee Rader}, \bibinfo{person}{Kelley Cotter}, {and} \bibinfo{person}{Janghee Cho}.} \bibinfo{year}{2018}\natexlab{}.
\newblock \showarticletitle{Explanations as Mechanisms for Supporting Algorithmic Transparency}. In \bibinfo{booktitle}{\emph{Proceedings of the 2018 CHI Conference on Human Factors in Computing Systems}} (Montreal QC, Canada) \emph{(\bibinfo{series}{CHI '18})}. \bibinfo{publisher}{Association for Computing Machinery}, \bibinfo{address}{New York, NY, USA}, \bibinfo{pages}{1–13}.
\newblock
\showISBNx{9781450356206}
\urldef\tempurl%
\url{https://doi.org/10.1145/3173574.3173677}
\showDOI{\tempurl}


\bibitem[Rattan and Eberhardt(2010)]%
        {rattan_role_2010}
\bibfield{author}{\bibinfo{person}{Aneeta Rattan} {and} \bibinfo{person}{Jennifer~L Eberhardt}.} \bibinfo{year}{2010}\natexlab{}.
\newblock \showarticletitle{The role of social meaning in inattentional blindness: When the gorillas in our midst do not go unseen}.
\newblock \bibinfo{journal}{\emph{Journal of Experimental Social Psychology}} \bibinfo{volume}{46}, \bibinfo{number}{6} (\bibinfo{year}{2010}), \bibinfo{pages}{1085--1088}.
\newblock


\bibitem[Reid et~al\mbox{.}(2022)]%
        {10.1145/3549498}
\bibfield{author}{\bibinfo{person}{Elizabeth Reid}, \bibinfo{person}{Regan~L. Mandryk}, \bibinfo{person}{Nicole~A. Beres}, \bibinfo{person}{Madison Klarkowski}, {and} \bibinfo{person}{Julian Frommel}.} \bibinfo{year}{2022}\natexlab{}.
\newblock \showarticletitle{Feeling Good and In Control: In-Game Tools to Support Targets of Toxicity}.
\newblock \bibinfo{journal}{\emph{Proc. ACM Hum.-Comput. Interact.}} \bibinfo{volume}{6}, \bibinfo{number}{CHI PLAY}, Article \bibinfo{articleno}{235} (\bibinfo{date}{oct} \bibinfo{year}{2022}), \bibinfo{numpages}{27}~pages.
\newblock
\urldef\tempurl%
\url{https://doi.org/10.1145/3549498}
\showDOI{\tempurl}


\bibitem[Reyes(2017)]%
        {hoffman_review_2012}
\bibfield{author}{\bibinfo{person}{Angela Reyes}.} \bibinfo{year}{2017}\natexlab{}.
\newblock \bibinfo{booktitle}{\emph{{Language, identity, and stereotype among Southeast Asian American youth: The other Asian}}}.
\newblock \bibinfo{publisher}{Routledge}.
\newblock


\bibitem[Rho et~al\mbox{.}(5 02)]%
        {rho_class_2017}
\bibfield{author}{\bibinfo{person}{Eugenia Ha~Rim Rho}, \bibinfo{person}{Oliver~L. Haimson}, \bibinfo{person}{Nazanin Andalibi}, \bibinfo{person}{Melissa Mazmanian}, {and} \bibinfo{person}{Gillian~R. Hayes}.} \bibinfo{year}{2017-05-02}\natexlab{}.
\newblock \showarticletitle{Class Confessions: Restorative Properties in Online Experiences of Socioeconomic Stigma}. In \bibinfo{booktitle}{\emph{Proceedings of the 2017 {CHI} Conference on Human Factors in Computing Systems}} (New York, {NY}, {USA}) \emph{(\bibinfo{series}{{CHI} '17})}. \bibinfo{publisher}{Association for Computing Machinery}, \bibinfo{pages}{3377--3389}.
\newblock
\showISBNx{978-1-4503-4655-9}
\urldef\tempurl%
\url{https://doi.org/10.1145/3025453.3025921}
\showDOI{\tempurl}


\bibitem[Rho et~al\mbox{.}(2018)]%
        {rho_fostering_2018}
\bibfield{author}{\bibinfo{person}{Eugenia Ha~Rim Rho}, \bibinfo{person}{Gloria Mark}, {and} \bibinfo{person}{Melissa Mazmanian}.} \bibinfo{year}{2018}\natexlab{}.
\newblock \showarticletitle{Fostering civil discourse online: Linguistic behavior in comments of\# metoo articles across political perspectives}.
\newblock \bibinfo{journal}{\emph{Proceedings of the ACM on human-computer interaction}} \bibinfo{volume}{2}, \bibinfo{number}{CSCW} (\bibinfo{year}{2018}), \bibinfo{pages}{1--28}.
\newblock


\bibitem[Rho and Mazmanian(2019)]%
        {rho_hashtag_2019}
\bibfield{author}{\bibinfo{person}{Eugenia Ha~Rim Rho} {and} \bibinfo{person}{Melissa Mazmanian}.} \bibinfo{year}{2019}\natexlab{}.
\newblock \showarticletitle{Hashtag burnout? a control experiment investigating how political hashtags shape reactions to news content}.
\newblock \bibinfo{journal}{\emph{Proceedings of the ACM on human-computer interaction}} \bibinfo{volume}{3}, \bibinfo{number}{CSCW} (\bibinfo{year}{2019}), \bibinfo{pages}{1--25}.
\newblock


\bibitem[Rho and Mazmanian(4 21)]%
        {rho_political_2020}
\bibfield{author}{\bibinfo{person}{Eugenia Ha~Rim Rho} {and} \bibinfo{person}{Melissa Mazmanian}.} \bibinfo{year}{2020-04-21}\natexlab{}.
\newblock \showarticletitle{Political Hashtags \& the Lost Art of Democratic Discourse}. In \bibinfo{booktitle}{\emph{Proceedings of the 2020 {CHI} Conference on Human Factors in Computing Systems}} (Honolulu {HI} {USA}). \bibinfo{publisher}{{ACM}}, \bibinfo{pages}{1--13}.
\newblock
\showISBNx{978-1-4503-6708-0}
\urldef\tempurl%
\url{https://doi.org/10.1145/3313831.3376542}
\showDOI{\tempurl}


\bibitem[Ribeiro et~al\mbox{.}(2016)]%
        {ribeiro-etal-2016-trust}
\bibfield{author}{\bibinfo{person}{Marco Ribeiro}, \bibinfo{person}{Sameer Singh}, {and} \bibinfo{person}{Carlos Guestrin}.} \bibinfo{year}{2016}\natexlab{}.
\newblock \showarticletitle{{``}Why Should {I} Trust You?{''}: Explaining the Predictions of Any Classifier}. In \bibinfo{booktitle}{\emph{Proceedings of the 2016 Conference of the North {A}merican Chapter of the Association for Computational Linguistics: Demonstrations}}. \bibinfo{publisher}{Association for Computational Linguistics}, \bibinfo{address}{San Diego, California}, \bibinfo{pages}{97--101}.
\newblock
\urldef\tempurl%
\url{https://doi.org/10.18653/v1/N16-3020}
\showDOI{\tempurl}


\bibitem[Riddle and Sinclair(2019)]%
        {riddle_racial_2019}
\bibfield{author}{\bibinfo{person}{Travis Riddle} {and} \bibinfo{person}{Stacey Sinclair}.} \bibinfo{year}{2019}\natexlab{}.
\newblock \showarticletitle{Racial disparities in school-based disciplinary actions are associated with county-level rates of racial bias}.
\newblock \bibinfo{journal}{\emph{Proceedings of the National Academy of Sciences}} \bibinfo{volume}{116}, \bibinfo{number}{17} (\bibinfo{year}{2019}), \bibinfo{pages}{8255--8260}.
\newblock


\bibitem[Roberts(2018)]%
        {99-sarah}
\bibfield{author}{\bibinfo{person}{Sarah Roberts}.} \bibinfo{year}{2018}\natexlab{}.
\newblock \showarticletitle{Digital detritus: `Error' and the logic of opacity in social media content moderation}.
\newblock \bibinfo{journal}{\emph{First Monday}}  \bibinfo{volume}{23} (\bibinfo{date}{03} \bibinfo{year}{2018}).
\newblock
\urldef\tempurl%
\url{https://doi.org/10.5210/fm.v23i3.8283}
\showDOI{\tempurl}


\bibitem[Sap et~al\mbox{.}(9 07)]%
        {sap_risk_2019}
\bibfield{author}{\bibinfo{person}{Maarten Sap}, \bibinfo{person}{Dallas Card}, \bibinfo{person}{Saadia Gabriel}, \bibinfo{person}{Yejin Choi}, {and} \bibinfo{person}{Noah~A. Smith}.} \bibinfo{year}{2019-07}\natexlab{}.
\newblock \showarticletitle{The Risk of Racial Bias in Hate Speech Detection}. In \bibinfo{booktitle}{\emph{Proceedings of the 57th Annual Meeting of the Association for Computational Linguistics}} (Florence, Italy). \bibinfo{publisher}{Association for Computational Linguistics}, \bibinfo{pages}{1668--1678}.
\newblock
\urldef\tempurl%
\url{https://doi.org/10.18653/v1/P19-1163}
\showDOI{\tempurl}


\bibitem[Scheuerman et~al\mbox{.}(2018)]%
        {10.1145/3274424}
\bibfield{author}{\bibinfo{person}{Morgan~Klaus Scheuerman}, \bibinfo{person}{Stacy~M. Branham}, {and} \bibinfo{person}{Foad Hamidi}.} \bibinfo{year}{2018}\natexlab{}.
\newblock \showarticletitle{Safe Spaces and Safe Places: Unpacking Technology-Mediated Experiences of Safety and Harm with Transgender People}.
\newblock \bibinfo{journal}{\emph{Proc. ACM Hum.-Comput. Interact.}} \bibinfo{volume}{2}, \bibinfo{number}{CSCW}, Article \bibinfo{articleno}{155} (\bibinfo{date}{nov} \bibinfo{year}{2018}), \bibinfo{numpages}{27}~pages.
\newblock
\urldef\tempurl%
\url{https://doi.org/10.1145/3274424}
\showDOI{\tempurl}


\bibitem[Schneider et~al\mbox{.}(2018)]%
        {schneider2018digital}
\bibfield{author}{\bibinfo{person}{Christoph Schneider}, \bibinfo{person}{Markus Weinmann}, {and} \bibinfo{person}{Jan Vom~Brocke}.} \bibinfo{year}{2018}\natexlab{}.
\newblock \showarticletitle{{Digital nudging: guiding online user choices through interface design}}.
\newblock \bibinfo{journal}{\emph{Commun. ACM}} \bibinfo{volume}{61}, \bibinfo{number}{7} (\bibinfo{year}{2018}), \bibinfo{pages}{67--73}.
\newblock


\bibitem[Simons and Chabris(1999)]%
        {simons_gorillas_1999}
\bibfield{author}{\bibinfo{person}{Daniel~J Simons} {and} \bibinfo{person}{Christopher~F Chabris}.} \bibinfo{year}{1999}\natexlab{}.
\newblock \showarticletitle{Gorillas in our midst: Sustained inattentional blindness for dynamic events}.
\newblock \bibinfo{journal}{\emph{perception}} \bibinfo{volume}{28}, \bibinfo{number}{9} (\bibinfo{year}{1999}), \bibinfo{pages}{1059--1074}.
\newblock


\bibitem[Smith et~al\mbox{.}(2021)]%
        {10.1145/3479561}
\bibfield{author}{\bibinfo{person}{C.~Estelle Smith}, \bibinfo{person}{William Lane}, \bibinfo{person}{Hannah Miller~Hillberg}, \bibinfo{person}{Daniel Kluver}, \bibinfo{person}{Loren Terveen}, {and} \bibinfo{person}{Svetlana Yarosh}.} \bibinfo{year}{2021}\natexlab{}.
\newblock \showarticletitle{{Effective Strategies for Crowd-Powered Cognitive Reappraisal Systems: A Field Deployment of the Flip*Doubt Web Application for Mental Health}}.
\newblock \bibinfo{journal}{\emph{Proc. ACM Hum.-Comput. Interact.}} \bibinfo{volume}{5}, \bibinfo{number}{CSCW2}, Article \bibinfo{articleno}{417} (\bibinfo{date}{oct} \bibinfo{year}{2021}), \bibinfo{numpages}{37}~pages.
\newblock
\urldef\tempurl%
\url{https://doi.org/10.1145/3479561}
\showDOI{\tempurl}


\bibitem[Song et~al\mbox{.}(2023)]%
        {modsandbox}
\bibfield{author}{\bibinfo{person}{Jean~Y. Song}, \bibinfo{person}{Sangwook Lee}, \bibinfo{person}{Jisoo Lee}, \bibinfo{person}{Mina Kim}, {and} \bibinfo{person}{Juho Kim}.} \bibinfo{year}{2023}\natexlab{}.
\newblock \showarticletitle{ModSandbox: Facilitating Online Community Moderation Through Error Prediction and Improvement of Automated Rules}. In \bibinfo{booktitle}{\emph{Proceedings of the 2023 CHI Conference on Human Factors in Computing Systems}} (Hamburg, Germany) \emph{(\bibinfo{series}{CHI '23})}. \bibinfo{publisher}{Association for Computing Machinery}, \bibinfo{address}{New York, NY, USA}, Article \bibinfo{articleno}{107}, \bibinfo{numpages}{20}~pages.
\newblock
\showISBNx{9781450394215}
\urldef\tempurl%
\url{https://doi.org/10.1145/3544548.3581057}
\showDOI{\tempurl}


\bibitem[Spencer(2017)]%
        {article-66}
\bibfield{author}{\bibinfo{person}{Michael Spencer}.} \bibinfo{year}{2017}\natexlab{}.
\newblock \showarticletitle{Microaggressions and Social Work Practice, Education, and Research}.
\newblock \bibinfo{journal}{\emph{Journal of Ethnic \& Cultural Diversity in Social Work}}  \bibinfo{volume}{26} (\bibinfo{date}{02} \bibinfo{year}{2017}), \bibinfo{pages}{1--5}.
\newblock
\urldef\tempurl%
\url{https://doi.org/10.1080/15313204.2016.1268989}
\showDOI{\tempurl}


\bibitem[Steiger et~al\mbox{.}(2021)]%
        {10.1145/3411764.3445092}
\bibfield{author}{\bibinfo{person}{Miriah Steiger}, \bibinfo{person}{Timir~J Bharucha}, \bibinfo{person}{Sukrit Venkatagiri}, \bibinfo{person}{Martin~J. Riedl}, {and} \bibinfo{person}{Matthew Lease}.} \bibinfo{year}{2021}\natexlab{}.
\newblock \showarticletitle{The Psychological Well-Being of Content Moderators: The Emotional Labor of Commercial Moderation and Avenues for Improving Support}. In \bibinfo{booktitle}{\emph{Proceedings of the 2021 CHI Conference on Human Factors in Computing Systems}} (Yokohama, Japan) \emph{(\bibinfo{series}{CHI '21})}. \bibinfo{publisher}{Association for Computing Machinery}, \bibinfo{address}{New York, NY, USA}, Article \bibinfo{articleno}{341}, \bibinfo{numpages}{14}~pages.
\newblock
\showISBNx{9781450380966}
\urldef\tempurl%
\url{https://doi.org/10.1145/3411764.3445092}
\showDOI{\tempurl}


\bibitem[Stieglitz and Dang-Xuan(2013)]%
        {stieglitz_emotions_2013}
\bibfield{author}{\bibinfo{person}{Stefan Stieglitz} {and} \bibinfo{person}{Linh Dang-Xuan}.} \bibinfo{year}{2013}\natexlab{}.
\newblock \showarticletitle{Emotions and information diffusion in social media—sentiment of microblogs and sharing behavior}.
\newblock \bibinfo{journal}{\emph{Journal of management information systems}} \bibinfo{volume}{29}, \bibinfo{number}{4} (\bibinfo{year}{2013}), \bibinfo{pages}{217--248}.
\newblock


\bibitem[Sue et~al\mbox{.}(2019)]%
        {sue2019disarming}
\bibfield{author}{\bibinfo{person}{Derald~Wing Sue}, \bibinfo{person}{Sarah Alsaidi}, \bibinfo{person}{Michael~N Awad}, \bibinfo{person}{Elizabeth Glaeser}, \bibinfo{person}{Cassandra~Z Calle}, {and} \bibinfo{person}{Narolyn Mendez}.} \bibinfo{year}{2019}\natexlab{}.
\newblock \showarticletitle{{Disarming racial microaggressions: Microintervention strategies for targets, White allies, and bystanders.}}
\newblock \bibinfo{journal}{\emph{American Psychologist}} \bibinfo{volume}{74}, \bibinfo{number}{1} (\bibinfo{year}{2019}), \bibinfo{pages}{128}.
\newblock


\bibitem[Sue et~al\mbox{.}(2007)]%
        {sue_racial_2007}
\bibfield{author}{\bibinfo{person}{Derald~Wing Sue}, \bibinfo{person}{Christina~M Capodilupo}, \bibinfo{person}{Gina~C Torino}, \bibinfo{person}{Jennifer~M Bucceri}, \bibinfo{person}{Aisha Holder}, \bibinfo{person}{Kevin~L Nadal}, {and} \bibinfo{person}{Marta Esquilin}.} \bibinfo{year}{2007}\natexlab{}.
\newblock \showarticletitle{Racial microaggressions in everyday life: implications for clinical practice.}
\newblock \bibinfo{journal}{\emph{American psychologist}} \bibinfo{volume}{62}, \bibinfo{number}{4} (\bibinfo{year}{2007}), \bibinfo{pages}{271}.
\newblock


\bibitem[Sue and Spanierman(2020)]%
        {sue_microaggressions_2020}
\bibfield{author}{\bibinfo{person}{Derald~Wing Sue} {and} \bibinfo{person}{Lisa~Beth Spanierman}.} \bibinfo{year}{2020}\natexlab{}.
\newblock \bibinfo{booktitle}{\emph{Microaggressions in everyday life, 2nd ed}}.
\newblock \bibinfo{publisher}{John Wiley \& Sons, Inc.}
\newblock
\showISBNx{978-1-119-51379-7 978-1-119-51380-3 978-1-119-51381-0}
\newblock
\shownote{Pages: xx, 349}.


\bibitem[Sundar et~al\mbox{.}(2015)]%
        {sundar2015toward}
\bibfield{author}{\bibinfo{person}{S~Shyam Sundar}, \bibinfo{person}{Haiyan Jia}, \bibinfo{person}{T~Franklin Waddell}, {and} \bibinfo{person}{Yan Huang}.} \bibinfo{year}{2015}\natexlab{}.
\newblock \showarticletitle{{Toward a theory of interactive media effects (TIME) four models for explaining how interface features affect user psychology}}.
\newblock \bibinfo{journal}{\emph{The handbook of the psychology of communication technology}} (\bibinfo{year}{2015}), \bibinfo{pages}{47--86}.
\newblock


\bibitem[Sundararajan et~al\mbox{.}(2017)]%
        {Sundararajan2017AxiomaticAF}
\bibfield{author}{\bibinfo{person}{Mukund Sundararajan}, \bibinfo{person}{Ankur Taly}, {and} \bibinfo{person}{Qiqi Yan}.} \bibinfo{year}{2017}\natexlab{}.
\newblock \showarticletitle{Axiomatic Attribution for Deep Networks}. In \bibinfo{booktitle}{\emph{International Conference on Machine Learning}}.
\newblock
\urldef\tempurl%
\url{https://api.semanticscholar.org/CorpusID:16747630}
\showURL{%
\tempurl}


\bibitem[Suzor et~al\mbox{.}(2019)]%
        {quteprints126386}
\bibfield{author}{\bibinfo{person}{Nicolas Suzor}, \bibinfo{person}{Sarah~Myers West}, \bibinfo{person}{Andrew Quodling}, {and} \bibinfo{person}{Jillian York}.} \bibinfo{year}{2019}\natexlab{}.
\newblock \showarticletitle{{What do we mean when we talk about transparency? Towards meaningful transparency in commercial content moderation}}.
\newblock \bibinfo{journal}{\emph{International Journal of Communication}}  \bibinfo{volume}{13} (\bibinfo{year}{2019}), \bibinfo{pages}{1526--1543}.
\newblock
\urldef\tempurl%
\url{https://eprints.qut.edu.au/126386/}
\showURL{%
\tempurl}


\bibitem[Swart et~al\mbox{.}(2020)]%
        {10.1145/3313831.3376178}
\bibfield{author}{\bibinfo{person}{Michael Swart}, \bibinfo{person}{Ylana Lopez}, \bibinfo{person}{Arunesh Mathur}, {and} \bibinfo{person}{Marshini Chetty}.} \bibinfo{year}{2020}\natexlab{}.
\newblock \showarticletitle{{Is This An Ad?: Automatically Disclosing Online Endorsements On YouTube With AdIntuition}}. In \bibinfo{booktitle}{\emph{Proceedings of the 2020 CHI Conference on Human Factors in Computing Systems}} (Honolulu, HI, USA) \emph{(\bibinfo{series}{CHI '20})}. \bibinfo{publisher}{Association for Computing Machinery}, \bibinfo{address}{New York, NY, USA}, \bibinfo{pages}{1–12}.
\newblock
\showISBNx{9781450367080}
\urldef\tempurl%
\url{https://doi.org/10.1145/3313831.3376178}
\showDOI{\tempurl}


\bibitem[Tausczik and Pennebaker(2010)]%
        {doi:10.1177/0261927X09351676}
\bibfield{author}{\bibinfo{person}{Yla~R. Tausczik} {and} \bibinfo{person}{James~W. Pennebaker}.} \bibinfo{year}{2010}\natexlab{}.
\newblock \showarticletitle{{The Psychological Meaning of Words: LIWC and Computerized Text Analysis Methods}}.
\newblock \bibinfo{journal}{\emph{Journal of Language and Social Psychology}} \bibinfo{volume}{29}, \bibinfo{number}{1} (\bibinfo{year}{2010}), \bibinfo{pages}{24--54}.
\newblock
\urldef\tempurl%
\url{https://doi.org/10.1177/0261927X09351676}
\showDOI{\tempurl}
\showeprint{https://doi.org/10.1177/0261927X09351676}


\bibitem[Thach et~al\mbox{.}(0)]%
        {doi:10.1177/14614448221109804}
\bibfield{author}{\bibinfo{person}{Hibby Thach}, \bibinfo{person}{Samuel Mayworm}, \bibinfo{person}{Daniel Delmonaco}, {and} \bibinfo{person}{Oliver Haimson}.} \bibinfo{year}{0}\natexlab{}.
\newblock \showarticletitle{{(In)visible moderation: A digital ethnography of marginalized users and content moderation on Twitch and Reddit}}.
\newblock \bibinfo{journal}{\emph{New Media \& Society}} \bibinfo{volume}{0}, \bibinfo{number}{0} (\bibinfo{year}{0}), \bibinfo{pages}{14614448221109804}.
\newblock
\urldef\tempurl%
\url{https://doi.org/10.1177/14614448221109804}
\showDOI{\tempurl}
\showeprint{https://doi.org/10.1177/14614448221109804}


\bibitem[Timmer et~al\mbox{.}(1 12)]%
        {timmer_can_2021}
\bibfield{author}{\bibinfo{person}{Roelien~C. Timmer}, \bibinfo{person}{David Liebowitz}, \bibinfo{person}{Surya Nepal}, {and} \bibinfo{person}{Salil~S. Kanhere}.} \bibinfo{year}{2021-12}\natexlab{}.
\newblock \showarticletitle{Can pre-trained {Transformers} be used in detecting complex sensitive sentences? - A {Monsanto} case study}. In \bibinfo{booktitle}{\emph{2021 Third {IEEE} International Conference on Trust, Privacy and Security in Intelligent Systems and Applications ({TPS}-{ISA})}}. \bibinfo{pages}{90--97}.
\newblock
\urldef\tempurl%
\url{https://doi.org/10.1109/TPSISA52974.2021.00010}
\showDOI{\tempurl}


\bibitem[To et~al\mbox{.}(2021)]%
        {10.1145/3411764.3445590}
\bibfield{author}{\bibinfo{person}{Alexandra To}, \bibinfo{person}{Hillary Carey}, \bibinfo{person}{Geoff Kaufman}, {and} \bibinfo{person}{Jessica Hammer}.} \bibinfo{year}{2021}\natexlab{}.
\newblock \showarticletitle{Reducing Uncertainty and Offering Comfort: Designing Technology for Coping with Interpersonal Racism}. In \bibinfo{booktitle}{\emph{Proceedings of the 2021 CHI Conference on Human Factors in Computing Systems}} (Yokohama, Japan) \emph{(\bibinfo{series}{CHI '21})}. \bibinfo{publisher}{Association for Computing Machinery}, \bibinfo{address}{New York, NY, USA}, Article \bibinfo{articleno}{398}, \bibinfo{numpages}{17}~pages.
\newblock
\showISBNx{9781450380966}
\urldef\tempurl%
\url{https://doi.org/10.1145/3411764.3445590}
\showDOI{\tempurl}


\bibitem[To et~al\mbox{.}(2020)]%
        {to_they_2020}
\bibfield{author}{\bibinfo{person}{Alexandra To}, \bibinfo{person}{Wenxia Sweeney}, \bibinfo{person}{Jessica Hammer}, {and} \bibinfo{person}{Geoff Kaufman}.} \bibinfo{year}{2020}\natexlab{}.
\newblock \showarticletitle{{``They Just Don't Get It": Towards Social Technologies for Coping with Interpersonal Racism}}.
\newblock \bibinfo{journal}{\emph{Proceedings of the ACM on Human-Computer Interaction}} \bibinfo{volume}{4}, \bibinfo{number}{CSCW1} (\bibinfo{year}{2020}), \bibinfo{pages}{1--29}.
\newblock


\bibitem[Utych(2018)]%
        {utych_negative_2018}
\bibfield{author}{\bibinfo{person}{Stephen~M Utych}.} \bibinfo{year}{2018}\natexlab{}.
\newblock \showarticletitle{Negative affective language in politics}.
\newblock \bibinfo{journal}{\emph{American Politics Research}} \bibinfo{volume}{46}, \bibinfo{number}{1} (\bibinfo{year}{2018}), \bibinfo{pages}{77--102}.
\newblock


\bibitem[Utz and Breuer(2016)]%
        {utz_informational_2016}
\bibfield{author}{\bibinfo{person}{Sonja Utz} {and} \bibinfo{person}{Johannes Breuer}.} \bibinfo{year}{2016}\natexlab{}.
\newblock \showarticletitle{Informational benefits from social media use for professional purposes: Results from a longitudinal study}.
\newblock \bibinfo{journal}{\emph{Cyberpsychology: Journal of Psychosocial Research on Cyberspace}} \bibinfo{volume}{10}, \bibinfo{number}{4} (\bibinfo{year}{2016}).
\newblock


\bibitem[Veale et~al\mbox{.}(2018)]%
        {10.1145/3173574.3174014}
\bibfield{author}{\bibinfo{person}{Michael Veale}, \bibinfo{person}{Max Van~Kleek}, {and} \bibinfo{person}{Reuben Binns}.} \bibinfo{year}{2018}\natexlab{}.
\newblock \showarticletitle{{Fairness and Accountability Design Needs for Algorithmic Support in High-Stakes Public Sector Decision-Making}}. In \bibinfo{booktitle}{\emph{Proceedings of the 2018 CHI Conference on Human Factors in Computing Systems}} (Montreal QC, Canada) \emph{(\bibinfo{series}{CHI '18})}. \bibinfo{publisher}{Association for Computing Machinery}, \bibinfo{address}{New York, NY, USA}, \bibinfo{pages}{1–14}.
\newblock
\showISBNx{9781450356206}
\urldef\tempurl%
\url{https://doi.org/10.1145/3173574.3174014}
\showDOI{\tempurl}


\bibitem[Vincent(2020)]%
        {vincent_2020}
\bibfield{author}{\bibinfo{person}{James Vincent}.} \bibinfo{year}{2020}\natexlab{}.
\newblock \bibinfo{title}{{Nextdoor CEO says it's `our fault' moderators deleted black lives matter posts}}.
\newblock
\newblock
\urldef\tempurl%
\url{https://www.theverge.com/2020/7/2/21311046/nextdoor-ceo-admits-fault-moderators-racial-bias-black-lives-matter}
\showURL{%
\tempurl}


\bibitem[Waseem et~al\mbox{.}(2017)]%
        {waseem-etal-2017-understanding}
\bibfield{author}{\bibinfo{person}{Zeerak Waseem}, \bibinfo{person}{Thomas Davidson}, \bibinfo{person}{Dana Warmsley}, {and} \bibinfo{person}{Ingmar Weber}.} \bibinfo{year}{2017}\natexlab{}.
\newblock \showarticletitle{Understanding Abuse: A Typology of Abusive Language Detection Subtasks}. In \bibinfo{booktitle}{\emph{Proceedings of the First Workshop on Abusive Language Online}}. \bibinfo{publisher}{Association for Computational Linguistics}, \bibinfo{address}{Vancouver, BC, Canada}, \bibinfo{pages}{78--84}.
\newblock
\urldef\tempurl%
\url{https://doi.org/10.18653/v1/W17-3012}
\showDOI{\tempurl}


\bibitem[Welch(2007)]%
        {doi:10.1177/1043986207306870}
\bibfield{author}{\bibinfo{person}{Kelly Welch}.} \bibinfo{year}{2007}\natexlab{}.
\newblock \showarticletitle{Black Criminal Stereotypes and Racial Profiling}.
\newblock \bibinfo{journal}{\emph{Journal of Contemporary Criminal Justice}} \bibinfo{volume}{23}, \bibinfo{number}{3} (\bibinfo{year}{2007}), \bibinfo{pages}{276--288}.
\newblock
\urldef\tempurl%
\url{https://doi.org/10.1177/1043986207306870}
\showDOI{\tempurl}
\showeprint{https://doi.org/10.1177/1043986207306870}


\bibitem[West(2018)]%
        {doi:10.1177/1461444818773059}
\bibfield{author}{\bibinfo{person}{Sarah~Myers West}.} \bibinfo{year}{2018}\natexlab{}.
\newblock \showarticletitle{Censored, suspended, shadowbanned: User interpretations of content moderation on social media platforms}.
\newblock \bibinfo{journal}{\emph{New Media \& Society}} \bibinfo{volume}{20}, \bibinfo{number}{11} (\bibinfo{year}{2018}), \bibinfo{pages}{4366--4383}.
\newblock
\urldef\tempurl%
\url{https://doi.org/10.1177/1461444818773059}
\showDOI{\tempurl}
\showeprint{https://doi.org/10.1177/1461444818773059}


\bibitem[Williams et~al\mbox{.}(2021)]%
        {williams_after_2021-1}
\bibfield{author}{\bibinfo{person}{Monnica~T Williams}, \bibinfo{person}{Matthew~D Skinta}, {and} \bibinfo{person}{Ren{\'e}e Martin-Willett}.} \bibinfo{year}{2021}\natexlab{}.
\newblock \showarticletitle{After Pierce and Sue: A revised racial microaggressions taxonomy}.
\newblock \bibinfo{journal}{\emph{Perspectives on Psychological Science}} \bibinfo{volume}{16}, \bibinfo{number}{5} (\bibinfo{year}{2021}), \bibinfo{pages}{991--1007}.
\newblock


\bibitem[Wohn(2019)]%
        {10.1145/3290605.3300390}
\bibfield{author}{\bibinfo{person}{Donghee~Yvette Wohn}.} \bibinfo{year}{2019}\natexlab{}.
\newblock \showarticletitle{{Volunteer Moderators in Twitch Micro Communities: How They Get Involved, the Roles They Play, and the Emotional Labor They Experience}}. In \bibinfo{booktitle}{\emph{Proceedings of the 2019 CHI Conference on Human Factors in Computing Systems}} (Glasgow, Scotland Uk) \emph{(\bibinfo{series}{CHI '19})}. \bibinfo{publisher}{Association for Computing Machinery}, \bibinfo{address}{New York, NY, USA}, \bibinfo{pages}{1–13}.
\newblock
\showISBNx{9781450359702}
\urldef\tempurl%
\url{https://doi.org/10.1145/3290605.3300390}
\showDOI{\tempurl}


\bibitem[Wright et~al\mbox{.}(2021)]%
        {recast-2}
\bibfield{author}{\bibinfo{person}{Austin Wright}, \bibinfo{person}{Omar Shaikh}, \bibinfo{person}{Haekyu Park}, \bibinfo{person}{Will Epperson}, \bibinfo{person}{Muhammed Ahmed}, \bibinfo{person}{Stephane Pinel}, \bibinfo{person}{Duen Chau}, {and} \bibinfo{person}{Diyi Yang}.} \bibinfo{year}{2021}\natexlab{}.
\newblock \showarticletitle{{RECAST: Enabling User Recourse and Interpretability of Toxicity Detection Models with Interactive Visualization}}.
\newblock \bibinfo{journal}{\emph{Proceedings of the ACM on Human-Computer Interaction}}  \bibinfo{volume}{5} (\bibinfo{date}{04} \bibinfo{year}{2021}), \bibinfo{pages}{1--26}.
\newblock
\urldef\tempurl%
\url{https://doi.org/10.1145/3449280}
\showDOI{\tempurl}


\bibitem[Wright et~al\mbox{.}(2020)]%
        {10.1145/3403676.3403691}
\bibfield{author}{\bibinfo{person}{Austin~P Wright}, \bibinfo{person}{Omar Shaikh}, \bibinfo{person}{Haekyu Park}, \bibinfo{person}{Will Epperson}, \bibinfo{person}{Muhammed Ahmed}, \bibinfo{person}{Stephane Pinel}, \bibinfo{person}{Diyi Yang}, {and} \bibinfo{person}{Duen~Horng Chau}.} \bibinfo{year}{2020}\natexlab{}.
\newblock \showarticletitle{RECAST: Interactive Auditing of Automatic Toxicity Detection Models}. In \bibinfo{booktitle}{\emph{The Eighth International Workshop of Chinese CHI}} (Honolulu, HI, USA) \emph{(\bibinfo{series}{Chinese CHI 2020})}. \bibinfo{publisher}{Association for Computing Machinery}, \bibinfo{address}{New York, NY, USA}, \bibinfo{pages}{80–82}.
\newblock
\showISBNx{9781450388153}
\urldef\tempurl%
\url{https://doi.org/10.1145/3403676.3403691}
\showDOI{\tempurl}


\bibitem[Wu et~al\mbox{.}(2022)]%
        {10.1145/3512901}
\bibfield{author}{\bibinfo{person}{Qunfang Wu}, \bibinfo{person}{Louisa~Kayah Williams}, \bibinfo{person}{Ellen Simpson}, {and} \bibinfo{person}{Bryan Semaan}.} \bibinfo{year}{2022}\natexlab{}.
\newblock \showarticletitle{{Conversations About Crime: Re-Enforcing and Fighting Against Platformed Racism on Reddit}}.
\newblock \bibinfo{journal}{\emph{Proc. ACM Hum.-Comput. Interact.}} \bibinfo{volume}{6}, \bibinfo{number}{CSCW1}, Article \bibinfo{articleno}{54} (\bibinfo{date}{apr} \bibinfo{year}{2022}), \bibinfo{numpages}{38}~pages.
\newblock
\urldef\tempurl%
\url{https://doi.org/10.1145/3512901}
\showDOI{\tempurl}


\bibitem[Xia et~al\mbox{.}(2020)]%
        {10.1145/3415179}
\bibfield{author}{\bibinfo{person}{Yan Xia}, \bibinfo{person}{Haiyi Zhu}, \bibinfo{person}{Tun Lu}, \bibinfo{person}{Peng Zhang}, {and} \bibinfo{person}{Ning Gu}.} \bibinfo{year}{2020}\natexlab{}.
\newblock \showarticletitle{{Exploring Antecedents and Consequences of Toxicity in Online Discussions: A Case Study on Reddit}}.
\newblock \bibinfo{journal}{\emph{Proc. ACM Hum.-Comput. Interact.}} \bibinfo{volume}{4}, \bibinfo{number}{CSCW2}, Article \bibinfo{articleno}{108} (\bibinfo{date}{oct} \bibinfo{year}{2020}), \bibinfo{numpages}{23}~pages.
\newblock
\urldef\tempurl%
\url{https://doi.org/10.1145/3415179}
\showDOI{\tempurl}


\bibitem[Yang et~al\mbox{.}(2020)]%
        {10.1145/3313831.3376301}
\bibfield{author}{\bibinfo{person}{Qian Yang}, \bibinfo{person}{Aaron Steinfeld}, \bibinfo{person}{Carolyn Ros\'{e}}, {and} \bibinfo{person}{John Zimmerman}.} \bibinfo{year}{2020}\natexlab{}.
\newblock \showarticletitle{{Re-Examining Whether, Why, and How Human-AI Interaction Is Uniquely Difficult to Design}}. In \bibinfo{booktitle}{\emph{Proceedings of the 2020 CHI Conference on Human Factors in Computing Systems}} (Honolulu, HI, USA) \emph{(\bibinfo{series}{CHI '20})}. \bibinfo{publisher}{Association for Computing Machinery}, \bibinfo{address}{New York, NY, USA}, \bibinfo{pages}{1–13}.
\newblock
\showISBNx{9781450367080}
\urldef\tempurl%
\url{https://doi.org/10.1145/3313831.3376301}
\showDOI{\tempurl}


\bibitem[Zannettou et~al\mbox{.}(2020)]%
        {zannettou2020measuring}
\bibfield{author}{\bibinfo{person}{Savvas Zannettou}, \bibinfo{person}{Mai ElSherief}, \bibinfo{person}{Elizabeth Belding}, \bibinfo{person}{Shirin Nilizadeh}, {and} \bibinfo{person}{Gianluca Stringhini}.} \bibinfo{year}{2020}\natexlab{}.
\newblock \showarticletitle{{Measuring and characterizing hate speech on news websites}}. In \bibinfo{booktitle}{\emph{12th ACM Conference on Web Science}}. \bibinfo{pages}{125--134}.
\newblock


\end{thebibliography}
\appendix
\section{Appendix}
\subsection{Annotation Guidelines} \label{appendix}
We analyze 13 of 16 themes of racial microaggressions from the revised Sue et al. (2009)'s taxonomy \cite{williams_after_2021-1}. Three of the 16 themes (tokenism, environmental exclusion, and environmental attacks) were excluded as they were not applicable to online contexts: the three themes pertain to situations in which people are present in a physical environment. For each of the 13 themes, we provide the definition from \cite{williams_after_2021-1} and examples of racial microaggressions from our RAMA corpus in Table \ref{tab:tablea}. We discussed these examples and definitions with workshop participants as an annotation guideline for labeling instances of acts of online racial microaggressions.

\begin{center}
\begin{table}[]
\resizebox{0.9\columnwidth}{!}{
\begin{tabular}{|c|c|l|l}
\cline{1-3}
\textbf{Theme} &
  \textbf{Definition} &
  \multicolumn{1}{c|}{\textbf{Examples}} &
   \\ \cline{1-3}
\begin{tabular}[c]{@{}c@{}}THEME-1: \\ Alien in own land /\\ Not a True Citizen\end{tabular} &
  \begin{tabular}[c]{@{}c@{}}When a question, statement, or behavior indicates\\ that a person of color is not a real citizen or\\  a meaningful part of society because they are not\\ White; Questioning the legitimacy of their identity.\end{tabular} &
  \begin{tabular}[c]{@{}l@{}}{}\\$\bullet$“If Adam and Eve are the first people in the Earth and they\\   are white, why are there Black people?”\\{}\\ $\bullet$“God gave black people rights in all corners of the globe\\  and then he made the Earth round.”\\{}\end{tabular} &
   \\ \cline{1-3}
\begin{tabular}[c]{@{}c@{}}THEME-2: \\ Racial Categorization \\ and Sameness\end{tabular} &
  \begin{tabular}[c]{@{}c@{}}{}\\ When a person is compelled to disclose their racial\\  group to enable others to attach pathological\\ racial stereotypes to the person; includes the assumption\\  that all people from a particular group are alike \\{}\end{tabular} &
  $\bullet$“All Black people look alike.” &
   \\ \cline{1-3}
\begin{tabular}[c]{@{}c@{}}THEME-3: \\ Assumptions about intelligence,\\  competence, or status\end{tabular} &
  \begin{tabular}[c]{@{}c@{}}{}\\When behavior or statements are based on an\\  assumption about  a person’s intelligence, competence,\\  education, income, or social status derived\\  from racial stereotypes. \\{}\end{tabular} &
  \begin{tabular}[c]{@{}l@{}}$\bullet$“How did Black people get so good at science? and why are they so athletic?”\end{tabular} &
   \\ \cline{1-3}
\begin{tabular}[c]{@{}c@{}}THEME-4:\\ Connecting via stereotypes\end{tabular} &
  \begin{tabular}[c]{@{}c@{}} {} \\When a person tries to communicate or connect\\  with a person through the use of stereotyped\\  speech or behavior to be accepted or understood; can\\  include racist jokes and epitaphs as terms of endearment.\\{}\end{tabular} &
  \begin{tabular}[c]{@{}l@{}}$\bullet$“Why do Black people love fried chicken and watermelon?”\\ {}\\$\bullet$“Why don’t Black people tip?”\end{tabular} &
   \\ \cline{1-3}
\begin{tabular}[c]{@{}c@{}}THEME-5:\\ False color blindness/ invalidating\\  racial or ethnic identity\end{tabular} &
  \begin{tabular}[c]{@{}c@{}}{} \\Expressing that an individual’s racial or ethnic identity\\  should not be acknowledged, which can be invalidating\\  for people who are proud of their identity\\  or who have suffered because of it.\\{}\end{tabular} &
  \begin{tabular}[c]{@{}l@{}}$\bullet$“I don't care if they are black, gay, green objects.”\\ $\bullet$“I'm white and I don't care, we're all the same human race.”\end{tabular} &
   \\ \cline{1-3}
\begin{tabular}[c]{@{}c@{}}THEME-6:\\ Myth of meritocracy/ \\ race is irrelevant for success\end{tabular} &
  \begin{tabular}[c]{@{}c@{}}{}\\When someone makes statements about success being\\  rooted in personal efforts and denial of the existence\\  of racism/White privilege; Statements which assert that\\  race does not play a role in succeeding in career\\  advancement or education.\\{}\end{tabular} &
  \begin{tabular}[c]{@{}l@{}}$\bullet$“Role should go to the best performer regardless of race.”\\ {}\\$\bullet$“Rich black people don’t face any form of systematic racism.”\end{tabular} &
   \\ \cline{1-3}
\begin{tabular}[c]{@{}c@{}}THEME-7:\\ Reverse-racism hostility\end{tabular} &
  \begin{tabular}[c]{@{}c@{}}Expressions of jealousy or hostility surrounding the\\  notion that POC get unfair advantages and benefits\\  because of their race.\end{tabular} &
  \begin{tabular}[c]{@{}l@{}}{}\\$\bullet$“I was fully qualified for the job, but they gave it to a Black girl.”\\ {}\\$\bullet$“Oh wait you’re Black; they practically guarantee you’d get into\\  that college.”\\{}\end{tabular} &
   \\ \cline{1-3}
\begin{tabular}[c]{@{}c@{}}THEME-8:\\ Criminality or dangerousness\end{tabular} &
  \begin{tabular}[c]{@{}c@{}}Demonstrating belief in stereotypes that POC are dangerous, \\ untrustworthy, and likely to commit crimes or cause \\ bodily harm; A person of color is presumed to be \\ dangerous, criminal, or deviant on the basis of their race.\end{tabular} &
  \begin{tabular}[c]{@{}l@{}}{}\\$\bullet$“I held back because he was Black” [user is speaking in the \\ context of avoiding conflict with a Black person out of fear of \\ physical retaliation].\\{}\\ $\bullet$"Black men are dangerous." \\ {} \end{tabular} &
   \\ \cline{1-3}
\begin{tabular}[c]{@{}c@{}}THEME-9:\\ Avoidance and distancing\end{tabular} &
  \begin{tabular}[c]{@{}c@{}}{} \\When POC are avoided or measures are taken to\\  prevent physical contact or close proximity.\\{}\end{tabular} &
  $\bullet$“When I see a Black person approaching me, I cross the street.” &
   \\ \cline{1-3}
\begin{tabular}[c]{@{}c@{}}THEME-10: \\ Denial of individual racism\end{tabular} &
  \begin{tabular}[c]{@{}c@{}}{} \\When a person tries to make a case that they are not biased, \\ often by talking about antiracist things they have done\\ to deflect perceived scrutiny of their own biased behaviors;\\  A statement made when Whites renounce their racial biases.\\{}\end{tabular} &
  \begin{tabular}[c]{@{}l@{}}$\bullet$“I’m not a racist. I have several Black friends.”\\{}\\ $\bullet$“I'm not racist but Black people make me uncomfortable.”\end{tabular} &
   \\ \cline{1-3}
\begin{tabular}[c]{@{}c@{}}THEME-11:\\ Pathologizing minority \\ culture or appearance\end{tabular} &
  \begin{tabular}[c]{@{}c@{}}{} \\When people criticize others on the basis of perceived\\ or real cultural differences in appearance, traditions, behaviors, \\ or preferences; The notion that the values and communication\\ styles of the dominant culture are ideal.\\{} \end{tabular} &
  \begin{tabular}[c]{@{}l@{}}$\bullet$ "Black kids shouldn’t dress that way."\\ {}\\$\bullet$ You’re pretty for a Black girl.\end{tabular} &
   \\ \cline{1-3}
\begin{tabular}[c]{@{}c@{}}THEME-12:\\ Exoticization and \\ eroticization\end{tabular} &
  \begin{tabular}[c]{@{}c@{}}{}\\When a person of color is treated according to\\  sexualized stereotypes or attention to differences\\ that are characterized as exotic in some way.\\{}\end{tabular} &
  \begin{tabular}[c]{@{}l@{}}{}\\$\bullet$“Black women are exotic. I have a fetish for Black women, am I racist?”\\{}\end{tabular} &
   \\ \cline{1-3}
\begin{tabular}[c]{@{}c@{}}THEME-13:\\ Second-class citizen/\\ ignored and invisible\end{tabular} &
  \begin{tabular}[c]{@{}c@{}}{}\\When POC are treated with less respect, consideration, or\\  care than is normally expected or customary; may include\\ being ignored or being unseen/ invisible; Occurs when a \\ White person   is given preferential treatment as a consumer\\  over a person of color.\\{}\end{tabular} &
  \begin{tabular}[c]{@{}l@{}}$\bullet$“Oh, sorry we kept you waiting so long. From your surname\\  on the form, we thought you were Black!”\\ {}\\$\bullet$“Black people and LGBT are untouchable.”\end{tabular} &
   \\ \cline{1-3}
\end{tabular}%
}
\caption{Themes and examples of acts of racial microaggressions from our RAMA corpus.}
\label{tab:tablea}
\end{table}
\end{center}
\vspace{-10pt}
\received{January 2023}
\received[revised]{July 2023}
\received[accepted]{November 2023}


\end{document}